\documentclass[iop, twocolappendix]{emulateapj}

\usepackage{amsmath, amssymb}
\usepackage{aas_macros}

\usepackage[backref, breaklinks, colorlinks, citecolor=blue, linkcolor=blue, urlcolor=blue]{hyperref}
\usepackage[all]{hypcap}

\newcommand{\figurewidth}{0.69}

\newcommand{\msun}{~\mbox{M}_\odot}

\newcommand{\mpc}{~\mbox{Mpc}}
\newcommand{\hmpc}{\,h^{-1}\,\mbox{Mpc}}

\newcommand{\kpc}{~\mbox{kpc}}

\newcommand{\pc}{~\mbox{pc}}
\newcommand{\cm}{~\mbox{cm}}
\newcommand{\gyr}{~\mbox{Gyr}}
\newcommand{\myr}{~\mbox{Myr}}

\newcommand{\K}{~\mbox{K}}
\newcommand{\erg}{~\mbox{erg}}

\newcommand{\omegamatter}{\Omega_{\rm matter}}
\newcommand{\omegabaryon}{\Omega_{\rm baryon}}
\newcommand{\omegalambda}{\Omega_{\rm \Lambda}}
\newcommand{\omegadarkmatter}{\Omega_{\rm dark\,matter}}

\newcommand{\rvir}{R_{\rm vir}}

\newcommand{\mthm}{M_{200{\rm m}}}
\newcommand{\rthm}{R_{200{\rm m}}}
\newcommand{\vthm}{V_{200{\rm m}}}

\newcommand{\rthc}{R_{200{\rm c}}}

\newcommand{\mdelta}{M_\Delta}
\newcommand{\rdelta}{R_\Delta}

\newcommand{\mstar}{M_{\rm star}}

\newcommand{\vcirc}{v_{\rm circ}}
\newcommand{\vrad}{v_{\rm rad}}
\newcommand{\vradave}{\bar{v}_{\rm rad}}
\newcommand{\vradavedark}{\bar{v}_{\rm rad}^{\rm dark}}
\newcommand{\vradavegas}{\bar{v}_{\rm rad}^{\rm gas}}
\newcommand{\vtan}{v_{\rm tan}}
\newcommand{\vtanave}{\bar{v}_{\rm tan}}

\newcommand{\rsplashback}{R_{\rm splashback}}
\newcommand{\rinfall}{R_{\rm infall}}

\newcommand{\rhodark}{\rho_{\rm dark}}
\newcommand{\rhogas}{\rho_{\rm gas}}
\newcommand{\rhobaryon}{\rho_{\rm baryon}}
\newcommand{\fbaryon}{f_{\rm baryon}}
\newcommand{\mdark}{m_{\rm dark}}

\newcommand{\mbaryon}{m_{\rm baryon}}

\begin{document}

\journalinfo{The Astrophysical Journal, accepted}

\title{The physical nature of the cosmic accretion of baryons and dark matter into halos and their galaxies}
\shorttitle{Cosmic accretion into halos and galaxies}

\author{Andrew R. Wetzel \altaffilmark{1, 2, 3}}
\author{Daisuke Nagai \altaffilmark{3, 4, 5}}
\shortauthors{Wetzel \& Nagai}

\altaffiltext{1}{TAPIR, California Institute of Technology, Pasadena, CA, USA}
\altaffiltext{2}{Carnegie Observatories, Pasadena, CA, USA}
\altaffiltext{3}{Department of Astronomy, Yale University, New Haven, CT, USA}
\altaffiltext{4}{Department of Physics, Yale University, New Haven, CT, USA}
\altaffiltext{5}{Yale Center for Astronomy \& Astrophysics, Yale University, New Haven, CT, USA}

\begin{abstract}
The cosmic accretion of both dark matter and baryons into halos typically is measured using some evolving virial relation, but recent work suggests that most halo growth at late cosmic time ($z\lesssim2$) is not physical but is rather the by-product of an evolving virial radius (``pseudo-evolution'').
Using \textit{Omega25}, a suite of cosmological simulations that incorporate both dark matter and gas dynamics with differing treatments of gas cooling, star formation, and thermal feedback, we systematically explore the physics that governs cosmic accretion into halos and their galaxies.
Physically meaningful cosmic accretion of both dark matter and baryons occurs at $z\gtrsim1$ across our halo mass range: $\mthm=10^{11-14}\msun$.
However, dark matter, because it is dissipationless, is deposited (in a time-average sense) at $\gtrsim\rthm(z)$ in a shell-like manner, such that dark-matter mass and density experience little-to-no physical growth at \textit{any} radius within a halo at $z<1$.
In contrast, gas, because it is able to cool radiatively, experiences significant accretion at all radii, at a rate that roughly tracks the accretion rate at $\rthm$, at all redshifts.
Infalling gas starts to decouple from dark matter at $\approx2\,\rthm$ and continues to accrete to smaller radii until the onset of strong angular-momentum support at $\approx0.1\,\rthm$.
Thus, while the growth of dark matter is subject to pseudo-evolution, the growth of baryons is not.
The fact that the accretion rate of gas on galactic scales tracks the accretion rate near $\rthm$ provides insight into the tight relations between the masses/sizes of galaxies and those of their host halos across cosmic time.
\end{abstract}


\section{Introduction}

In the paradigm of cosmological structure formation, gravitationally bound halos form at the peaks of the primordial density field as dark matter and baryons undergo nonlinear gravitational collapse.
Dark matter, being collisionless and dissipationless, conserves its orbital energy, remaining in an extended dispersion-supported profile with overlapping inward- and outward-moving orbits \citep{GunnGott1972, Gott1975, Gunn1977, FillmoreGoldreich1984, ColeLacey1996}.
In contrast, gas collides, shocks, mixes, and eventually dissipates energy via radiative cooling, causing it to collapse to the minimum of a halo's potential well and seed the formation of stars and galaxies \citep{WhiteRees1978, FallEfstathiou1980, Blumenthal1986, Dubinski1994, Mo1998}.

Within this paradigm, debate persists about the most physically meaningful ways to describe the physical extent of a halo, the rate of cosmic accretion into a halo, and the amount of mass growth within a halo, including how these compare for dark matter versus baryons.
These are important questions because measurements of cosmic accretion and mass growth depend sensitively on how and where one measures them.
Thus, understanding the evolution of halos requires a detailed understanding of the relevant physical scales across cosmic time, including the physical meaning (if any) of a choice for a halo's virial boundary/edge.
Furthermore, because cosmic accretion into a halo feeds the growth of the galaxy inside, understanding the physics of all of these scales is necessary for developing a physical picture of galaxy evolution in a cosmological context.

Many works have examined the nature of dark-matter accretion into halos.
Most previous works measured cosmic accretion or mass growth according to some evolving virial radius, $\rvir$, that was linked to the background density of the universe \citep[for example,][]{Wechsler2002, Zhao2003}.
However, recent works have questioned the physical meaning of the commonly used halo virial radius and its implications for inferred cosmological accretion and mass growth \citep{Busha2005, Prada2006, Diemand2007, Cuesta2008}.
Specifically, these works found that low-mass halos experience little-to-no significant physical growth of dark matter at fixed physical radii, especially at late cosmic time ($z \lesssim 2$).
Moreover, the recent work by \citet{Diemer2013} found that most of the growth of dark-matter mass at halo masses $\lesssim 10 ^ {13} \msun$ arises because one measures halo mass within some virial radius that is tied to a reference background density that evolves with time, which in turn can lead to inferred growth of halo mass even if the physical density profile of the halo remains constant, an effect that they called ``pseudo-evolution''.
This suggests that there is little-to-no \textit{physical} accretion into lower-mass halos at late cosmic time, such that they effectively evolve as ``island'' halos, divorced from the cosmic background, at $z \lesssim 1$.

Pseudo-evolution has a number of important implications for understanding phenomenological links between galaxies and dark-matter halos as well as for models of galaxy evolution.
For example, observations are probing the relation between the stellar mass of a galaxy and the virial mass of its host halo, including its evolution with time \citep[for example,][]{Leauthaud2012, Yang2012, Behroozi2013c, Hudson2015}, finding that the relation between galaxy mass and halo mass evolves only weakly at $z < 1$ (and possibly at higher $z$), which suggests that galaxy mass evolves largely in sync with halo mass.
Similarly, \citet{Kravtsov2013} found a tight linear relation between the size of a galaxy and the size of its host halo across a wide range of masses at $z \approx 0$, despite the significantly varying ratio of stellar-to-halo mass across this range.
These studies suggest that the mass and size of a galaxy is set by, or at least responds to, that of its host halo, but these relations are meaningful only insofar as one uses a physically sensible radius (and thus mass) for a halo.
Moreover, many (semi-analytic) models of galaxy evolution try to link the accretion rate of a halo in simulations to its baryonic accretion rate and in turn to the star formation rate of the galaxy \citep[for example,][and references therein]{Bouche2010, Benson2010, Lilly2013, Lu2014}.
Halo mass growth that is incorrectly attributed from pseudo-evolution would affect all of these analyses.

However, it remains unclear what role pseudo-evolution plays in the cosmic accretion of baryons, because gas dynamics can be markedly different: gas is collisional, so it can shock and mix, and it also can dissipate energy via radiative cooling.
A number of works have examined the accretion rates of gas into galaxies \citep[for example,][]{Ocvirk2008, Keres2009} and halos \citep[for example,][]{vandeVoort2011, FaucherGiguere2011, Dekel2013, Woods2014, Nelson2015}, though in almost all cases they measured mass growth/flux at some predefined and evolving virial radius.
Few works have compared in detail the specific accretion rates of gas versus dark matter:
\citet{vandeVoort2011} and \citet{FaucherGiguere2011} both found that specific accretion rates near the virial radius were broadly similar for baryons and dark matter, with some (up to a factor of 2) reduction in the specific rate for baryons, depending on the model for stellar winds; but they both found significantly reduced baryon accretion rates at small radii near the galaxy compared to that near the virial radius.
However, all of these works used differing techniques to measure accretion and at somewhat different choices for virial radii.
Furthermore, many of these works focused on the role of feedback from stars and/or black holes on baryonic accretion rates, using various phenomenological models for driving stellar winds.
While there is general consensus that stellar winds can alter accretion rates into the galaxies, results are mixed regarding the regulation of gas accretion at larger radii within a halo.
However, all of these works found \textit{some} level of gas accretion into galaxies at late cosmic time, implying that the pseudo-evolution of dark matter may not extend to gas.
Indeed, \citet{Dekel2013}, examined baryonic accretion rates in a suite of cosmological zoom-in simulations at $z > 1$ and found that the mass inflow rate at $0.1 \, \rvir$ is broadly similar to ($\sim 50\%$) that at $\rvir$; they found similar trends examining the rate at fixed physical radii of 10 and $100 \kpc$, implying that baryon accretion to small radii is indeed physical.

More generally, the physical nature of cosmic accretion has a variety of implications for the evolution of gas in halos and galaxies at late cosmic time, especially for low-mass galaxies.
For example, galaxies at $\mstar \lesssim 10 ^ {10.5} \msun$ formed $> 60\%$ of their mass since $z = 1$ \citep{Leitner2012}, and the rate of decline of the cosmic density of star formation broadly mimics the decline of dark-matter accretion rates into halos at fixed mass \citep[for example,][]{Bouche2010, Lilly2013}.
However, it is not clear how much star formation and galaxy growth at $z < 1$ is linked to cosmic accretion, as opposed to the consumption of gas that already is within galaxies, given that (molecular) gas fractions are observed to decrease over time \citep[for example][]{Bauermeister2013}, or the recycling of gas in galaxies from stellar winds \citep[for example,][]{Oppenheimer2010, LeitnerKravtsov2011}.
Additionally, cosmic gas accretion into halos drives the evolution of extended gas around galaxies, referred to as the circumgalactic medium, as many observations and surveys now are probing \citep[for example,][]{Rudie2012, Tumlinson2013}.
For such studies it is important to understand both the most physically meaningful virial definition to use for a halo as well as the rate of accretion of (relatively unenriched) gas and how it propagates to smaller radii.

The primary goal of this work is to understand the physical nature of cosmic accretion into halos and how it propagates down to scales of the galaxy inside, for both baryons and dark matter.
In particular, we aim to bridge the gap between detailed studies of halo growth, typically based on dark-matter-only simulations, and detailed studies of cosmic gas accretion into galaxies.
More specifically, we seek to understand the significance of pseudo-evolution not only for dark-matter accretion, but also its role in baryon accretion and hence galaxy growth.
We focus on halos of mass $10 ^ {11 - 13} \msun$ at late cosmic time ($z < 2$).
Such halos are observed to host galaxies with $\mstar \approx 10 ^ {9 - 11} \msun$ ($\mstar$ in our simulation with star formation is $\approx 2 - 4 \times$ higher because of overcooling; see \autoref{sec:simulation}).
These mass and redshift regimes are where the effects of pseudo-evolution are particularly strong \citep{Diemer2013}, where observations constrain well the relation between galaxies and their host halos as well as gas in/around galaxies.
We use simulations with varying treatments of gas physics, some including star formation and thermal feedback, though our simulations only marginally resolve the scales within galaxies, and our prescription for stellar feedback does not drive particularly strong winds out of galaxies, which play a strong role in regulating accretion into the galaxy itself.
Thus, we focus on cosmic accretion and mass growth on scales within a halo, but we do not investigate accretion into the galaxy, or stellar mass growth, directly.
We defer such work to a follow-up analysis.

Throughout, we cite all masses using $h = 0.7$ for the dimensionless Hubble parameter.

\section{Theory of Halo Collapse}
\label{sec:halo_collapse}

We first review the basic theoretical framework for \textit{spherical} collapse and virialization of a halo, to set the stage for and aid in the interpretation of our numerical results.
In the standard model \citep{GunnGott1972, Gunn1977, FillmoreGoldreich1984, ColeLacey1996}, if a spherical region is sufficiently overdense, its gravitational self-attraction overcomes the initial cosmological expansion, such that a mass shell will reach a maximum radius and then collapse.
Specifically, for flat $\Lambda$CDM cosmology, the radial acceleration around some overdense region is
\begin{equation}
\frac{{\rm d} ^ 2 r}{\rm {d} t ^ 2} = - \frac{G \, m(< r)}{r ^ 2} + \frac{8 \pi G}{3} \rho_\Lambda r
\label{eq:acceleration}
\end{equation}
in which $r$ is the physical radius from the center of the overdensity, $m(< r)$ is the enclosed mass, $G$ is the gravitational constant, and $\rho_\Lambda$ is the (constant) physical density of dark energy.
Around the overdense region, this acceleration counteracts the initial outward-moving velocity that is set by the Hubble expansion, and if a shell experiences turn-around, it decouples from the Hubble flow, at which point it no longer ``feels'' the expansion of the universe, modulo the acceleration from dark energy in \autoref{eq:acceleration}.
We refer to the radius of this first turn-around as $r_{\rm ta}$.

In Einstein-de Sitter cosmology ($\omegamatter = 1$, $\omegalambda = 0$), turn-around occurs when the average density within the sphere is $\simeq 5.6 \times$ that of the background.
Assuming the virial theorem for the equilibrium state of the halo, such that $K = -1 / 2 \, U$, for which $K$ and $U$ are the kinetic and potential energies, its final $r$ (density) will be $2 (8) \times$ that at $r_{\rm ta}$.
Over this period of collapse, the background universe has expanded to become $4 \times$ less dense, so a ``virialized'' halo has an average density of $18 \pi ^ 2 = 178 \times$ that of background, a value that many authors round to 200.
(See \citealt{BryanNorman1998} for a generalization of this model to $\omegalambda > 0$.)

More generally, one can define a halo's virial radius, $\rdelta$, such that the average interior density is $\Delta$ times some reference density, $\rho_{\rm ref}$: $\mdelta = \Delta \frac{4}{3} \pi \rho_{\rm ref} \rdelta ^ 3$.
In this work, we use $\Delta = 200$m, that is, we define halos as containing $200 \times$ the average matter density of the universe.
(We will compare other virial definitions in the context of the results of this paper in future work.)

While the above model describes halo collapse in terms of instantaneous energetics, the actual physics of collapse is more complicated, given that halos experience ongoing accretion, so they almost never are well-relaxed virialized systems, especially at $r \sim \rthm$.
First we consider dark matter.
After reaching $r_{\rm ta}$, a shell continues to collapse until reaching its first pericenter, after which it orbits back out to its apocenter, or secondary turn-around radius, $r_{\rm ta,\,2}$.
If there were no change in $m(< r_{\rm ta,\,2})$ throughout this full orbit, energy conservation implies that $r_{\rm ta,\,2} = r_{\rm ta}$.
In reality, the continued accretion of mass shells from larger $r$ causes the potential to deepen, so $r_{\rm ta,\,2} < r_{\rm ta}$: the higher the accretion rate, the more that $r_{\rm ta,\,2} < r_{\rm ta}$ \citep{GunnGott1972, Gott1975, Gunn1977, FillmoreGoldreich1984, DiemerKravtsov2014, Adhikari2014}.
Throughout this orbit, a shell passes through other shells that collapsed at different times, because dark matter is collisionless.
Thus, a halo represents a superposition of shells of inward- and outward-moving orbits that have collapsed at different epochs.
At any time, there is an outermost $r_{\rm ta,\,2}$, which corresponds to the shell that is reaching its $r_{\rm ta,\,2}$ for the first time.
We refer to this as the splashback radius, $\rsplashback$, and it corresponds to the maximum $r$ of matter that has passed through the core of the halo.
Thus, a halo's density profile declines rapidly beyond $\rsplashback$, and typically $\rsplashback = (0.8 - 2) \, \rthm$; it is smaller for halos with higher rates of accretion \citep{DiemerKravtsov2014, Adhikari2014}.

The physics of gas accretion is different, because gas is a collisional fluid, so there are no shell crossings of orbits, and there is no splashback radius.
Thus, after reaching $r_{\rm ta}$, gas collapses (typically supersonically) until it encounters a previously collapsed gas shell, at which point it shocks and heats \citep{WhiteRees1978, Blumenthal1986, Dubinski1994, Mo1998}.
Thus, while gas has no splashback radius, it can have a virial-shock radius, although this can occur at $r$ much smaller (or larger) than $\rthm$, including near the galaxy itself \citep{DekelBirnboim2006, Ocvirk2008}.
If shocked, gas is heated to near the halo's virial temperature, at which point it is supported by thermal pressure.
If the timescale for gas cooling is longer than the dynamical time at the given radius, the gas will remain in near hydrostatic equilibrium, which typically is true at $\mthm \gtrsim 10 ^ {11.6} \msun$ \citep{DekelBirnboim2006, Ocvirk2008}.
In lower-mass halos, even if gas is shock-heated to near the virial temperature \citep{Joung2012, Nelson2015}, it will cool rapidly and advect to smaller $r$, while remaining relatively cold \citep{Ocvirk2008, Keres2009}.

We emphasize that the above picture is valid for halo collapse that is purely spherical and smooth.
In reality, cosmological collapse is triaxial and clumpy.
For dark matter, this means that $\rsplashback$ is smeared out \citep[for example,][]{Adhikari2014}.
For gas, triaxial collapse can drive turbulence with effective pressure support, and gas can mix via Kelvin-Helmholtz and Rayleigh-Taylor instabilities.
For both components, angular-momentum support will regulate radial advection, and infalling satellite halos can persist as bound subhalos that behave neither purely collisional nor collisionless.
Thus, while the above picture is informative, it is not necessarily true in detail, and one needs to use cosmological simulations to model these processes fully, as we now describe.

\section{Numerical Methods}
\label{sec:methods}

\subsection{Simulations}
\label{sec:simulation}

To study the accretion of dark matter and gas in a realistic cosmological context, we performed and analyzed a suite of cosmological simulations that we call {\it ``Omega25''} using the Adaptive Refinement Tree (ART) Eulerian $N$-body plus hydrodynamics code \citep{Kravtsov1999, Kravtsov2002, Rudd2008} and the {\it Omega} high-performance computing cluster at Yale University.
These simulations include collisionless dynamics of dark matter and stars as well as gas dynamics in a cubical volume of comoving size $25 \hmpc = 36 \mpc$ in a flat $\Lambda$CDM cosmology: $\omegamatter = 0.27$, $\omegabaryon = 0.047$, $h = 0.7$, $n_{\rm s} = 0.95$, and $\sigma_8 = 0.82$.

ART uses adaptive refinement in space and time to reach the high dynamic range required to resolve galaxies and their halos in cosmological simulations.
We use simulations at two different resolutions: $128 ^ 3$ and $256 ^ 3$ root mesh cells/dark-matter particles.
Both runs use a maximum of 8 levels of adaptive refinement in the mesh.
The dark matter particle masses are $6.6 \times 10 ^ 8, 8.2 \times 10 ^ 7 \msun$, and the minimum mesh cell sizes are $1.09, 0.54 \kpc$ comoving, respectively.
We generate initial conditions at $z = 81$, and the number of time steps in the root grid to $z = 0$ is $\sim 500$ and $\sim 1000$ for the $128 ^ 3$ and $256 ^ 3$ runs, respectively.
Each level of adaptive refinement is a factor of 1 - 4 (typically 2) higher in time resolution.
We save 60 snapshots spaced evenly in $\log(1 + z)$ from $z = 9$ to 0, leading to a snapshot time resolution of $130 - 250 \myr$.
We present results only from the higher-resolution $256 ^ 3$ simulations, having used the lower-resolution simulations to ensure that our results do not depend significantly on resolution within the ranges of halo mass ($\mthm > 10 ^ {11} \msun$) and radius ($r > 7 \kpc$) that we examine.

In order to assess the effects of gas cooling and star formation on the dynamics of baryon accretion, we conducted each simulation with four different prescriptions for the inclusion of gas dynamics, star formation, and feedback, as follows:
\begin{enumerate}
\renewcommand{\labelenumi}{(\alph{enumi})}
\item including only dark matter,
\item additionally including gas without radiative cooling,
\item additionally turning on radiative cooling for primordial gas in the presence of a cosmic ultraviolet background (without star formation),
\item additionally including star formation, thermal feedback, metal enrichment, and metal-line cooling, as detailed below.
\end{enumerate}

We use equilibrium gas cooling and heating rates that incorporate Compton heating and cooling, heating from a cosmic ultraviolet ionizing background \citep{HaardtMadau1996}, and atomic cooling including metals.
We use Cloudy \citep{Ferland1998} to tabulate these for the temperature range of $10 ^ {2 - 9} \K$ and a grid of metallicities and ultraviolet intensities.

Our model for star formation and feedback is an extension of that of previous works using ART \citep{Nagai2007b, LeitnerKravtsov2011}, based on an empirically motivated efficiency and dependence on gas density:
\begin{equation}
\dot{\rho}_{\rm star} = \frac{\rho_{\rm gas}}{\tau_{\rm star}} \left( \frac{\rho_{\rm gas}}{0.01 \msun \pc ^ {-3}} \right) ^ {0.5} \, ,
\end{equation}
with $\tau_{\rm star} = 3 \gyr$.
We allow star particles of minimum mass $5 \times 10 ^ 5 \msun$ to form in cells with number density $> 0.5 \cm ^ {-3}$ and temperature $< 9000 \K$.
Such temperature and density thresholds are reasonable for simulations at our moderate resolution \citep{Saitoh2008}.

We model each newly formed star particle as a single stellar population with an initial mass function from \citet{Chabrier2003} in the range of $0.1 - 100 \msun$.
All stars with $\mstar > 8 \msun$ immediately deposit $2 \times 10 ^ {51} \erg$ of thermal energy into the gas of their host cell, accounting for energy input by stellar winds and type II (core-collapse) supernovae.
Moreover, they deposit min(0.2, $0.01 \mstar / \msun - 0.06$) of their mass as metals into their host cell.
In addition, we account for delayed supernova Ia, assuming that a fraction of 0.015 of the stellar mass at $(3 - 8) \msun$ explodes over the entire history of the population, with each event dumping $2 \times 10 ^ {51} \erg$ of thermal energy and ejecting $1.3 \msun$ of metals into the gas of the host cell.
Note that we do not add explicit momentum flux from supernovae or stellar winds.

Finally, we model mass loss from stellar winds for each star particle assuming that the cumulative fraction of mass lost at time $t$ since its birth is $f(t) = 0.05 \ln \left[ (t / 5 \myr) + 1 \right]$.
At each time step, this mass loss is added to the gas mass of its host cell along with its energy and momentum.
With these parameters, about 40\% of initial stellar mass is lost over a Hubble time \citep[see][for more details]{LeitnerKravtsov2011}

Our goal in this paper is to explore systematically the effects of gas and stellar physics on cosmological accretion rates.
Our prescription for stellar feedback does not generate strong outflows that are encompassed in more detailed simulations of galaxy evolution \citep[for example,][]{Governato2010, AgertzKravtsov2014, Hopkins2014a}.
As a result, our galaxies experience overcooling and produce $2 - 4$ times too many stars compared to observational constraints at our halo masses.
However, the physics of these outflows and the coupling to the surrounding halo gas remain areas of active investigation and debate.
Instead, our goal in this work is to understand the importance (or lack thereof) of cosmic accretion for dark matter and baryons absent strong outflows.
In a follow-up analysis, we will use higher-resolution zoom-in simulations to explore the role of more detailed outflows in regulating gas accretion.

\subsection{Halo Finding and Tracking}

We identify halos using a variant of the method in \citet{Tinker2008}.
We first find peaks in the dark-matter distribution by measuring the local density of each particle using a smoothing kernel based on its 256 nearest neighboring particles.\footnote{We choose this smoothing based on extensive tests, ensuring that it accurately centers on the minimum of the halo's potential well and not spuriously on a smaller but denser subhalo.}
Starting with the densest peak in the simulation, we grow a sphere around it until the average density, including all matter components (dark matter, gas, and stars), interior to the radius $\rthm$ is $200 \times \rho_{\rm matter}(z)$, the average matter density of the universe.
Excluding all other density peaks within $\rthm$, we repeat the procedure for the next densest dark matter particle until we have identified all isolated centers.

For each halo, we identify its main progenitor at the previous simulation snapshot.
First, at all snapshots, we identify 30\% of the most bound dark-matter particles in each halo.
Then, we link halos across adjacent snapshots that share these particles.
We identify the main progenitor as the halo at the previous snapshot that shares the most number of particles, and we follow back this progenitor link back across each snapshot to identify the main progenitor at a given $z$.

\subsection{Halo Selection}

For all analyses, we first select halos at $z = 0$ in bins of $\mthm$.
We focus on $\mthm(z = 0) = 10 ^ {11 - 12}$ and $10 ^ {12 - 13} \msun$; each simulation contains $\approx 700$ and 100 such halos, respectively.
Each simulation also contains 12 group-mass halos with $\mthm(z = 0) = 10 ^ {13 - 14} \msun$.
While we show results at this mass for our dark-matter simulation, we choose not to examine this mass range for our simulations that include hydrodynamics, given that the feedback from supermassive black holes, which we do not model, likely plays an increasingly important role in gas thermodynamics.
All of our halos contain at least 1200 dark-matter particles within $\rthm$ at $z = 0$, and we show profiles down to a radius of $7 \kpc$, which corresponds to 14 mesh cells at the highest refinement.

We focus on the formation histories of isolated halos by selecting only halos at $z = 0$ whose center lies $> 2 \, \rthm$ from the center of any more massive halo (for which $\rthm$ is that of the neighboring halo).
Thus, we seek to exclude strong environmental effects from neighboring halos, including tidal stripping, which typically starts at $\sim 2 \, \rthm$ beyond a more massive halo \citep{Behroozi2014}, as well as contamination from halos that used to be satellite (sub)halos but orbited beyond $\rthm$ after infall and are highly stripped \citep{Wetzel2014}.

In summary, we examine how isolated halos at $z = 0$ have evolved since $z = 2$, when the effects of pseudo-evolution in dark matter are strongest \citep{Diemer2013}.

\section{Physical accretion of dark matter}
\label{sec:dark_matter}

\renewcommand{\figurewidth}{0.69}
\begin{figure*}
\centering
{\large Simulation with only Dark Matter} \\
\includegraphics[width = \figurewidth \columnwidth]{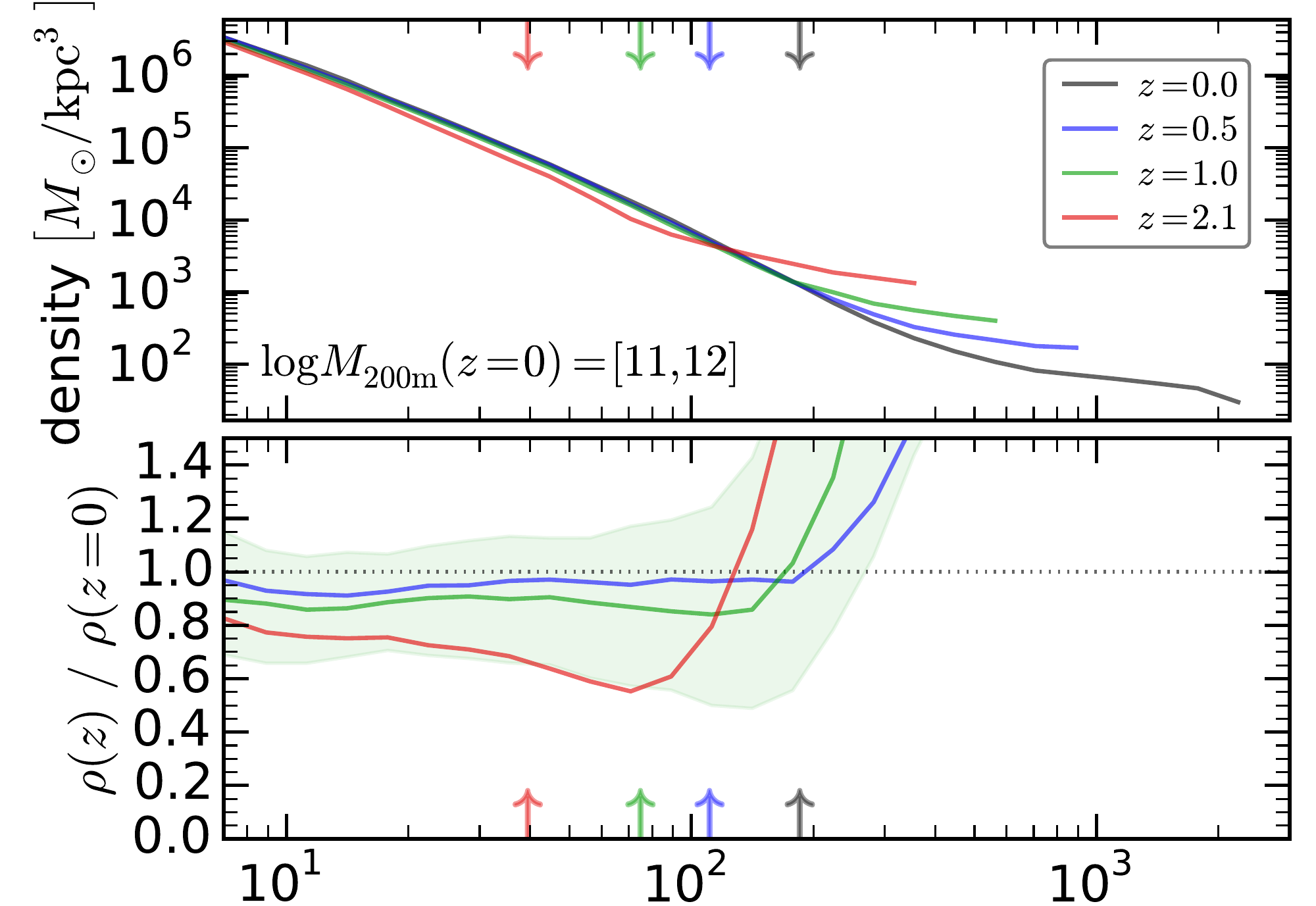}
\includegraphics[width = \figurewidth \columnwidth]{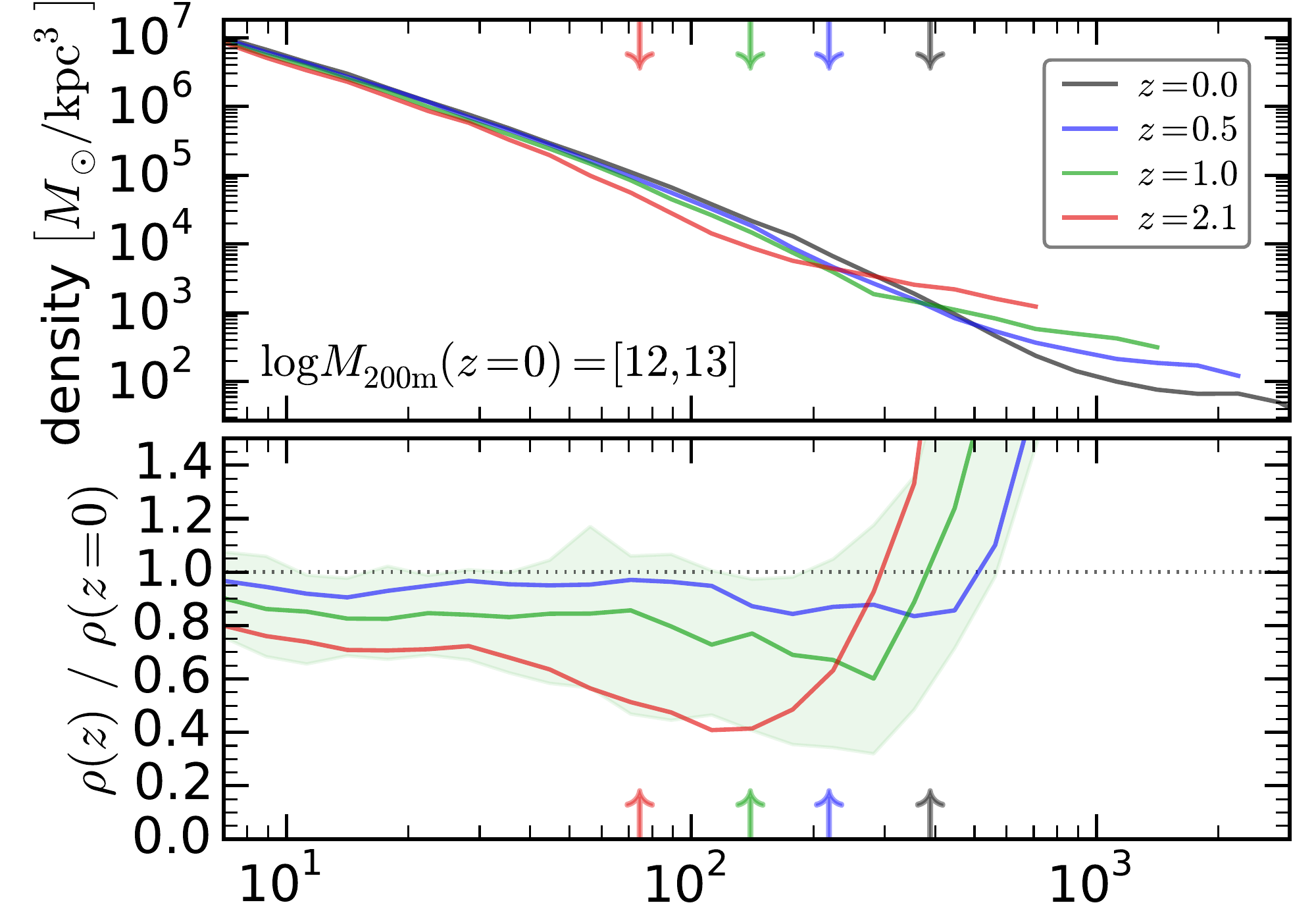}
\includegraphics[width = \figurewidth \columnwidth]{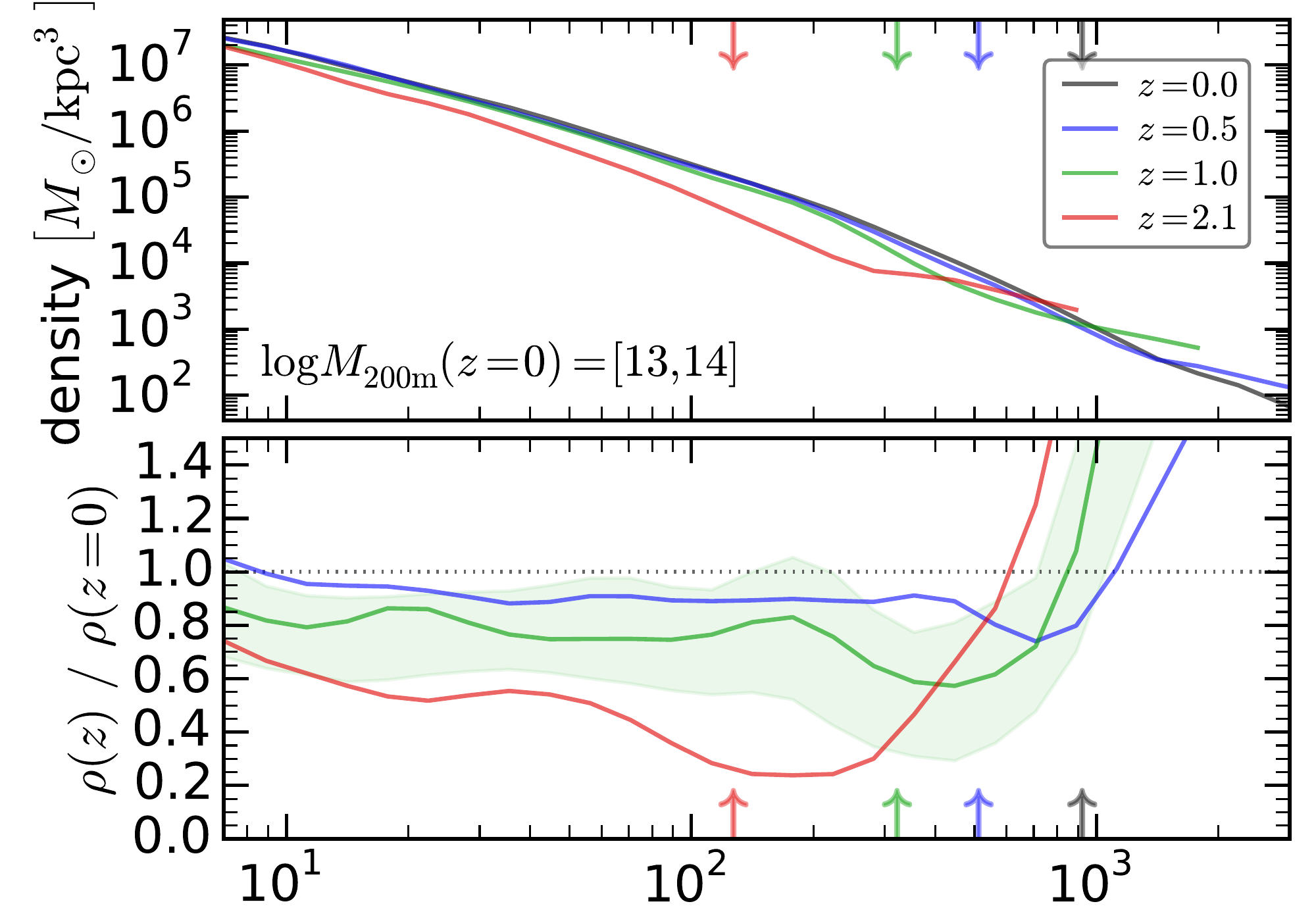}
\includegraphics[width = \figurewidth \columnwidth]{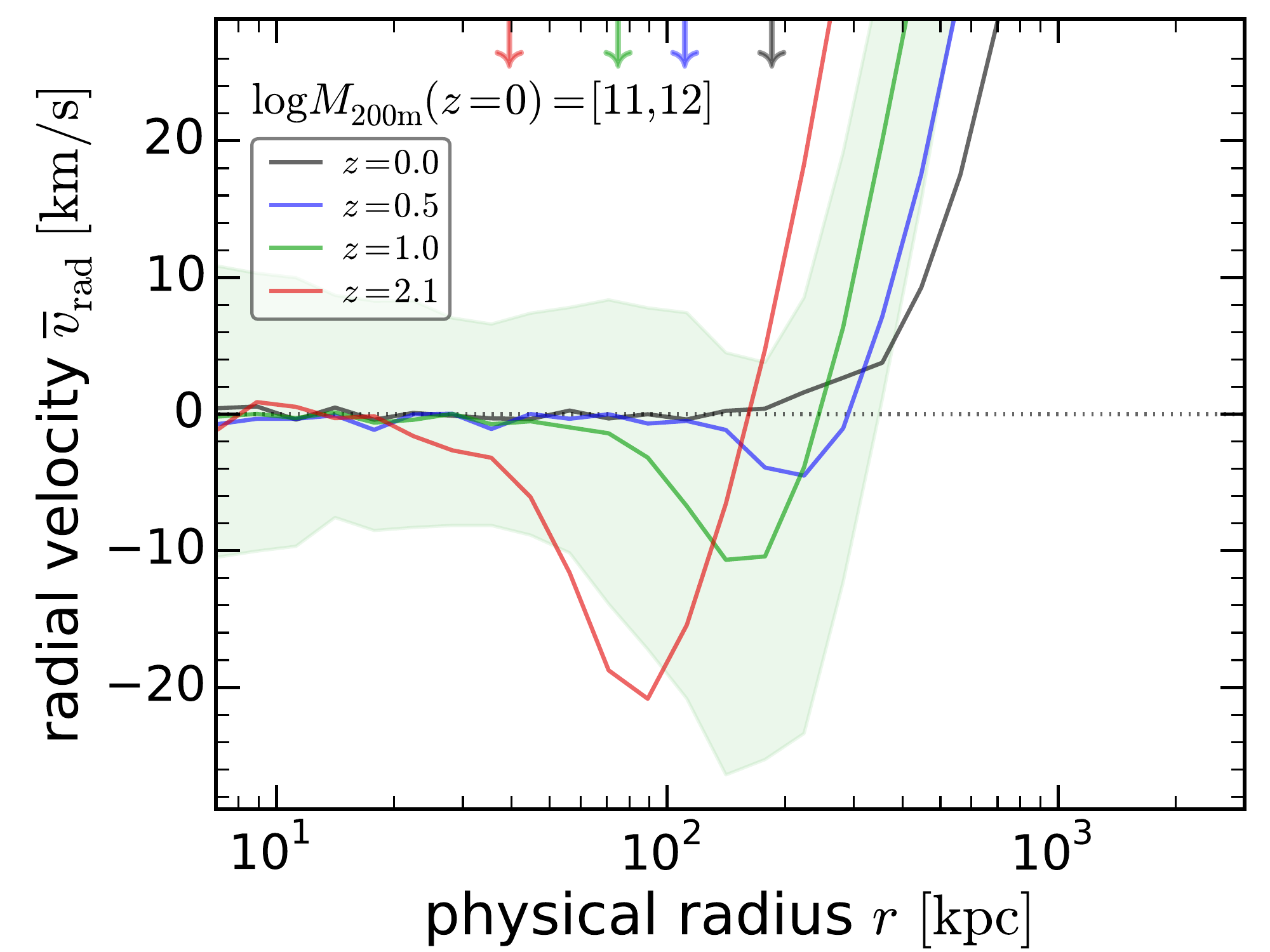}
\includegraphics[width = \figurewidth \columnwidth]{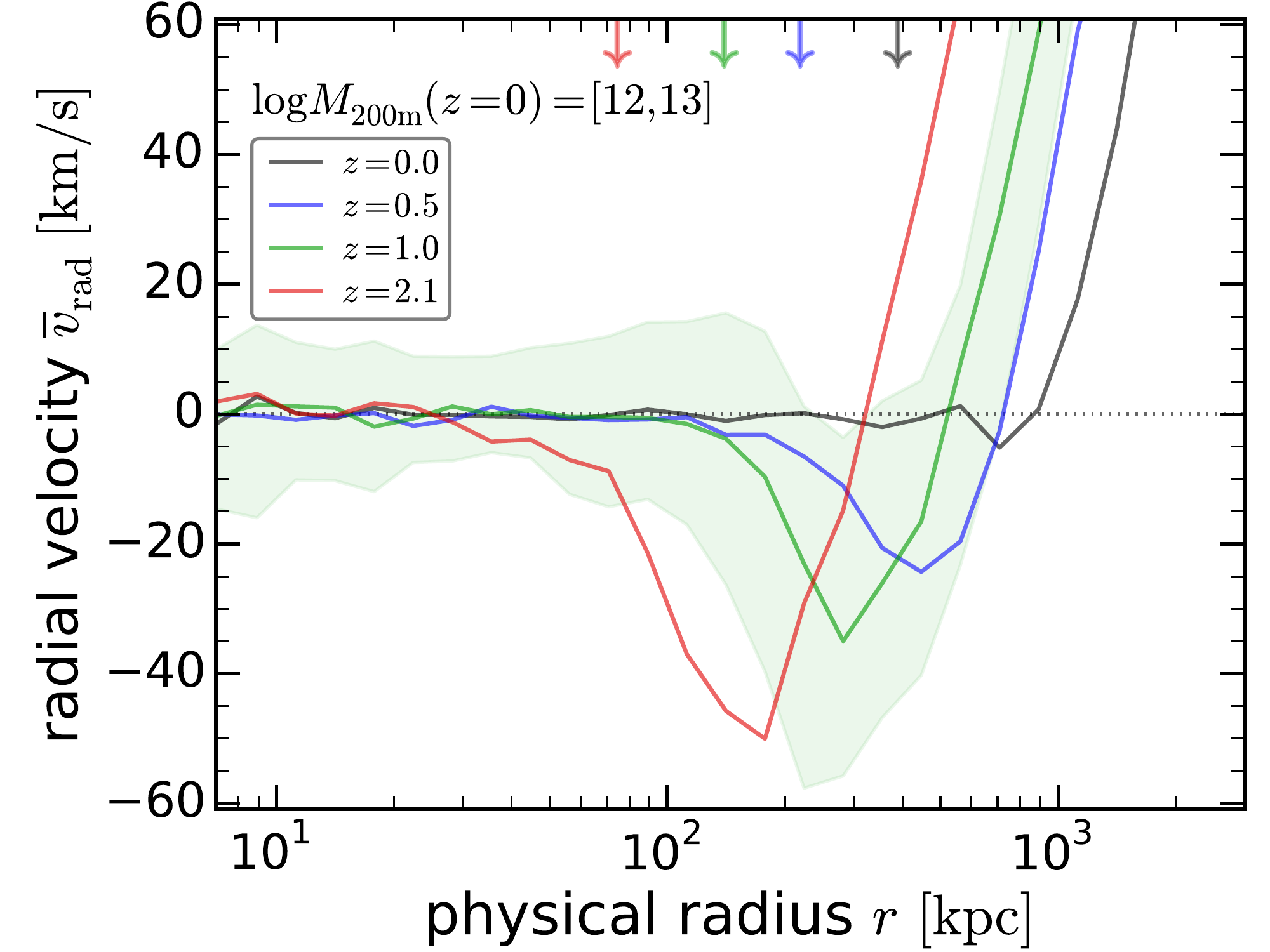}
\includegraphics[width = \figurewidth \columnwidth]{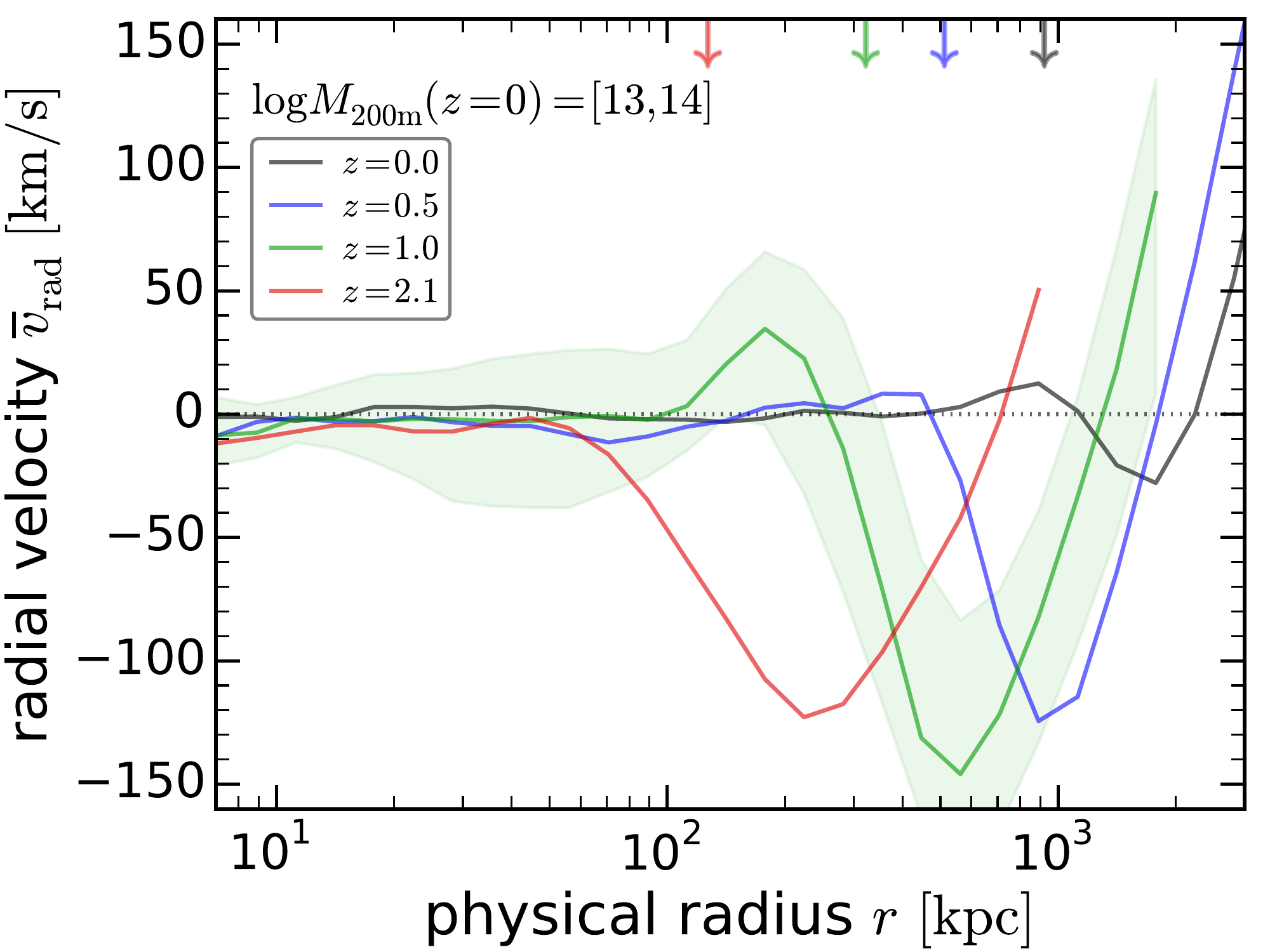}
\caption{
\textit{From the simulation with only dark matter}: profiles of dark matter versus physical radius, $r$, for isolated halos selected in three bins (left-to-right columns) of $\mthm$ at $z = 0$ and their progenitors.
For each quantity, we compute the average in bins of $r$ for each halo, and we show the median across the sample.
Shaded regions show the 68\% scatter of the value at/since $z = 1$.
Arrows show the median $\rthm$ at each $z$.
\textbf{Top row}: upper sub-panels show physical density, $\rho(r)$, while lower sub-panels show the ratio $\rho(r, z) / \rho(r, z = 0)$, that is, physical growth.
Since $z = 1$, $\rho(r)$ grows by $< 20\%$ within (and even beyond) $\rthm$, with more massive halos experiencing slightly more growth.
\textbf{Bottom row}: average radial velocity, $\vradave(r)$.
At $z > 1$, halos have pronounced virial infall regions, where $\vradave(r) < 0$, but by $z \sim 0$ such regions disappear at $\mthm < 10 ^ {13} \msun$, so these halos experience no physical cosmic accretion.
}
\label{fig:profile_dm}
\end{figure*}

\renewcommand{\figurewidth}{1.0}
\begin{figure}
\centering
{\large Simulation with only Dark Matter} \\
\includegraphics[width = \figurewidth \columnwidth]{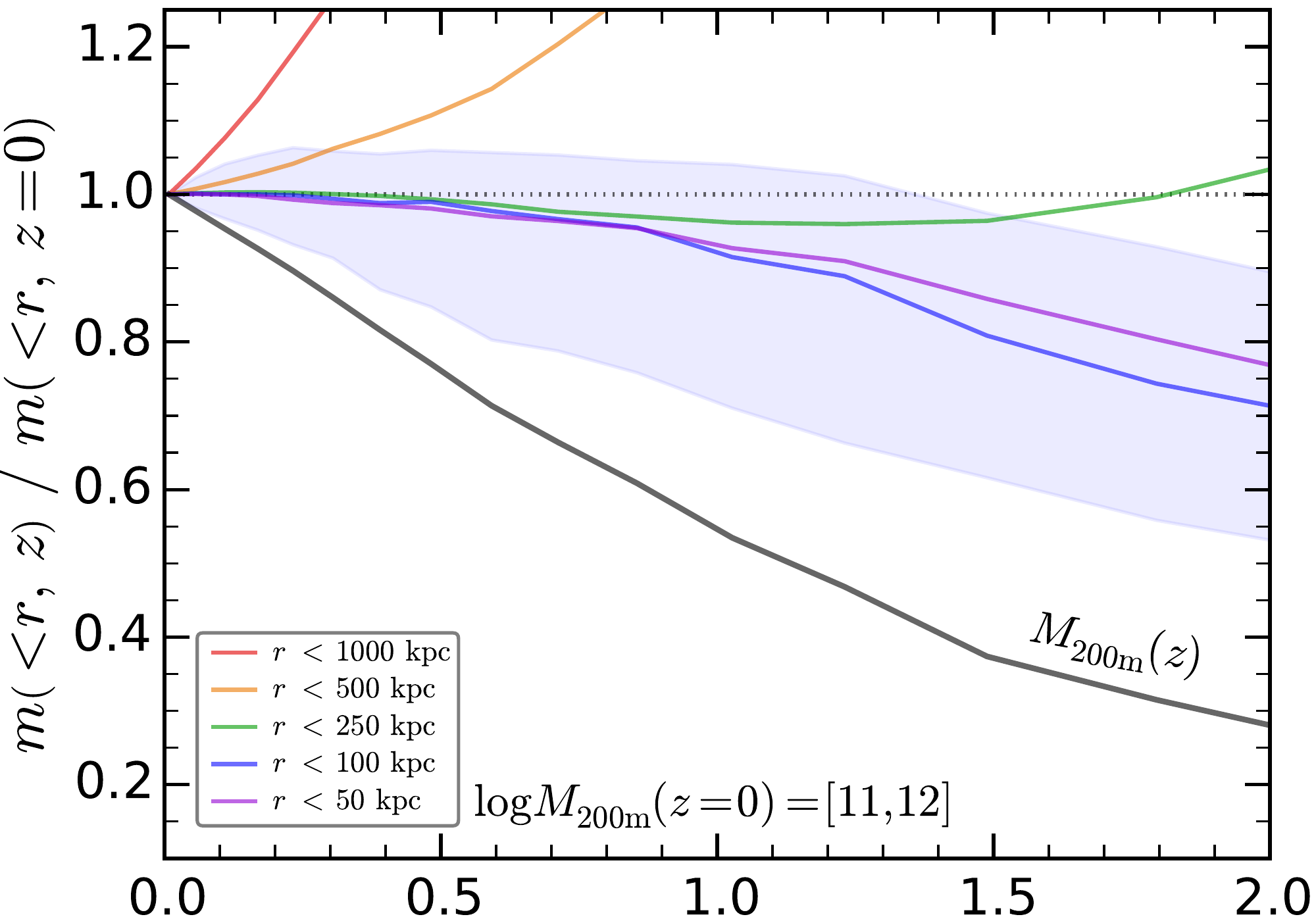}
\includegraphics[width = \figurewidth \columnwidth]{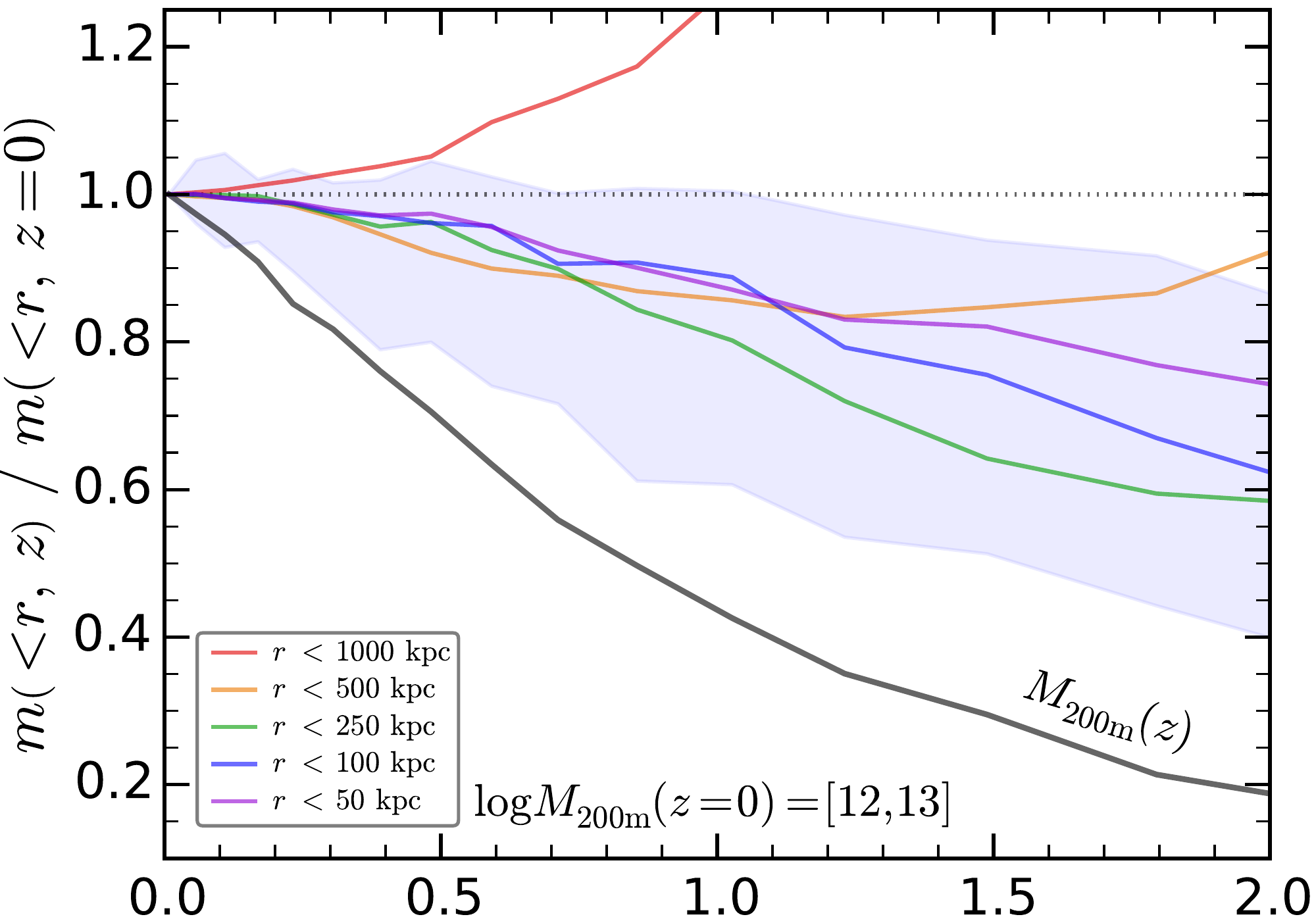}
\includegraphics[width = \figurewidth \columnwidth]{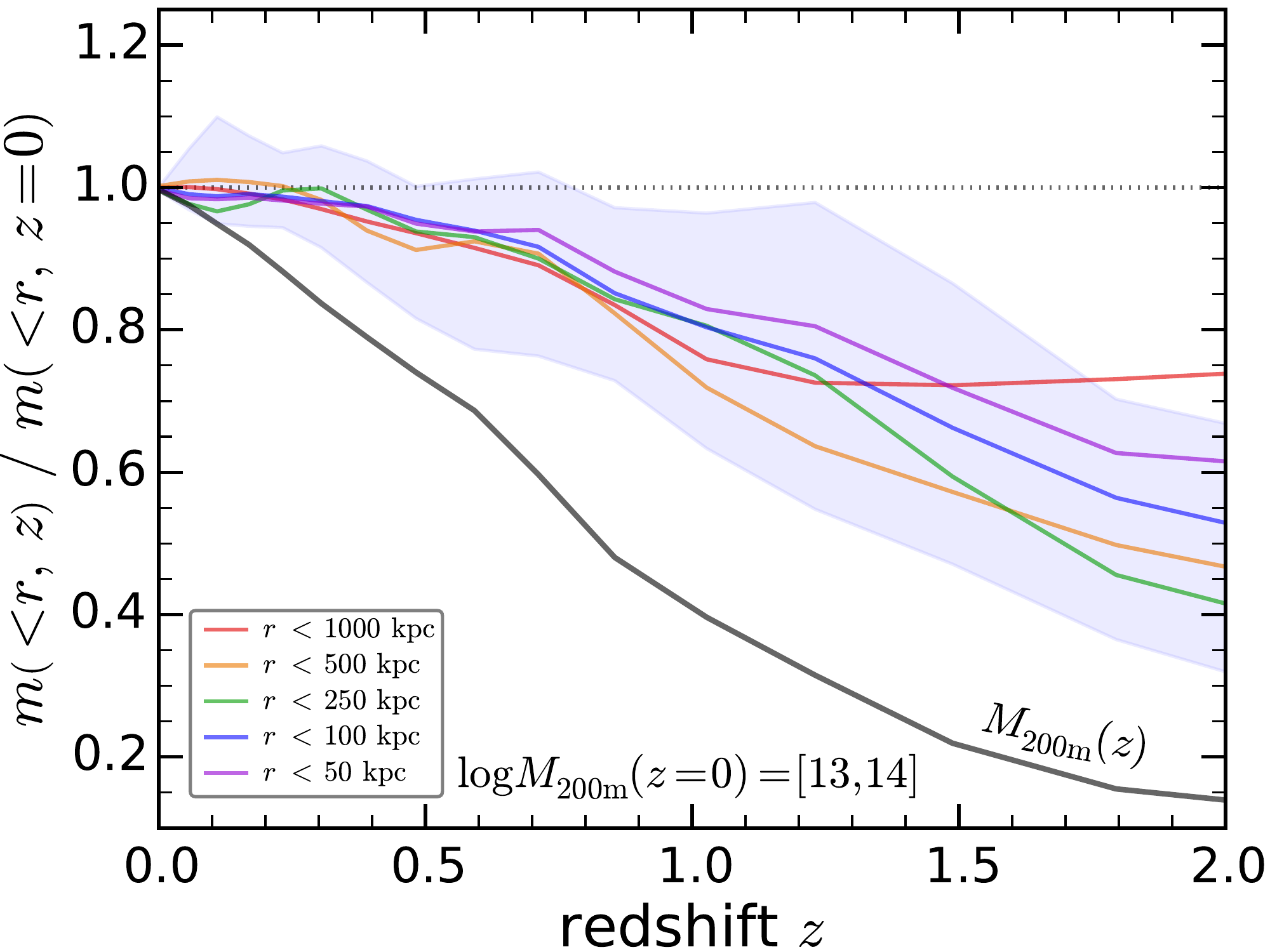}
\caption{
\textit{From the simulation with only dark matter}: mass of dark matter, $m$, within fixed physical radii, $r$, as a function of redshift, $z$, for isolated halos selected in three bins (top-to-bottom panels) of $\mthm$ at $z = 0$ and their progenitors.
At each $r$ and $z$, we compute the ratio $m(< r, z) / m(< r, z = 0)$ for each halo, and we show the median across each sample, with the shaded region showing the 68\% scatter at $r < 100 \kpc$.
Thick black curve shows \textit{inferred} growth of mass according to $\mthm(z) = m(< \rthm(z))$.
For reference, the range of median $\rthm(z)$ for halo progenitors from $z = 0$ to 2 is $185 - 55$, $390 - 90$, $920 - 130 \kpc$ physical for $\mthm(z = 0) = 10 ^ {11 - 12}$, $10 ^ {12 - 13}$, $10 ^ {13 - 14} \msun$, respectively.
At all $r$, physical mass growth ($10 - 20\%$ since $z = 1$) is \textit{significantly} less than that inferred from $\mthm(z)$ (doubling since $z = 1$).
}
\label{fig:mass_v_redshift_dm}
\end{figure}

We first examine the physical nature of cosmic accretion in the simulation with only dark matter, to set the stage to compare later with hydrodynamic simulations.
We select isolated halos in three bins of $\mthm$ at $z = 0$ and follow their progenitors back to $z = 2$.
For each quantity, we compute the average value in bins of physical radius for each halo, and we show the median value across the sample of halos at each $z$.
We also show the 68\% scatter for each quantity at/since $z = 1$.

\autoref{fig:profile_dm} (top) shows profiles of the physical density of dark matter, $\rhodark(r, z)$.
Upper sub-panels show $\rhodark$ versus physical radius, $r$, while lower sub-panels show the ratio $\rhodark(r, z) / \rhodark(r, z = 0)$ at fixed $r$, that is, physical growth.
For the latter, we compute this ratio for each halo and its progenitor, and panels show the median of this ratio across the sample.

At large $r$, $\rhodark(r)$ around halos declines over time, following the declining average matter density of the expanding universe.
Within this expanding background, gravitational attraction causes density to increase at smaller $r$, according to \autoref{eq:acceleration}.
However, the physical growth of $\rhodark(r)$ at any fixed $r$ is modest, being $< 40\%$ since $z = 2$ and only 10\% since $z = 1$ for $\mthm(z = 0) = 10 ^ {11 - 12} \msun$.
Higher-mass halos show slightly stronger physical growth.
Moreover, as the 68\% scatter in growth since $z = 1$ (shaded region) demonstrates, a significant fraction of halos experienced \textit{no} physical growth of dark matter over the last $\sim 8 \gyr$.

\autoref{fig:profile_dm} (bottom) shows the corresponding profiles of the average radial velocity of dark matter, $\vradavedark(r)$.
At large $r$, $\vradavedark(r)$ is positive (net outflowing) and increases with $r$, following the Hubble flow of the expanding universe.
The gravitational acceleration (\autoref{eq:acceleration}) causes $\vradavedark(r)$ to decline over time at fixed $r$.
The outermost $r$ where $\vradavedark = 0$ is where matter is experiencing its first turn-around at the given $z$.

Moving to a slightly smaller $r$, $\vradavedark(r) < 0$, that is, the majority of the mass is infalling.
However, $\vradavedark(r)$, which is an average over all masses at a given $r$, reaches a minimum value and turns up at a smaller $r$.
We refer to this radius of minimum $\vradave(r)$ as $\rinfall$.
Physically, $\rinfall$ represents the characteristic radius where the \textit{average} infall velocity is maximal, and it occurs because $\vradavedark(r)$ is an average of a juxtaposition of inward- and outward-moving orbits that overlap because dark matter is collisionless.
(Though a given mass shell will continue to collapse to smaller $r$ at increasingly negative $\vrad$.)
Thus, at $r > \rinfall$, most mass is infalling for the first time, while at $r < \rinfall$, most mass already has passed through the halo.
In other words, $\rinfall \approx \rsplashback$, corresponding to the outermost shell that is experiencing secondary turn-around at the given $z$ after passing through the halo core.\footnote{
Because cosmological accretion is triaxial and clumpy, $\rsplashback$ for a given shell is smeared over an extended range of $r$ \citep{Adhikari2014, DiemerKravtsov2014}.
Therefore, the way that one measures $\rsplashback$ can influence its value.
For example, using $r$ where the slope of the density profile is steepest, as those works did, can yield a $\rsplashback$ that is smaller than $\rinfall$.
Nevertheless, $\rinfall$ does represent the $r$ where the splashback mass starts to dominate over the infalling mass.}
At smaller $r$, $\vradavedark \approx 0$, where there is equal mass in inward- and outward-moving orbits and little change in mass over time.

$\rinfall$ increases monotonically with time, while the minimum of $\vradavedark(r)$ tends to weaken over time.
These trends are linked, as governed by \autoref{eq:acceleration}: infalling matter experiences a weaker halo potential at a larger $r$ (for these relatively static dark-matter halos), because dark energy causes increasingly stronger positive acceleration at larger $r$ over time.
This fact, of decreasing inflow velocity, combined with the decreasing density of inflowing mass at later times, means that the cosmic accretion rate, and thus the growth of $\rhodark(r)$ in the halo, declines over time, as \autoref{fig:profile_dm} (top) shows.
At any $z$, the most significant growth of $\rhodark(r)$ occurs at $r \approx \rinfall$.
However, \autoref{fig:profile_dm} (top) shows \textit{some} growth of $\rhodark(r)$ at smaller $r$, even where $\vradavedark(r) \approx 0$, because (collisionless and dissipationless) mass inflowing from a larger $r$ does deposit itself, in a time-average sense, at smaller $r$.

At $z \gtrsim 1$, halos in our mass range typically have a clear infall region, where $\vradavedark(r) < 0$, but at $z \approx 0$, this infall region has disappeared for halos with $\mthm \lesssim 10 ^ {13} \msun$, meaning that they experience \textit{no} net physical growth of dark matter from cosmological accretion.
Halos with $\mthm > 10 ^ {13} \msun$, on the other hand, experience weak physical accretion of dark matter even at $z = 0$.

In \autoref{fig:profile_dm}, vertical arrows show the median $\rthm(z)$ of halos and their progenitors at each $z$.
Across all masses and $z$, $\rthm(z)$ approximates the inner edge of the infall region, where $\vradavedark(r) \approx 0$, reasonably well.
Similarly, $\rthm(z)$ approximates the $r$ beyond which the growth of $\rhodark(r)$ is most significant.
We will explore in more detail such correlations with $\rthm(z)$ in \autoref{sec:virial_scaling}.

Solving \autoref{eq:acceleration} for an Navarro-Frenk-White \citep{Navarro1997} mass profile implies that the acceleration from dark energy equals the gravitational acceleration from the halo at $r \approx 3.5 \, \rthm$ at $z = 0$, independent of mass \citep[see also][]{Busha2005}.
Thus, dark energy largely accounts for the upturn in $\vradavedark(r)$ at large $r$ for the most massive halos in \autoref{fig:profile_dm}.
However, dark energy alone cannot account for the upturn in $\vradavedark(r)$ just beyond $\rthm$ in lower-mass halos at low $z$.
The radial extent of the infall region for low-mass halos at low $z$ is in fact set by the (3-dimensional) tidal motions and angular momenta around such halos (see \autoref{sec:angular_momentum}, also \citealt{Cuesta2008}).

Another way to see the nature of cosmic accretion is to examine the physical growth of cumulative mass, $\mdark(z)$, within fixed $r$.
This also allows us to compare with the commonly used mass growth that one infers from an evolving $\mthm(z) = m \left( r < \rthm(z) \right)$.
\autoref{fig:mass_v_redshift_dm} shows the median ratio $\mdark(< r, z) / \mdark(< r, z = 0)$ at various $r$ for the same halos as in \autoref{fig:profile_dm}.
At large $r$, $\mdark(< r)$ declines over time, in accord with the declining average density of matter in the expanding universe.
At smaller $r$, cumulative mass instead grows over time, though the mass growth eventually stalls, which occurs at larger $r$ over time.
Thus, the amount of physical growth within most $r$ is modest since $z = 2$, especially as compared with the mass growth inferred from $\mthm(z)$ (thick black curve).
Indeed, as inferred from $\mthm(z)$, our lowest-mass halos have doubled in dark-matter mass since $z = 1$, but the amount of physical growth at any $r$ is significantly lower ($10 - 30\%$, increasing with mass).
Furthermore, the shaded region indicates the scatter in physical growth at $r < 100 \kpc$, highlighting that a significant fraction of isolated halos experienced \textit{no} physical growth since $z \sim 1$.
We find similar scatter for the growth of $\mthm(z)$ or in using a narrower bin of $\mthm(z = 0)$ of 0.1 dex, so the scatter at $r < 100 \kpc$ is real and meaningful.

These trends are consistent with previous related works based on dark matter simulations \citep{Busha2005, Diemand2007, Cuesta2008, Diemer2013, Adhikari2014}.

\section{Physical accretion of baryons: impact of gas physics}
\label{sec:gas_physics}

\renewcommand{\figurewidth}{1.0}
\begin{figure*}
\hspace{1.4 cm} {\large Simulation with Gas - Non-Radiative} \hspace{1.6 cm} {\large Simulation with Gas - Radiative Cooling} \\
\centering
\includegraphics[width = \figurewidth \columnwidth]{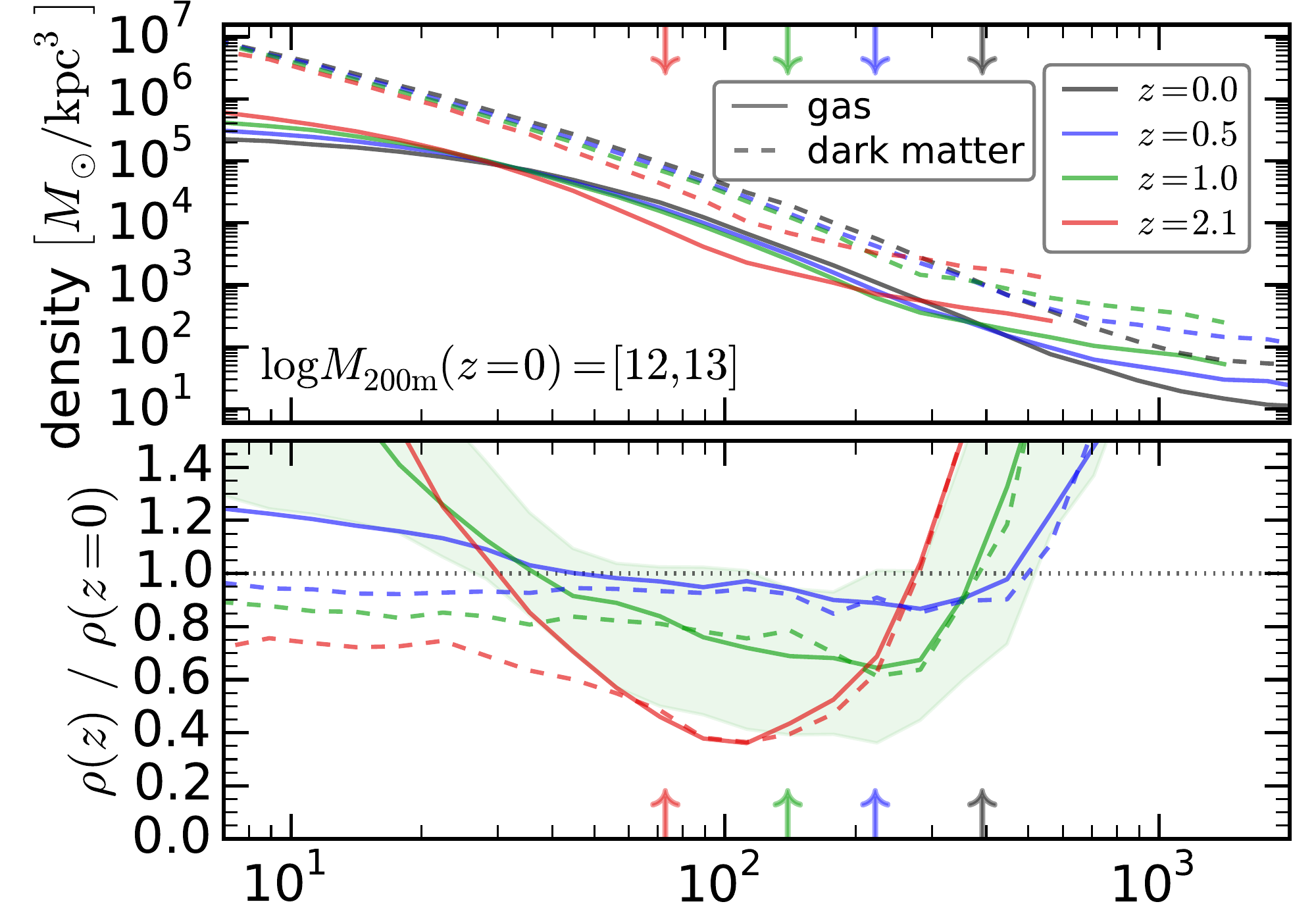}
\includegraphics[width = \figurewidth \columnwidth]{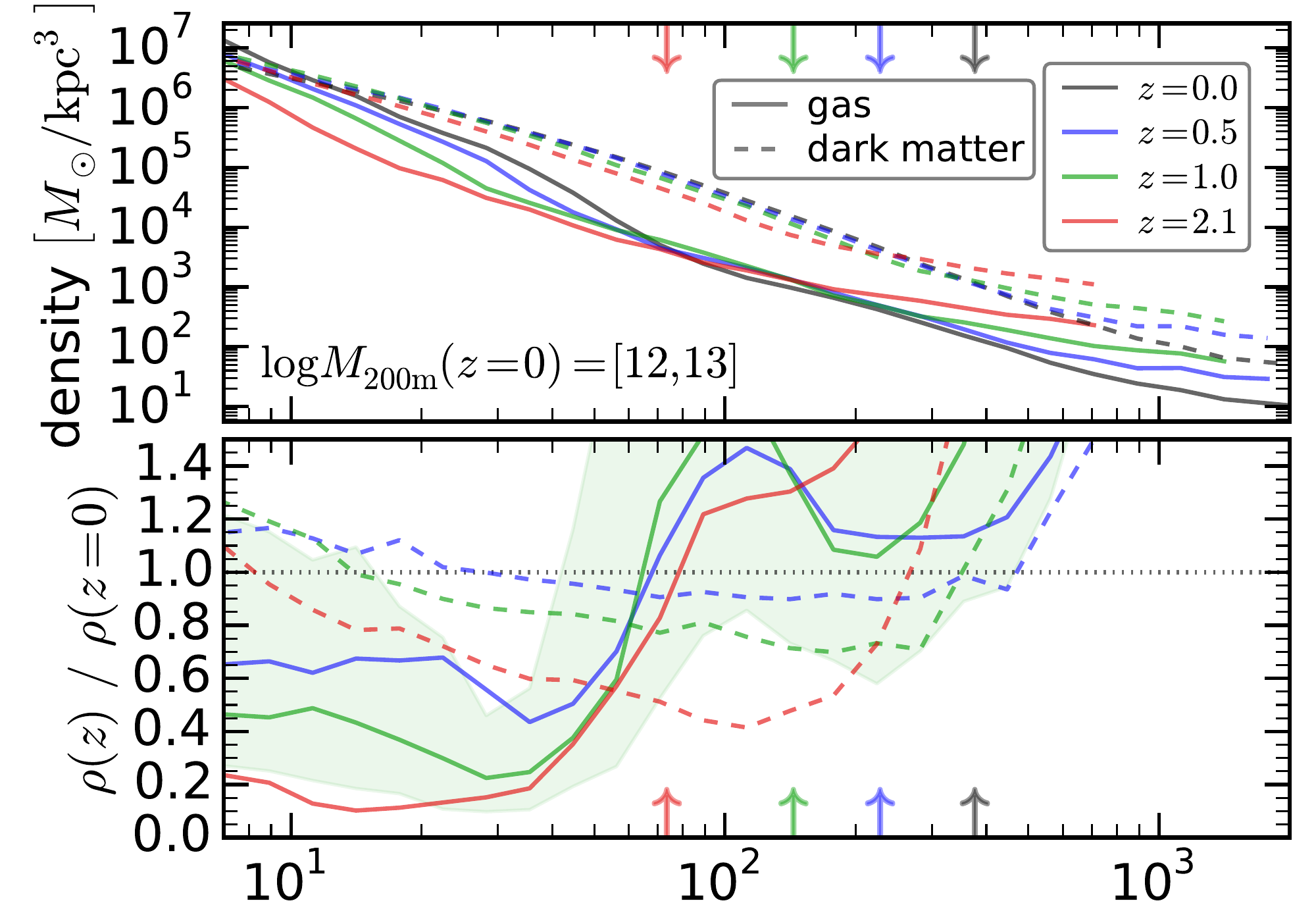}
\includegraphics[width = \figurewidth \columnwidth]{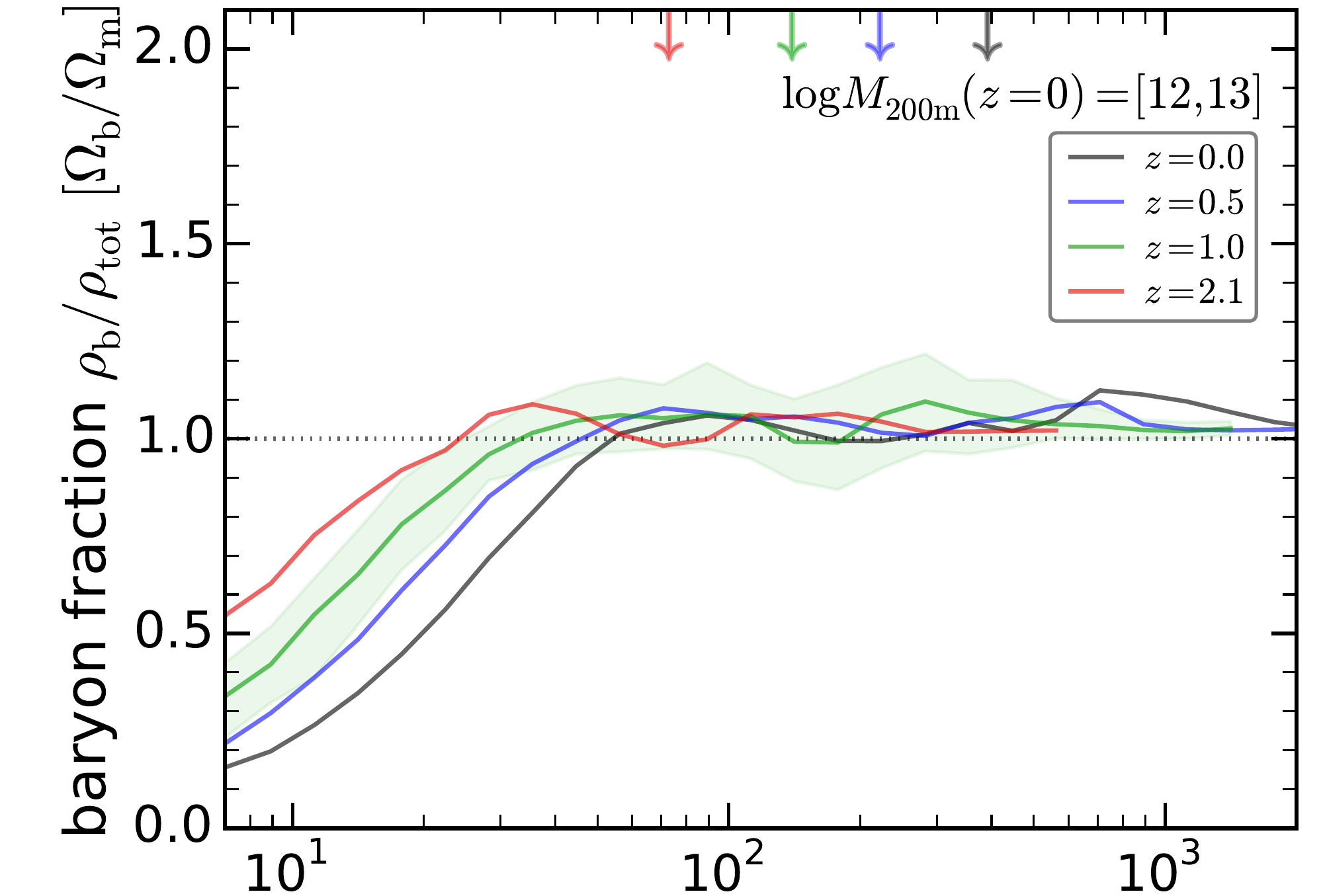}
\includegraphics[width = \figurewidth \columnwidth]{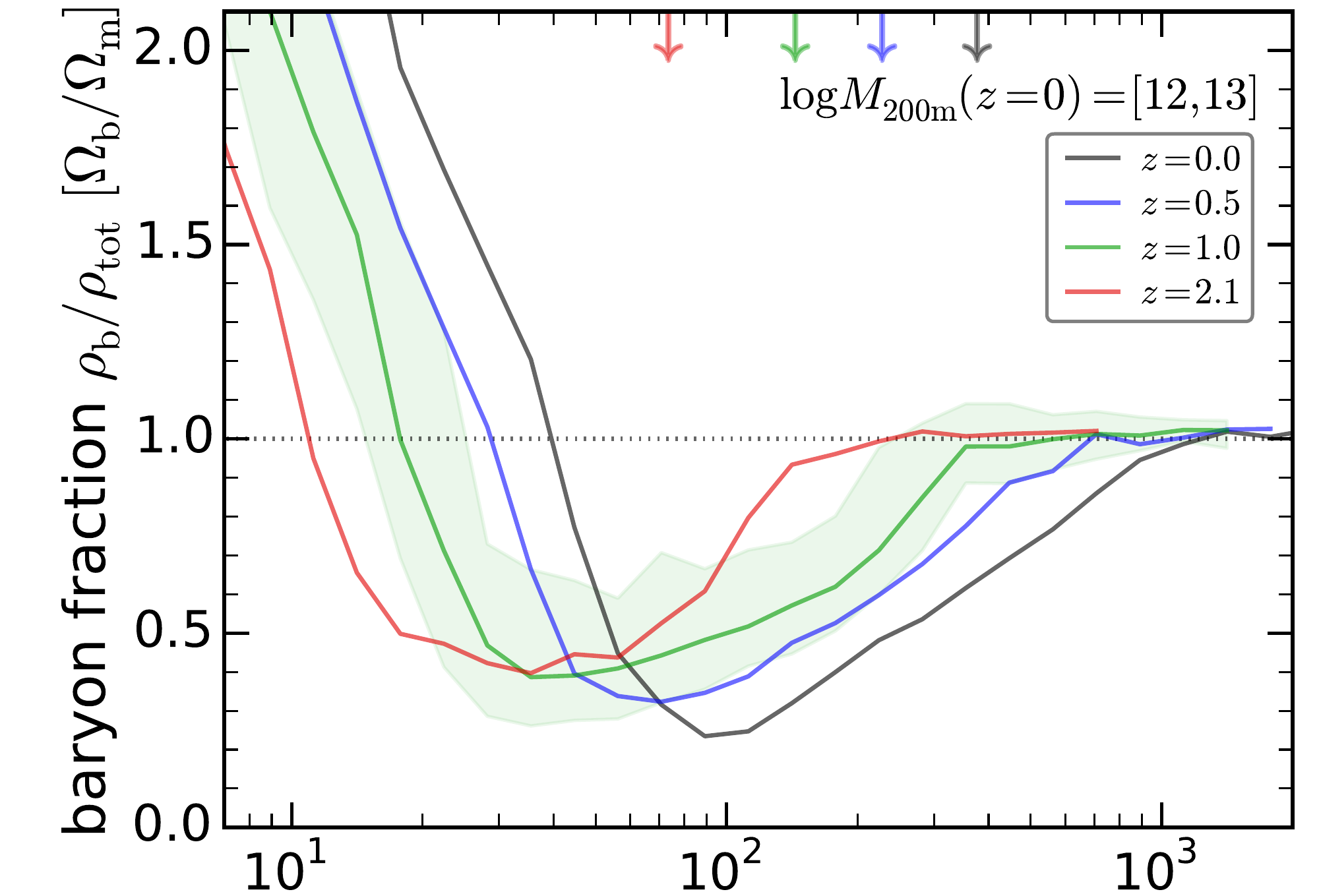}
\includegraphics[width = \figurewidth \columnwidth]{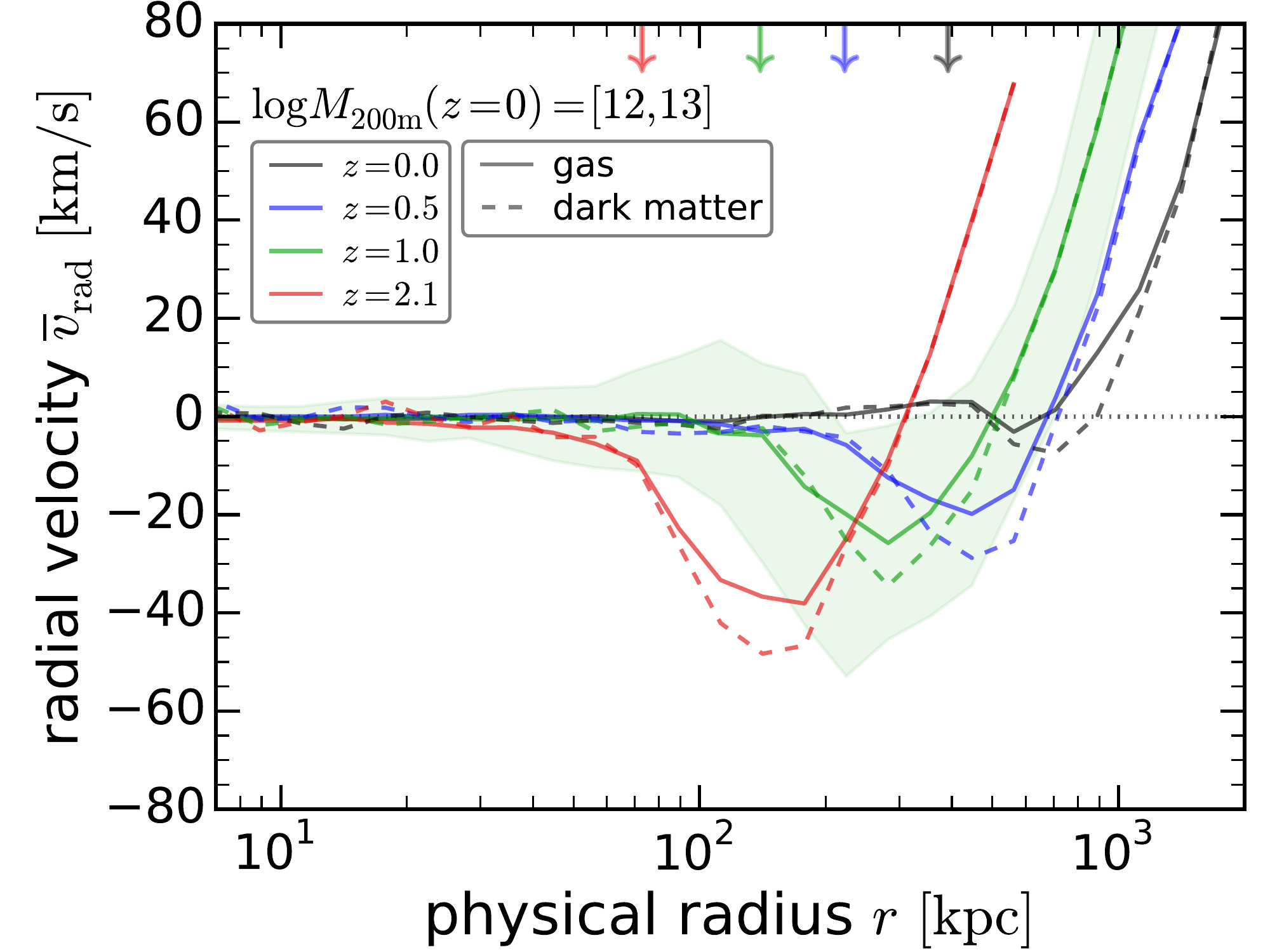}
\includegraphics[width = \figurewidth \columnwidth]{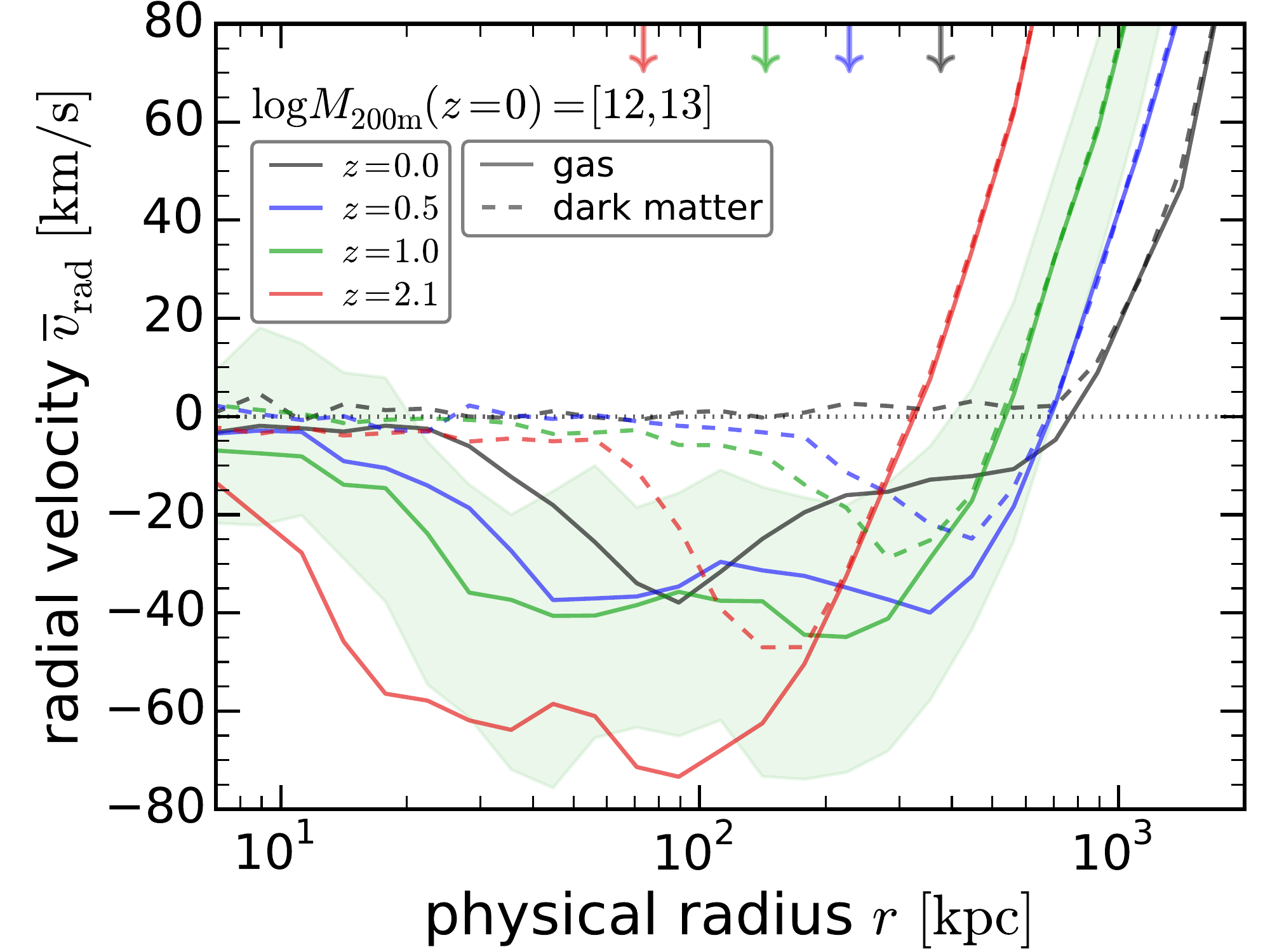}
\caption{
\textit{From the simulations with non-radiative (left) and radiatively cooling (right) gas}: profiles of gas (solid) and dark matter (dashed) versus physical radius, $r$, for isolated halos selected in a bin of $\mthm$ at $z = 0$ and their progenitors, similar to \autoref{fig:profile_dm}.
Shaded regions show the 68\% scatter of each value for gas at/since $z = 1$, and arrows show the median $\rthm$ at each $z$.
\textbf{Top row}: upper sub-panels show physical density, $\rho(r)$, while lower sub-panels show the ratio $\rho(r, z) / \rho(r, z = 0)$, that is, physical growth.
\textbf{Middle row}: baryon fraction, $\fbaryon(r) = \rhobaryon(r) / \left( \rhobaryon(r) + \rhodark(r) \right)$, in units of the cosmic value, $\omegabaryon / \omegamatter$.
\textbf{Bottom row}: average radial velocity, $\vradave(r)$.
Non-radiative gas (left column) closely tracks dark matter, though the additional thermal and turbulent pressure of gas causes a decrease in $\rhogas(r)$ at the core and reduced inflow velocity beyond $\rthm(z)$.
Gas with radiative cooling (right column) readily advects to small $r$, increasing the growth of $\rhogas(r)$ significantly.
}
\label{fig:profile_gas_physics}
\end{figure*}

Having explored the physical nature of cosmic accretion in the simulation with only dark matter, we now turn to examining both dark matter and baryons in simulations that model both.
In this section, we first explore systematically how gas physics alone affects cosmic accretion, to obtain a basic understanding of the underlying physics.
Thus, we first examine two set of simulations with no star formation that are identical, except that the first includes non-radiative gas, while the second employs radiative cooling in a ultraviolet background.
In the next section, we will show the simulation that additionally includes star formation and feedback.
Here, we report results for a single bin of $\mthm(z = 0) = 10 ^ {12 - 13} \msun$ and note that trends with gas physics are qualitatively similar across our range of $\mthm = 10 ^ {11 - 14} \msun$.

\autoref{fig:profile_gas_physics} shows the profiles of gas (solid) and dark matter (dashed) as a function of physical radius, $r$, similar to \autoref{fig:profile_dm}.
Top row shows $\rho(r)$ and its growth history; middle row shows the baryon fraction, $\fbaryon(r) = \rhobaryon(r) / \left( \rhobaryon(r) + \rhodark(r) \right)$, in units of the cosmic value, $\omegabaryon / \omegamatter$; and bottom row shows the average radial velocity, $\vradave(r)$.

First, \autoref{fig:profile_gas_physics} (left column) shows gas without radiative cooling.
While this is an unphysical scenario, especially for low-mass halos with virial temperatures that correspond to short cooling times compared to their dynamical times, it provides a benchmark for understanding the subsequent impact of gas cooling.

The inclusion of non-radiative gas imparts little change to the behavior of dark matter, which remains similar to that in the simulation with only dark matter (\autoref{fig:profile_dm}).
For gas accretion, \autoref{fig:profile_gas_physics} shows that at large $r$, $\rhogas(r)$ and $\vradavegas(r)$ closely track those of dark matter, as may be expected for freely infalling, supersonic gas whose dynamics are governed primarily by gravity.

\autoref{fig:profile_gas_physics} shows some decoupling of $\vradave(r)$ between non-radiative gas and dark matter just beyond $\rthm$, where $\vradavegas(r)$ is slightly weaker than $\vradavedark(r)$.
Such decoupling is not surprising, because the underlying dynamics of dark matter and non-radiative gas are distinct, as outlined in \autoref{sec:halo_collapse}: $\rinfall$ for (collisionless) dark matter is set by $\rsplashback$, while for non-radiative gas, whose collisional nature means that orbits cannot overlap, it is set by where gas shocks and/or starts to become supported by both thermal and turbulent pressure.
(Though accretion can be clumpy, and some gas can orbit within a satellite subhalo in a neither purely collisional nor collisionless manner.)
Thus, the addition of this pressure support is likely what causes weaker $\vradave(r)$ for gas as compared with dark matter.
At similar $r$, $\fbaryon(r)$ shows a corresponding, though weak, enhancement.

At $r \ll \rthm$, non-radiative gas decouples from dark matter, as the gas density profile flattens in the core and $\fbaryon(r)$ decreases, a result of strong thermal and turbulent pressure support.
Here, $\rhogas(r)$ decreases over time, because the higher rate of accretion at higher $z$ causes the non-radiative gas to be slightly overpressurized in the core.
This gas then expands to a somewhat lower density.

Given the strongly \textit{differing} physics that govern dark matter and non-radiative gas, we emphasize the strikingly \textit{similar} behavior that they display at essentially \textit{all} $r$.
This simply may be a reflection that the dynamics of both components, though different, conserve energy and respond to the same underlying halo potential.
The similar behavior of non-radiative gas and dark matter means that essentially all subsequent differential effects between them are driven not by the collisional nature of gas, but rather, by its ability to cool radiatively.

\autoref{fig:profile_gas_physics} (right column) shows the same profiles, but from the simulation with radiative cooling.
This represents the opposite extreme from non-radiative, because gas is able to experience runaway cooling and collapse to small $r$ without any feedback.
Thus, this case represents the maximal amount of gas cooling and cosmic accretion that is feasible for gas with primordial metallicity.

First, examining the impact on dark matter, the inclusion of radiative cooling does not change significantly the growth of $\rhodark(r)$ at $r \gtrsim 40 \kpc$, but it reduces the physical growth rate at smaller $r$, which may be because the runaway cooling of gas accelerates the growth of $\rhodark(r)$ in the core at early times.\footnote{
However, the severe overcooling in this simulation can create an artificially dense gas clump(s) near the halo core that can be offset from the (otherwise) primary dark-matter density peak by $\sim 10 \kpc$.
This introduces uncertainties in the exact center position, so one should interpret the evolution of the dark-matter density profile at $r \lesssim 20 \kpc$ in \autoref{fig:profile_gas_physics} (right) with care.}
However, $\vradavedark(r)$ does not change significantly.

For gas, the addition of radiative cooling causes it to decouple from dark matter starting at $r \approx 2 \, \rthm$, because this cooling prevents the formation of strong virial shocks and pressure support \citep{DekelBirnboim2006} at this $r$.
At $r \sim 100 \kpc$, the evolution of $\rhogas(r)$ is non-monotonic, as driven by the relative efficiency of cosmic accretion and cooling.
At $r \lesssim 70 \kpc$, $\rhogas(r)$ increases significantly over time, at a rate many times that of dark matter.
At $r \approx 70 \kpc$, $\rhogas(r)$ is nearly constant because of the balance of cosmic accretion and gas cooling.

The second row shows $\fbaryon(r)$, as scaled to the cosmic value, $\omegabaryon / \omegadarkmatter$.
At large $r$, baryons trace dark matter, but at $r \lesssim 2 \, \rthm$, $\fbaryon(r)$ decreases, as gas cools and advects to smaller $r$ in a relatively static profile of dark matter.
$\fbaryon(r)$ reaches a minimum at intermediate $r$, which moves outward over time, driven primarily by the changing $\rhobaryon(r)$.
At small $r$, $\fbaryon(r)$ rises rapidly as gas cools among relatively static dark matter.

As with non-radiative gas, $\vradavegas(r)$ for radiative gas tracks $\vradavedark(r)$ beyond dark matter's $\rinfall \approx 2 \, \rthm$.
However, in this case, infalling gas cools and continues to advect to smaller $r$ with relatively little pressure support, so $\rinfall$ is much smaller and less well-defined for gas.
Nevertheless, gas does not experience runaway advection to small $r$, because some shocking occurs at smaller $r$ that imparts some pressure support (as we have checked explicitly), and gas starts to feel significant angular-momentum support, as we explore in the next section.
In combination, these processes decrease the magnitude of $\vradavegas(r)$ at fixed $r$ over time.

To summarize, the cosmic accretion of gas is distinct from that of dark matter at $r \lesssim 2 \rthm$.
While the collisional nature of gas leads to some differences, this effect is modest.
Instead, the dissipational radiative cooling of gas drives strong differences, which lead to a significant physical advection/accretion of gas at all $r$ and $z$ atop a relatively static profile of dark matter.

\section{Physical accretion of baryons with star formation and feedback}
\label{sec:baryon}

\renewcommand{\figurewidth}{1.0}
\begin{figure*}
\centering
{\large Simulation with Star Formation \& Feedback} \\
\includegraphics[width = \figurewidth \columnwidth]{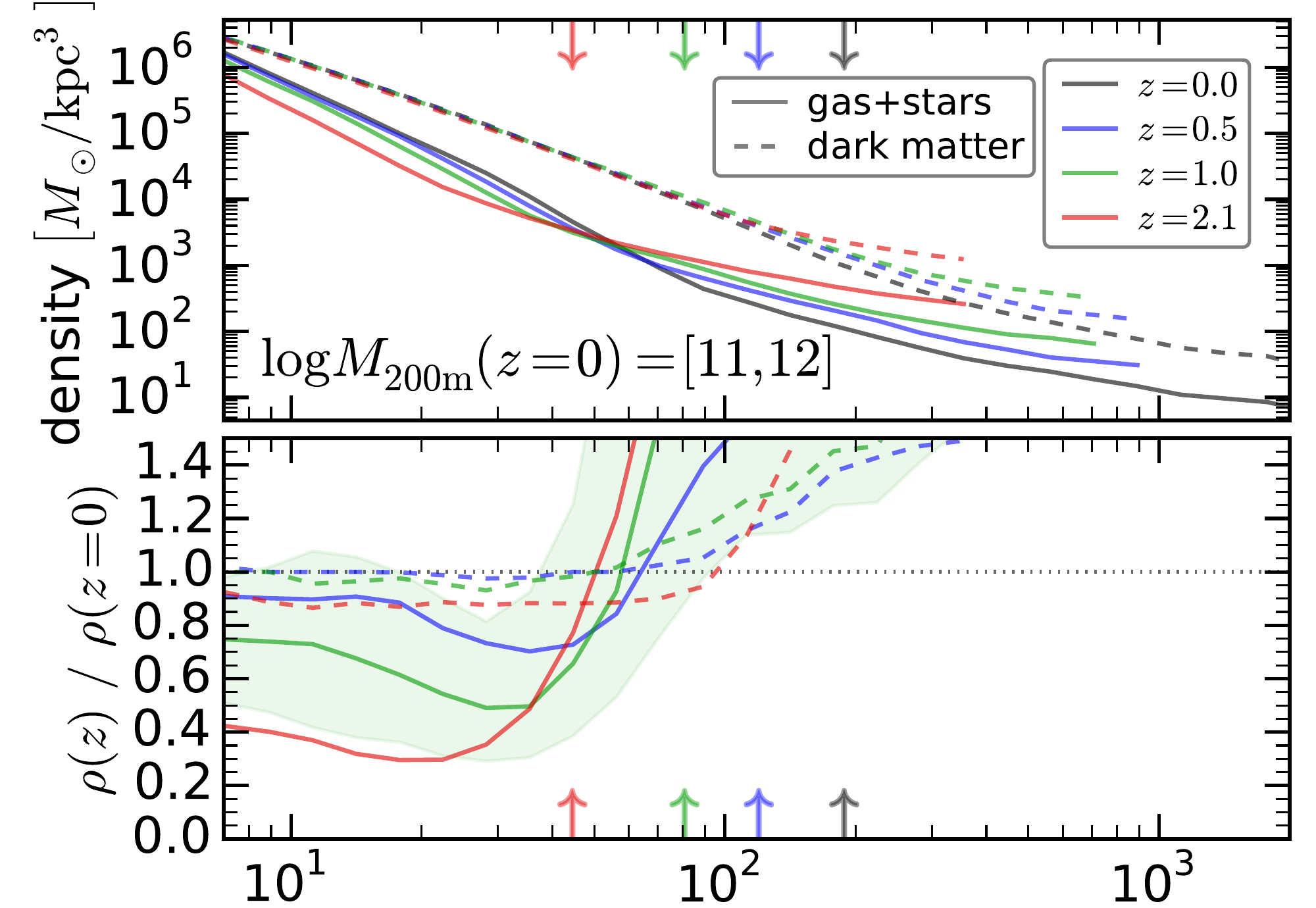}
\includegraphics[width = \figurewidth \columnwidth]{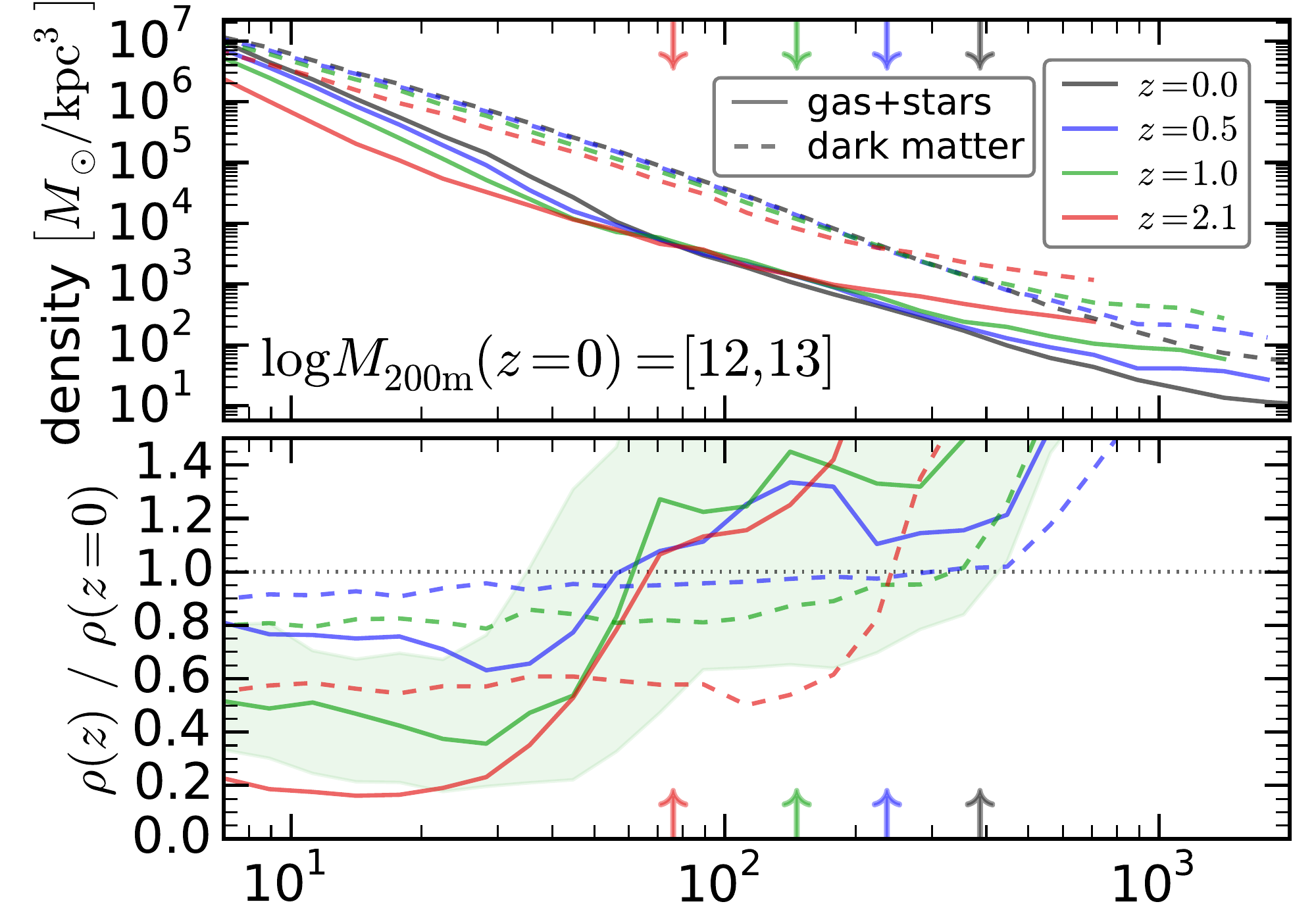}
\includegraphics[width = \figurewidth \columnwidth]{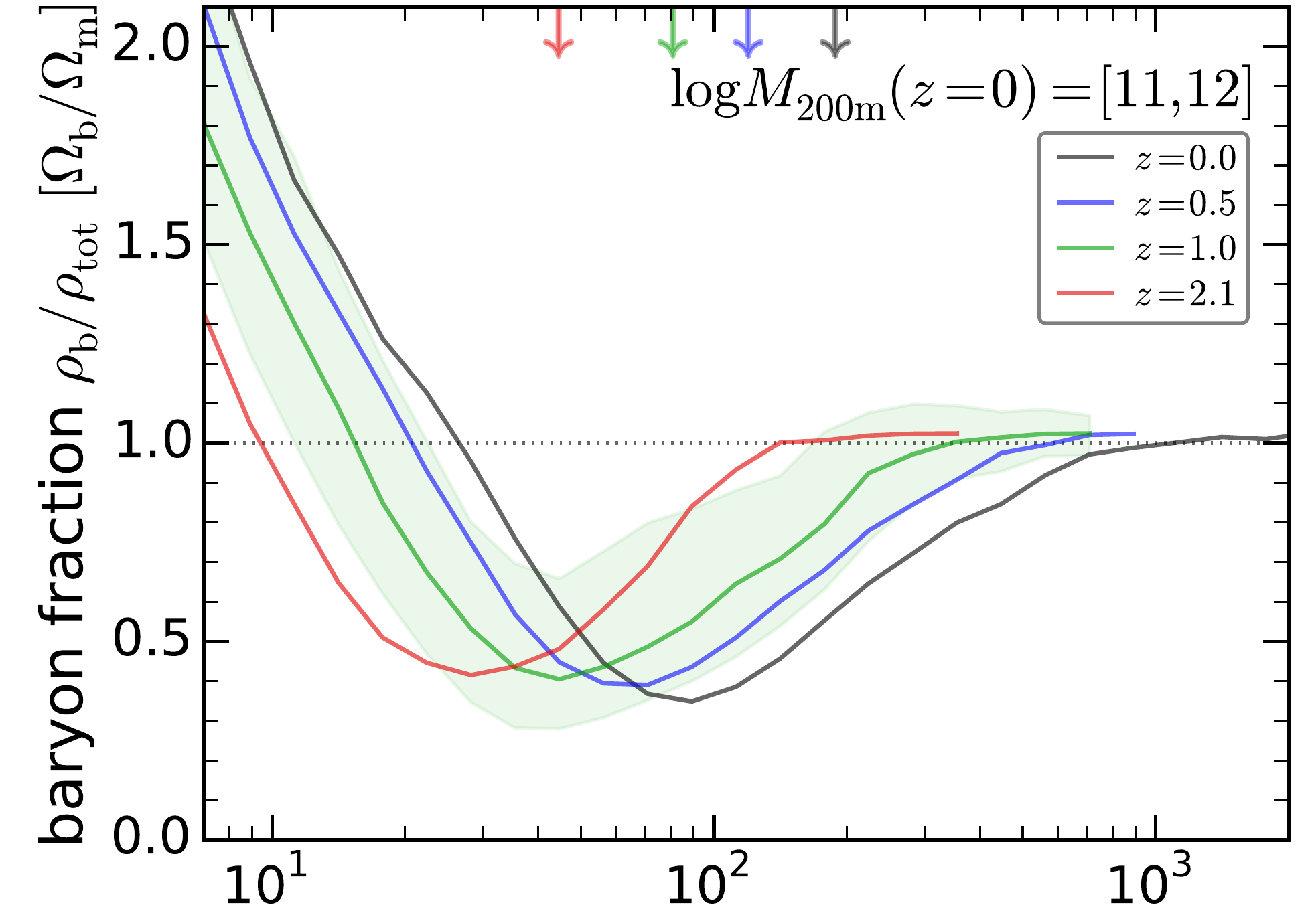}
\includegraphics[width = \figurewidth \columnwidth]{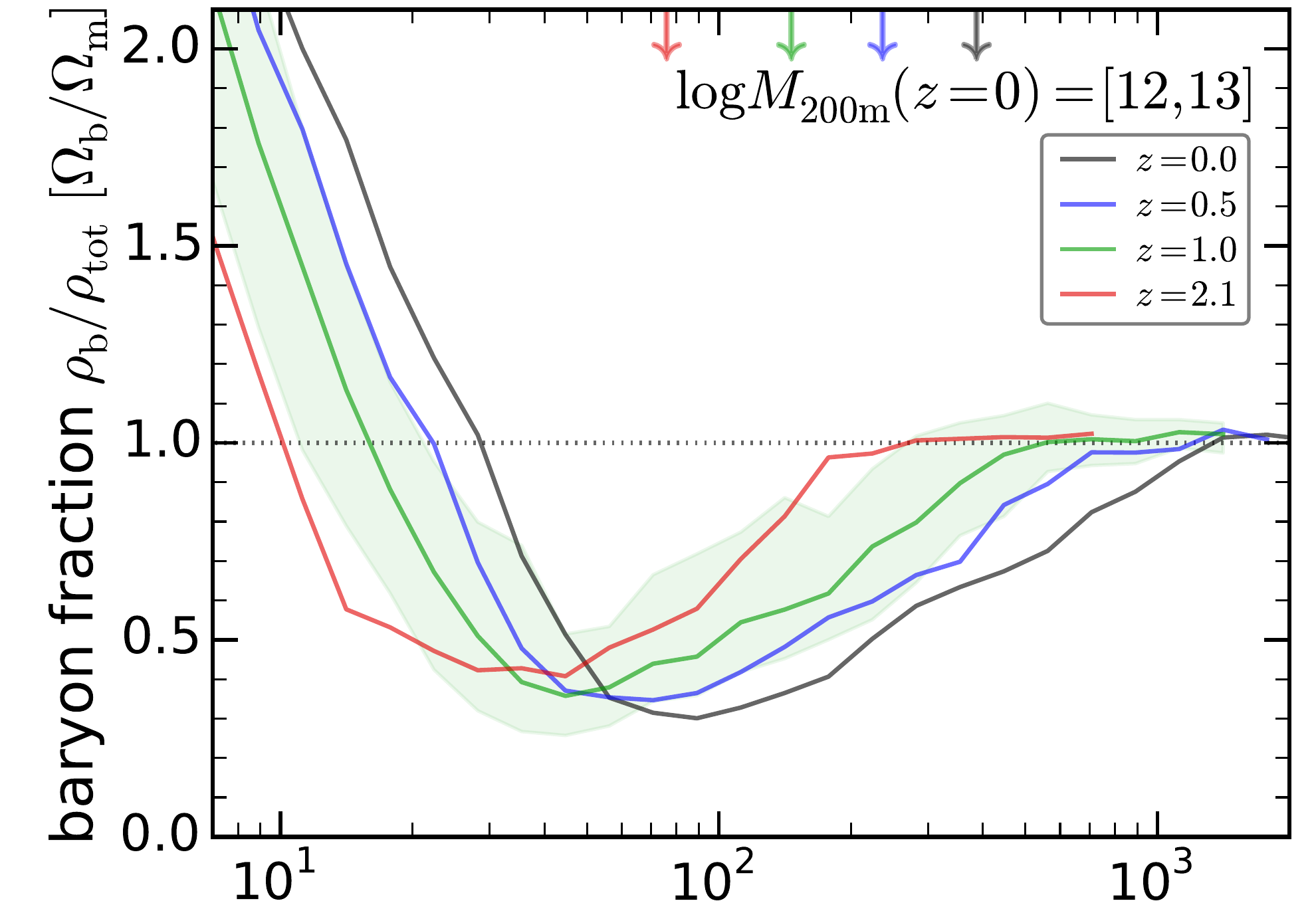}
\includegraphics[width = \figurewidth \columnwidth]{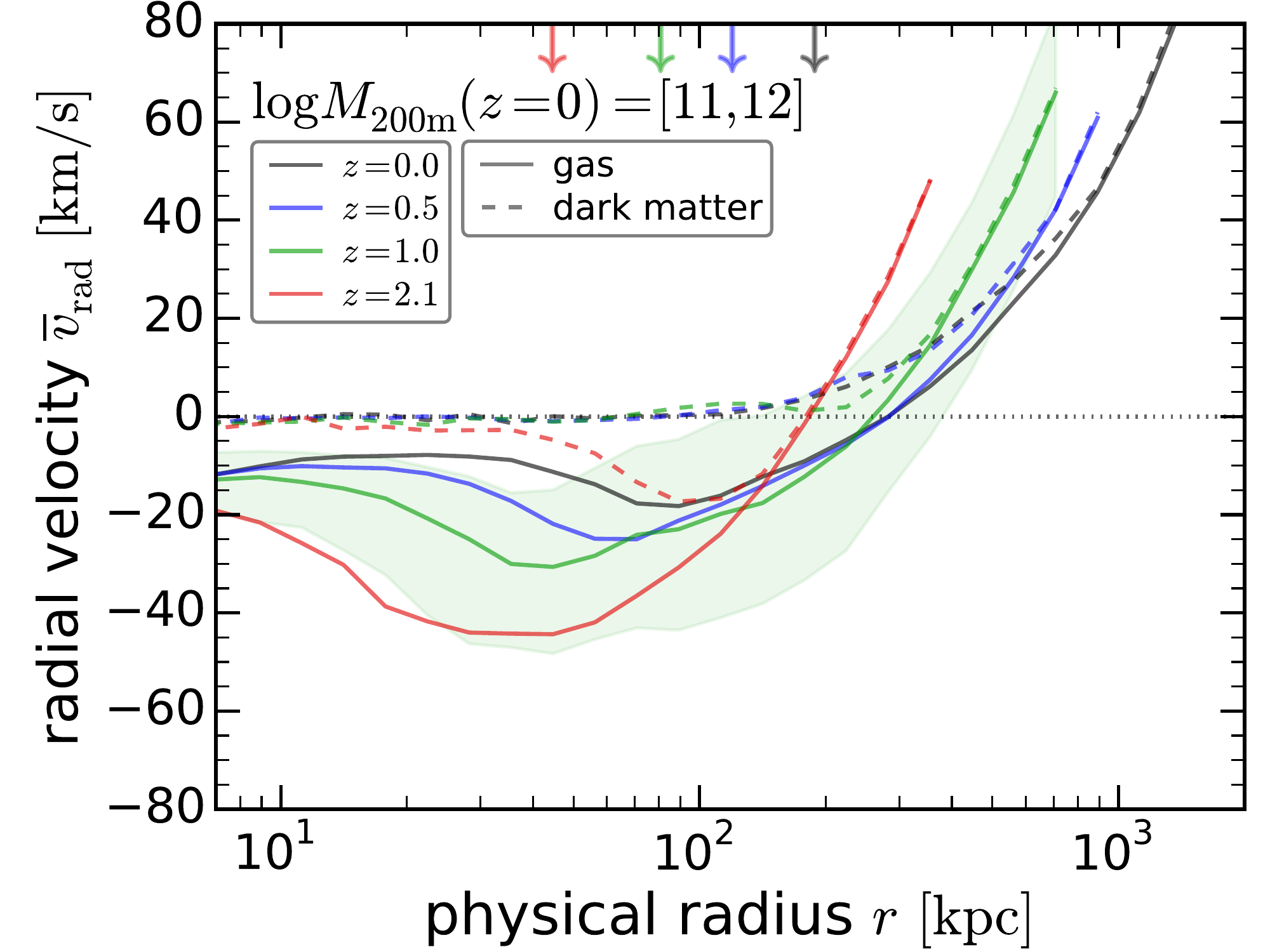}
\includegraphics[width = \figurewidth \columnwidth]{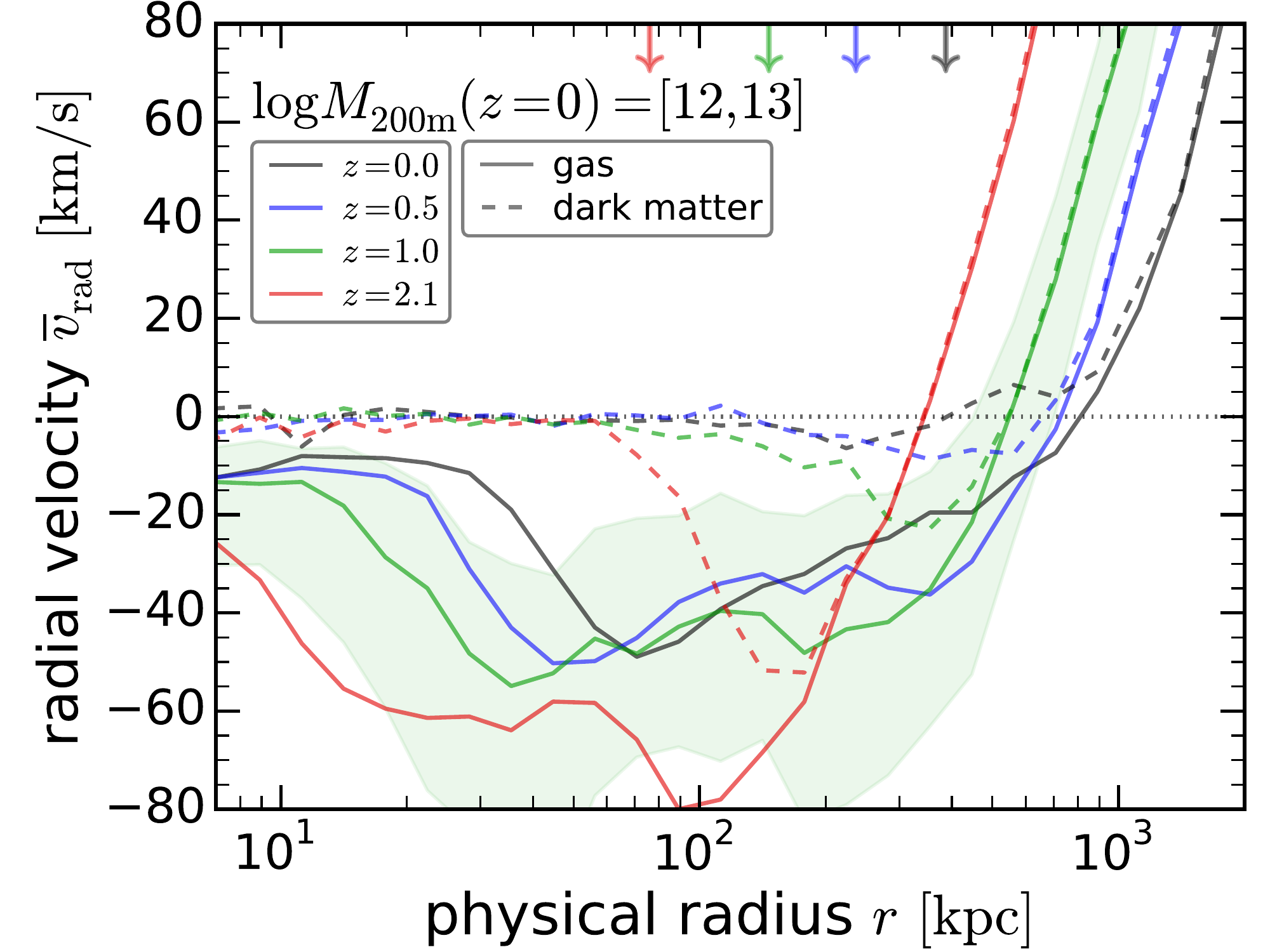}
\caption{
\textit{From the simulation with star formation and feedback}: profiles of baryons (gas + stars; solid) and dark matter (dashed) versus physical radius, $r$, for isolated halos selected in two bins (left and right columns) of $\mthm$ at $z = 0$ and their progenitors, similar to \autoref{fig:profile_dm}.
Shaded regions show the 68\% scatter of each value for baryons at/since $z = 1$, while arrows show median $\rthm$ at each $z$.
\textbf{Top row}: upper sub-panels show physical density, $\rho(r)$, while lower sub-panels show the ratio $\rho(r, z) / \rho(r, z = 0)$, that is, physical growth.
\textbf{Middle row}: baryon fraction, $\fbaryon(r) = \rhobaryon(r) / \left( \rhobaryon(r) + \rhodark(r) \right)$, in units of the cosmic value, $\omegabaryon / \omegamatter$.
\textbf{Bottom row}: average radial velocity, $\vradave(r)$.
At $r > \rthm(z)$, baryons start to decouple from dark matter as gas cools within the nearly static dark-matter potential.
At intermediate $r$, where gas cooling is efficient and there is no significant rotational support (see \autoref{fig:angular_momentum_profile_sf}), the dynamics of baryons and dark matter decouple the most, as the infall velocity of baryons reaches a maximum, and $\fbaryon(r)$ reaches a minimum.
This $r$ occurs well within $\rthm(z)$ and increases over time.
At small $r$, because of gas cooling, $\rhobaryon(r)$ increases significantly while $\rhodark(r)$ remains nearly unchanged.
}
\label{fig:profile_sf}
\end{figure*}

Having explored cosmic accretion in simulations with only dark matter, with non-radiative gas, and with radiatively cooling gas, we now examine our simulation that additionally includes star formation and feedback.
Here, the primary differences compared with the simulation with only cooling are (1) thermal energy injection from supernovae and stellar winds, which heat the gas at small radii; (2) advection of metals into the halo gas, which enhances the efficiency of cooling;\footnote{
For our halos, the gas metallicities at $r = 0.1 - 1 \rthm$ are $\sim 5 - 10\%$ of solar, which implies at most a modest (factors of a few) increase to the cooling rate compared to primoridal gas in the previous section.} and (3) the transition of some gas to stars, which behave as a collisionless fluid, similar to dark matter.

Henceforth, we compute all baryon masses including both gas and stars, to simplify the interpretation of cosmic accretion without the ambiguities of conversions between the two from star formation and stellar mass loss and to allow us to compare with our simulations without star formation.
However, we continue to compute velocities separately for gas.

\subsection{Density and radial velocity}

\autoref{fig:profile_sf} shows profiles of baryons and dark matter as a function of $r$, for halos selected in two bins of $\mthm$ at $z = 0$ and their progenitors, similar to Figures~\ref{fig:profile_dm} and \ref{fig:profile_gas_physics}.
\autoref{fig:profile_sf} (top row) shows $\rhodark(r)$ and $\rhobaryon(r)$ (upper sub-panels), as well as $\rhodark(r, z) / \rhodark(r, z = 0)$ and $\rhobaryon(r, z) / \rhobaryon(r, z = 0)$, that is, physical growth (lower sub-panels).

The injection of energy from feedback mitigates the effects of gas overcooling on dark matter at small $r$, compared with the cooling-only simulation (\autoref{fig:profile_gas_physics} right).
However, this feedback also reduces the infall velocity of dark matter at $r > \rthm(z)$ at $z \lesssim 1$ at both masses, as compared with simulation with only dark matter (\autoref{fig:profile_dm}).
We think that this is caused by angular-momentum transfer from cooling gas to dark matter, because the angular momentum of dark matter at $r \gtrsim \rthm(z)$ is $10 - 20\%$ higher in this simulation (see \autoref{sec:angular_momentum}) than with only dark matter.
Furthermore, this boost in dark-matter angular momentum is higher in lower-mass halos, likely because of their lower potential energy, which also explains the stronger differences in infall velocity at $r > \rthm(z)$ for lower-mass halos in \autoref{fig:profile_sf}.

Compared to $\rhodark(r)$ (dashed curves), $\rhobaryon(r)$ (solid curves) increases much more over time, again because of the ability of gas to cool radiatively and advect to smaller $r$.
Specifically, at small $r$, $\rhobaryon(r)$ increases monotonically with time, $30 - 50\%$ since $z = 1$, with more growth at higher mass, although thermal feedback has reduced the growth of $\rhobaryon(r)$ at $r \lesssim 40 \kpc$ noticeably compared to the case with only cooling.

\autoref{fig:profile_sf} (second row) shows $\fbaryon(r)$, as scaled to the cosmic value, $\omegabaryon / \omegadarkmatter$.
The trends are similar to those in \autoref{fig:profile_gas_physics} (right).
At large $r$, baryons trace dark matter at the cosmic value.
At $r \lesssim 2 \, \rthm$, $\fbaryon(r)$ decreases, as gas cools and advects to smaller $r$ in a relatively static profile of dark matter.
$\fbaryon(r)$ reaches a minimum at intermediate $r$, where gas dynamics are most decoupled from dark matter, and this $r$ moves outward over time, driven primarily by the changing $\rhobaryon(r)$, while $\rhodark(r)$ remains nearly static.
At small $r$, $\fbaryon(r)$ rises rapidly as cooled gas settles near the galaxy.
Thus, baryons advect to smaller $r$ even in the absence of the physical growth of dark matter.

\autoref{fig:profile_sf} (third row) shows $\vradavedark(r)$ and $\vradavegas(r)$.
Again, the trends are similar to those in \autoref{fig:profile_gas_physics} (right).
At $r \gtrsim 2 \, \rthm$, beyond $\rinfall$ of dark matter, $\vradavegas(r)$ closely tracks $\vradavedark(r)$.
As stated above, the $r$ of this decoupling is set by $\rsplashback \approx \rinfall$ of dark matter, whereas radiative gas continues to cool and advect to smaller $r$.
(Though the stars, being collisionless, can experience splashback similar to dark matter.)
Gas reaches stronger $\vradavegas$ at smaller $\rinfall$, which corresponds to the minimum of $\fbaryon(r)$.
This behavior is governed by the efficiency of gas cooling and the onset of rotational support, and indeed, at the smallest $r$, $\vradavegas(r)$ moves toward 0, where gas becomes strongly supported by angular momentum (see below).
These trends for gas persist at all $z$, though as with dark matter, the inflow velocity at fixed $r$ becomes weaker at later times.

Finally, we emphasize that with the inclusion of gas, star formation, and feedback, halos at $\mthm < 10 ^ {12} \msun$ stop having physically meaningful infall regions for dark matter at $z < 1$.
Furthermore, while meaningful gas inflow does persist at all $r < r_{\rm ta}$ (the largest $r$ where $\vradavegas(r) < 0$), note that $r_{\rm ta}$ does not increase at $z < 1$.
In other words, at $z < 1$ the physical size of the region from which low-mass halos accrete gas no longer grows, because these halos cannot overcome both dark-energy acceleration (\autoref{eq:acceleration}) and large-scale tidal motions (see \autoref{sec:angular_momentum}).
In this sense, low-mass halos \textit{have} decoupled from the cosmic background, but (residual) gas infall persists at all $r < r_{\rm ta}$ because of radiative cooling.
This trend is similar in the simulation with only cooling.

\subsection{Angular-momentum support}
\label{sec:angular_momentum}

\renewcommand{\figurewidth}{1.0}
\begin{figure}
\centering
{\large Simulation with Star Formation \& Feedback} \\
\includegraphics[width = \figurewidth \columnwidth]{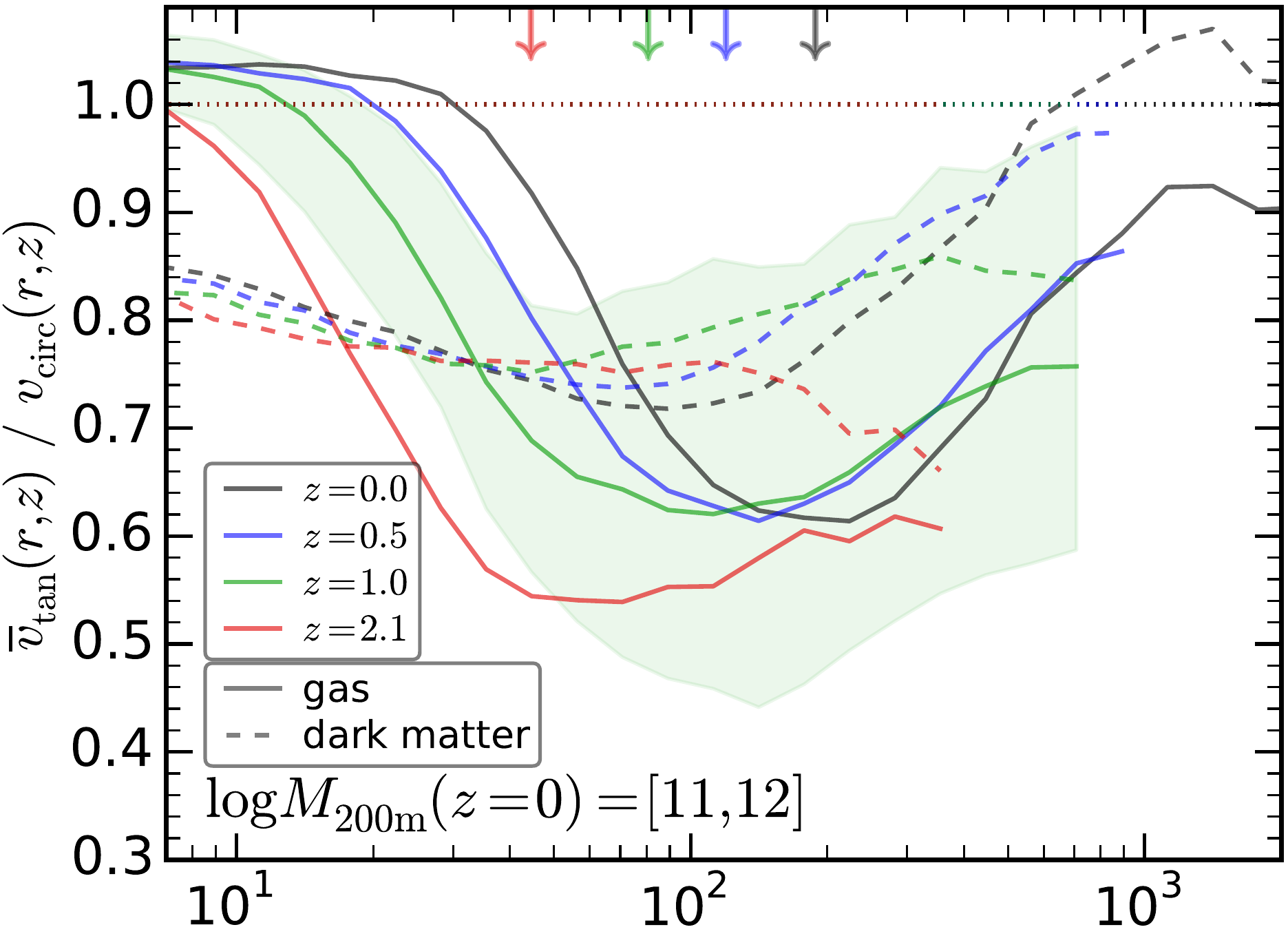}
\includegraphics[width = \figurewidth \columnwidth]{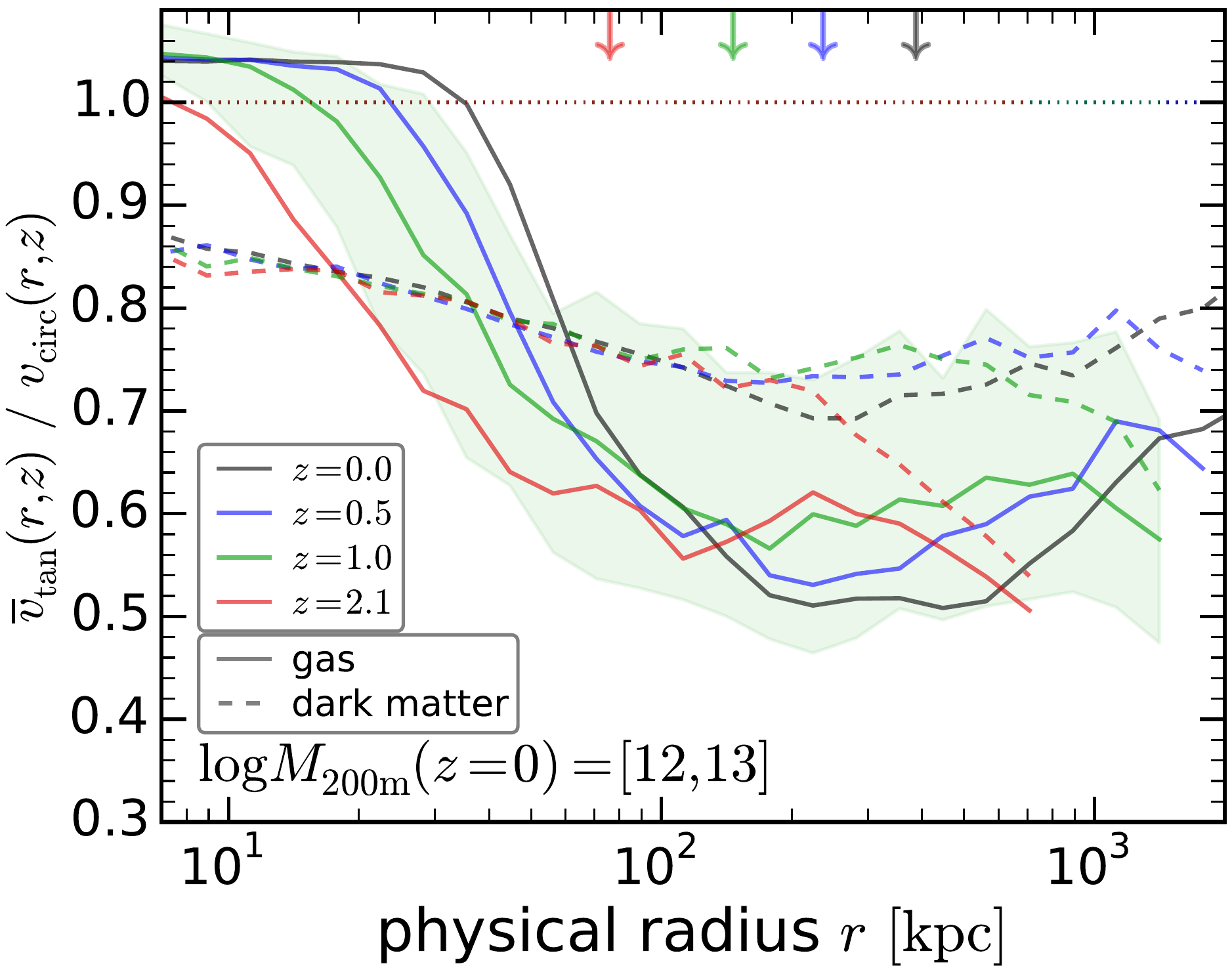}
\caption{
\textit{From the simulation with star formation and feedback}: profile of the ratio of the average tangential velocity, $\vtanave(r)$ to the circular velocity, $\vcirc(r) = \sqrt{G m_{\rm total}(< r) / r}$, which indicates the level of angular-momentum support, for baryons (gas + stars; solid) and dark matter (dashed), for isolated halos selected in two bins (top and bottom panels) of $\mthm$ at $z = 0$ and their progenitors.
Shaded region shows the 68\% scatter for baryons at $z = 1$, while arrows show the median $\rthm$ at each $z$.
Dark matter is not fully rotationally supported at any $r$, and the level is nearly constant over time.
By contrast, baryons are fully rotationally supported at a small $r$ ($< 40 \kpc$) that increases over time.
}
\label{fig:angular_momentum_profile_sf}
\end{figure}

For halos across the mass range that we examine, the cooling time of the gas is similar to or shorter than their dynamical times, especially at our low-mass end \citep{DekelBirnboim2006}, which naively implies that gas should be able to cool and advect efficiently to $r \sim 0$.
However, we have shown that the gas density profiles in even low-mass halos remain extended instead of accreting to $r \sim 0$ on a cooling timescale.
What then regulates the rate of gas accretion into the galaxy and causes the weaker infall velocity at small $r$ in \autoref{fig:profile_sf} (also in \autoref{fig:profile_gas_physics})?

The answer is that cosmic accretion plus gas cooling are not the only drivers of gas flow into the galaxy, but rather, angular-momentum support also plays a critical role.
To examine the level of angular-momentum support, we compute the \textit{magnitude} of tangential velocity (regardless of direction) for each cell/particle, and we sum this \textit{scalar} value to compute the average tangential velocity, $\vtanave(r)$, in bins of $r$.
We then scale $\vtanave(r)$ for each species to the circular velocity, $\vcirc(r) = \sqrt{G m_{\rm total}(< r) / r}$, for which $m_{\rm total}$ is the total mass within $r$.
\autoref{fig:angular_momentum_profile_sf} shows $\vtanave(r) / \vcirc(r)$ separately for baryons and dark matter.
Virialized orbits are rotationally supported by angular momentum if $\vtan(r) / \vcirc(r) = 1$.
\autoref{fig:angular_momentum_profile_sf} shows that dark matter is not fully rotationally supported at any $r$.
In addition, the level of rotational support within $\approx \rthm$ is nearly constant over time, analogous to the lack of evolution of $\rhodark(r)$.
By contrast, gas, which cools while largely conserving angular momentum, becomes strongly rotationally supported at small $r$ ($< 40 \kpc$), which moves outward over time.\footnote{
At the smallest $r$, $\vtan(r) / \vcirc(r) > 1$, because some gas and stars at the core are on high-energy unbound orbits for which $\vtan(r) > \vcirc(r)$.
Furthermore, some halos experience mergers and not fully relaxed, which leads to slight offsets in halo centering.}
Note that these $r$ are much larger than the galactic stellar disk, whose size is $\sim 0.01 \, \rthm$ \citep{Kravtsov2013}.
Thus, the rate of gas accretion into the disks of galaxies at late cosmic time is governed not just by cosmic accretion near $\rthm$, but rather by angular-momentum support and transport at $r < 40 \kpc$.

While the cooling of gas causes it to have a higher angular-momentum than dark matter at small $r$, note that at larger $r$, dark matter has systematically higher angular-momentum support than gas.
This is because gas experiences thermal pressure, mixing, and shocking that cause it to settle into a more coherent velocity field (stream), while dark matter, being collisionless, retains a higher dispersion in tangential velocity.
Indeed, we find that the scatter in $\vtan(r)$ for dark matter is higher than that of gas at all $r$ \citep[see also][]{Stewart2013}.

Other analyses \citep[e.g.,][]{Kimm2011, Stewart2013} found that gas accretes into a halo with systematically \textit{higher} angular momentum than dark matter.
These results seemingly contradict \autoref{fig:angular_momentum_profile_sf}, but the reason lies in the way that one computes/examines angular momentum.
\citet{Kimm2011} and \citet{Stewart2013} examined the net angular-momentum \textit{vector} of a halo, that is, they computed the angular-momentum vector of each particle, summed these vectors, and then took the magnitude of this sum.
By contrast, we compute the \textit{magnitude} of angular momentum for each particle/cell and sum this scalar.
Our choice is motivated by examining the angular-momentum support for each cell/particle, as opposed to the net angular momentum of the halo.

All of the above trends are similar in the simulation with only cooling.
However, in the non-radiative simulation, gas has less angular-momentum support than dark matter at all $r$, because gas shocking and pressure support reduce bulk orbital motions, and even gas with low angular momentum cannot cool and advect to small $r$.

\subsection{Physical Significance of $\rthm$}
\label{sec:virial_scaling}

\renewcommand{\figurewidth}{1.0}
\begin{figure*}
\centering
{\large Simulation with Star Formation \& Feedback} \\
\includegraphics[width = \figurewidth \columnwidth]{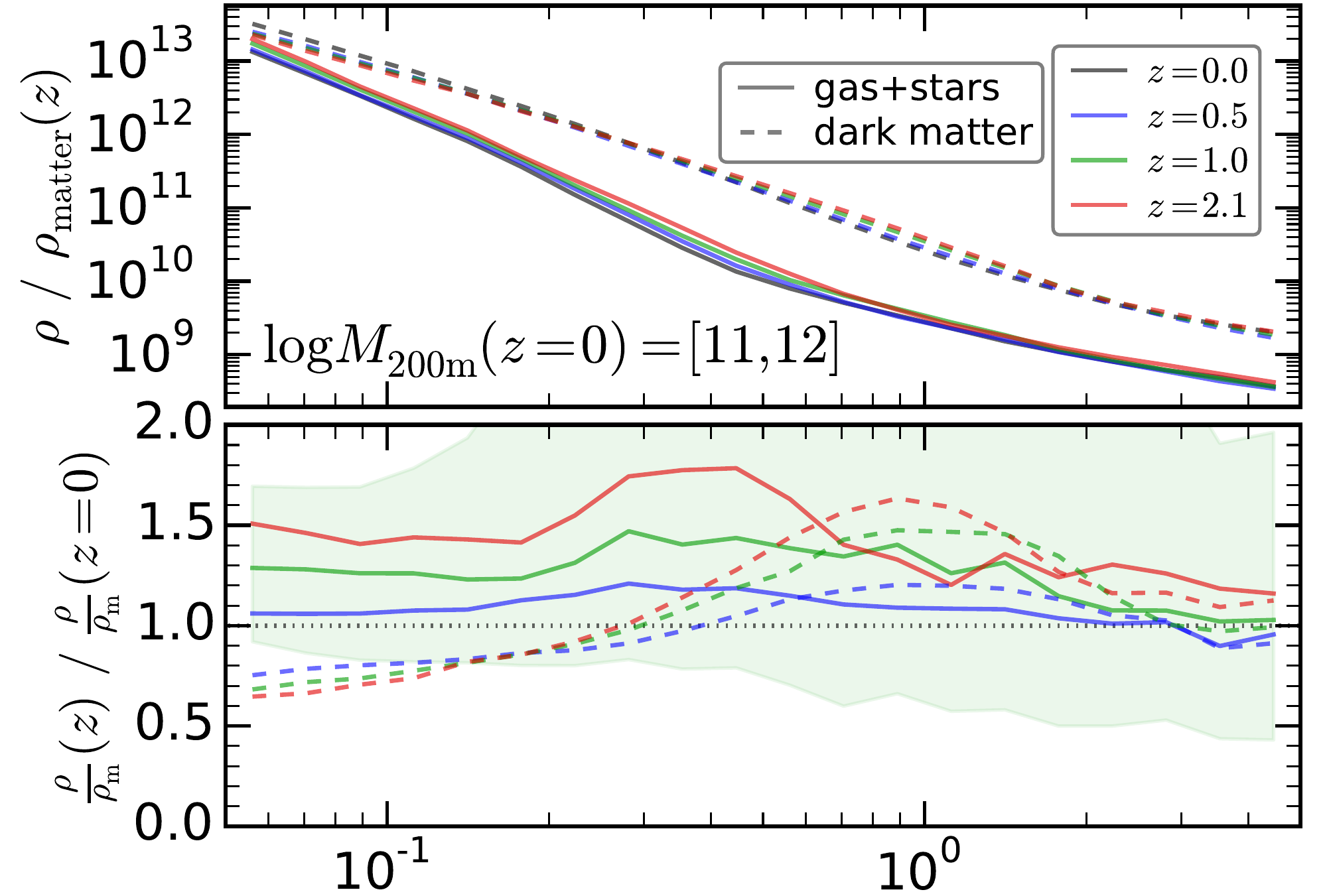}
\includegraphics[width = \figurewidth \columnwidth]{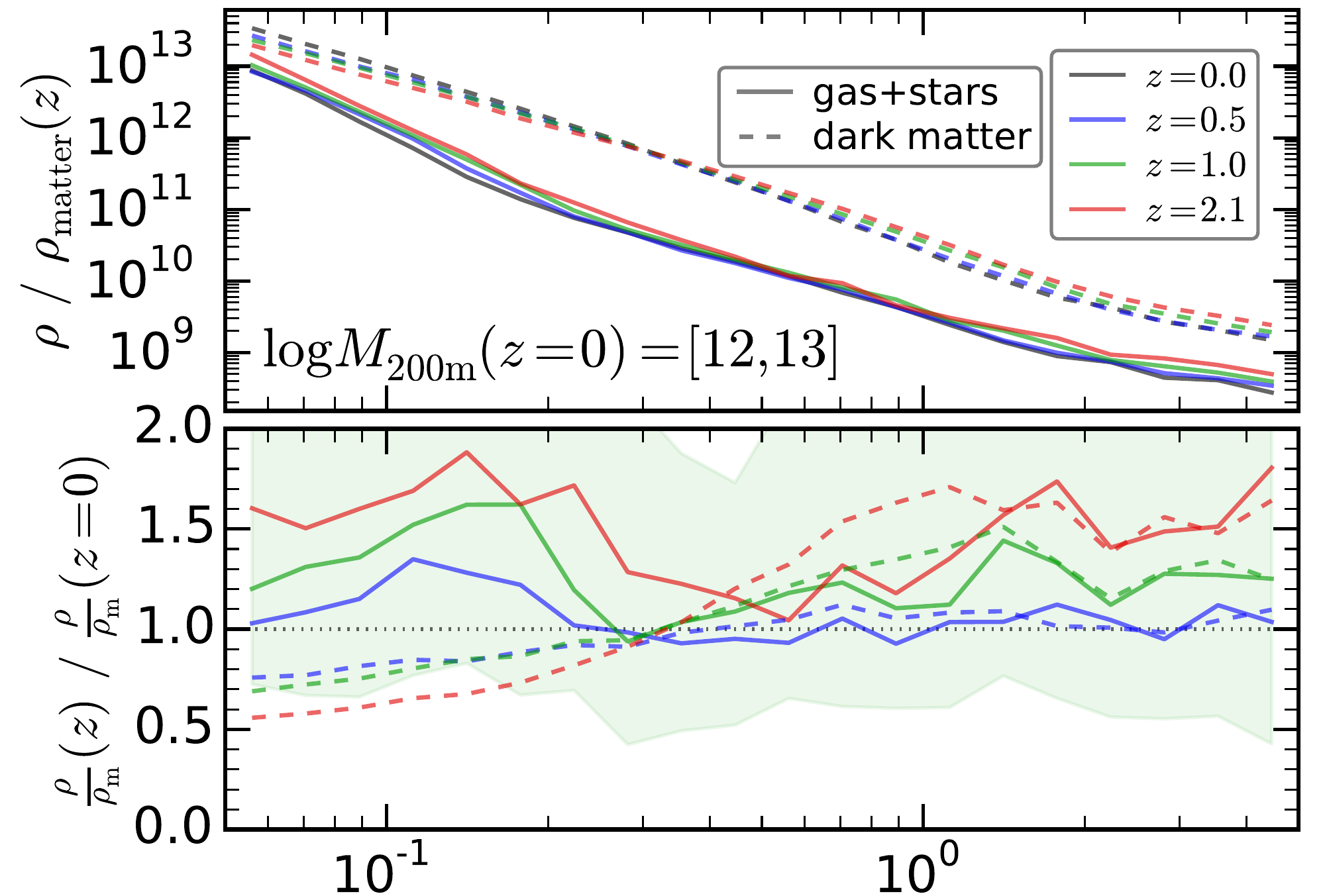}
\includegraphics[width = \figurewidth \columnwidth]{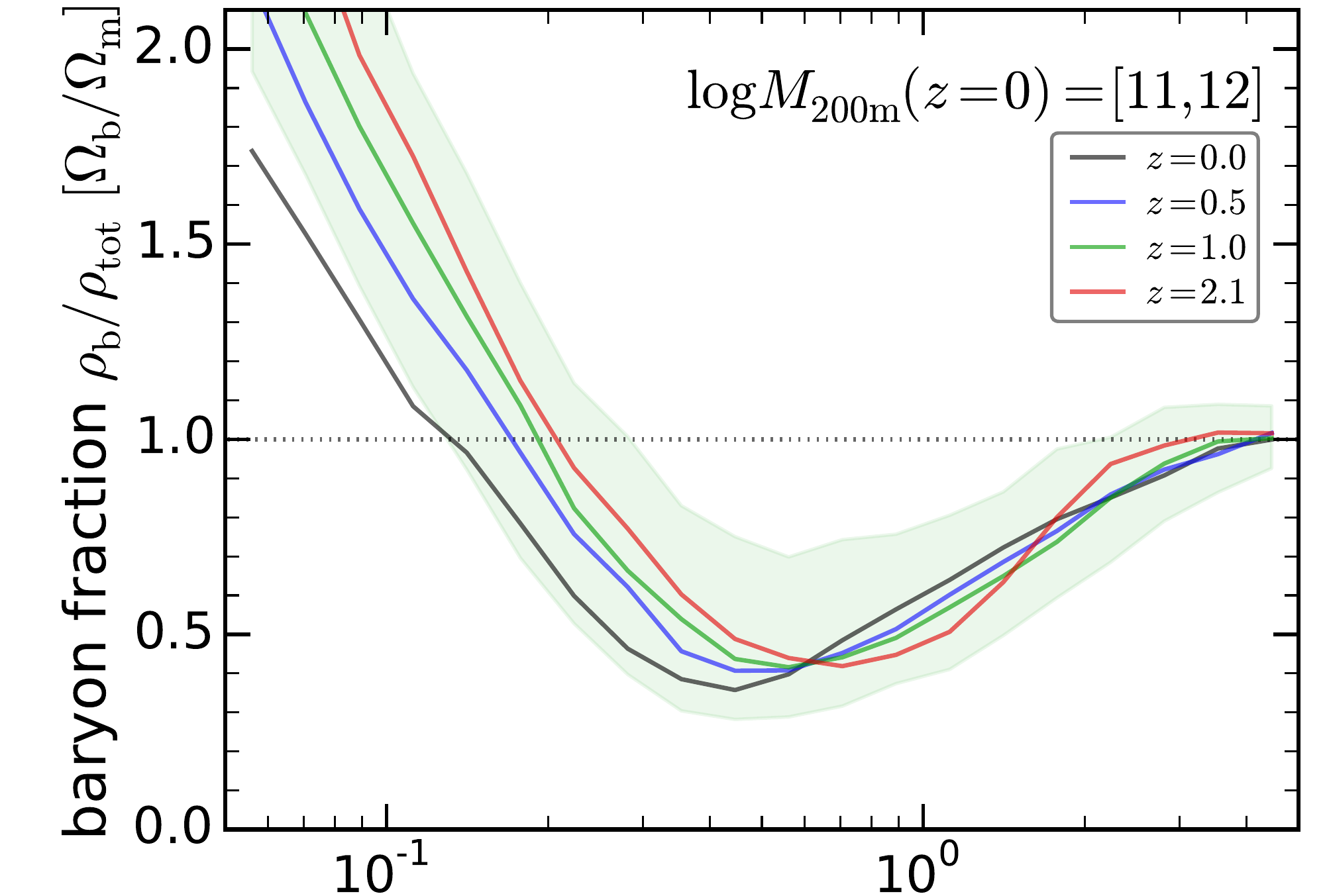}
\includegraphics[width = \figurewidth \columnwidth]{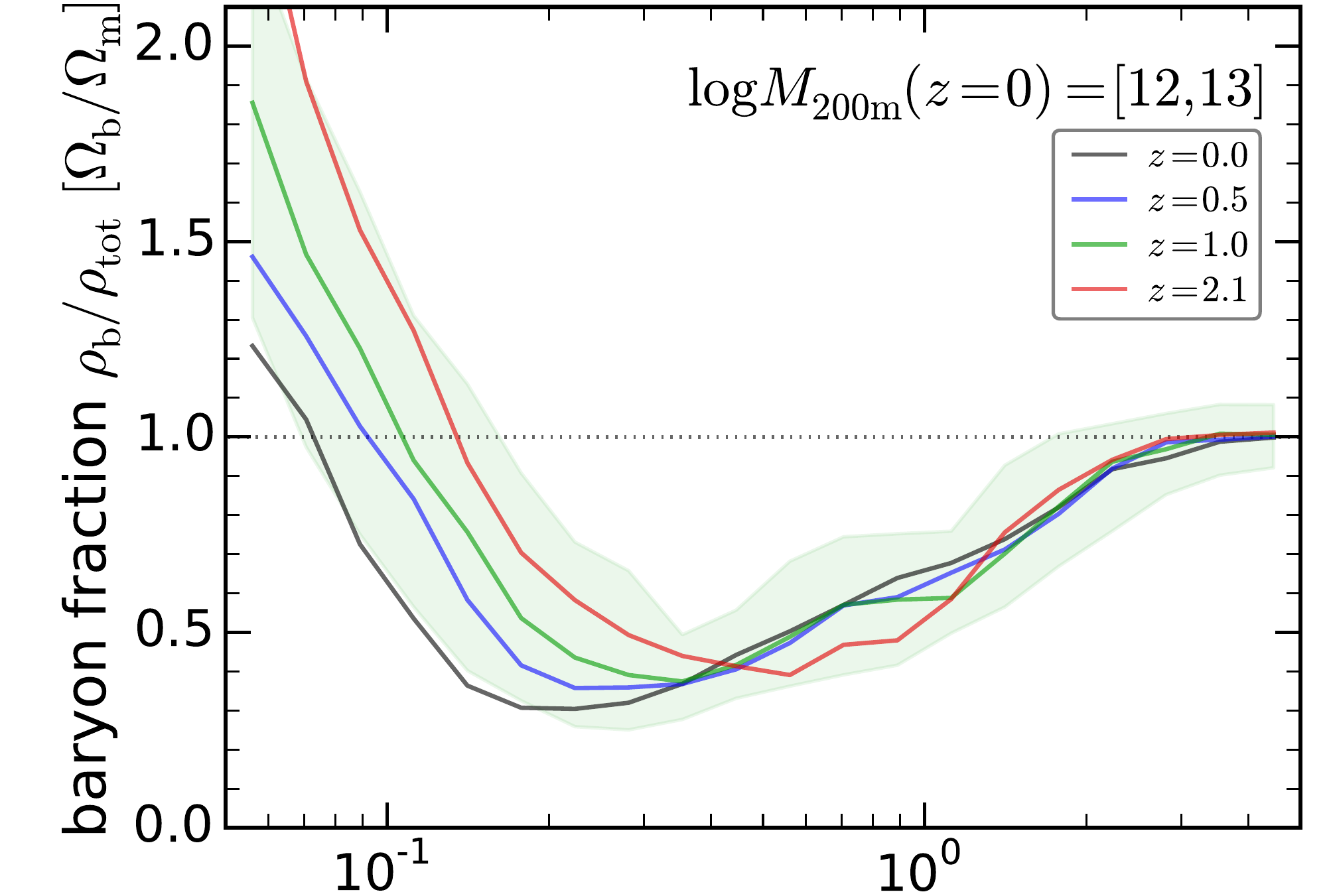}
\includegraphics[width = \figurewidth \columnwidth]{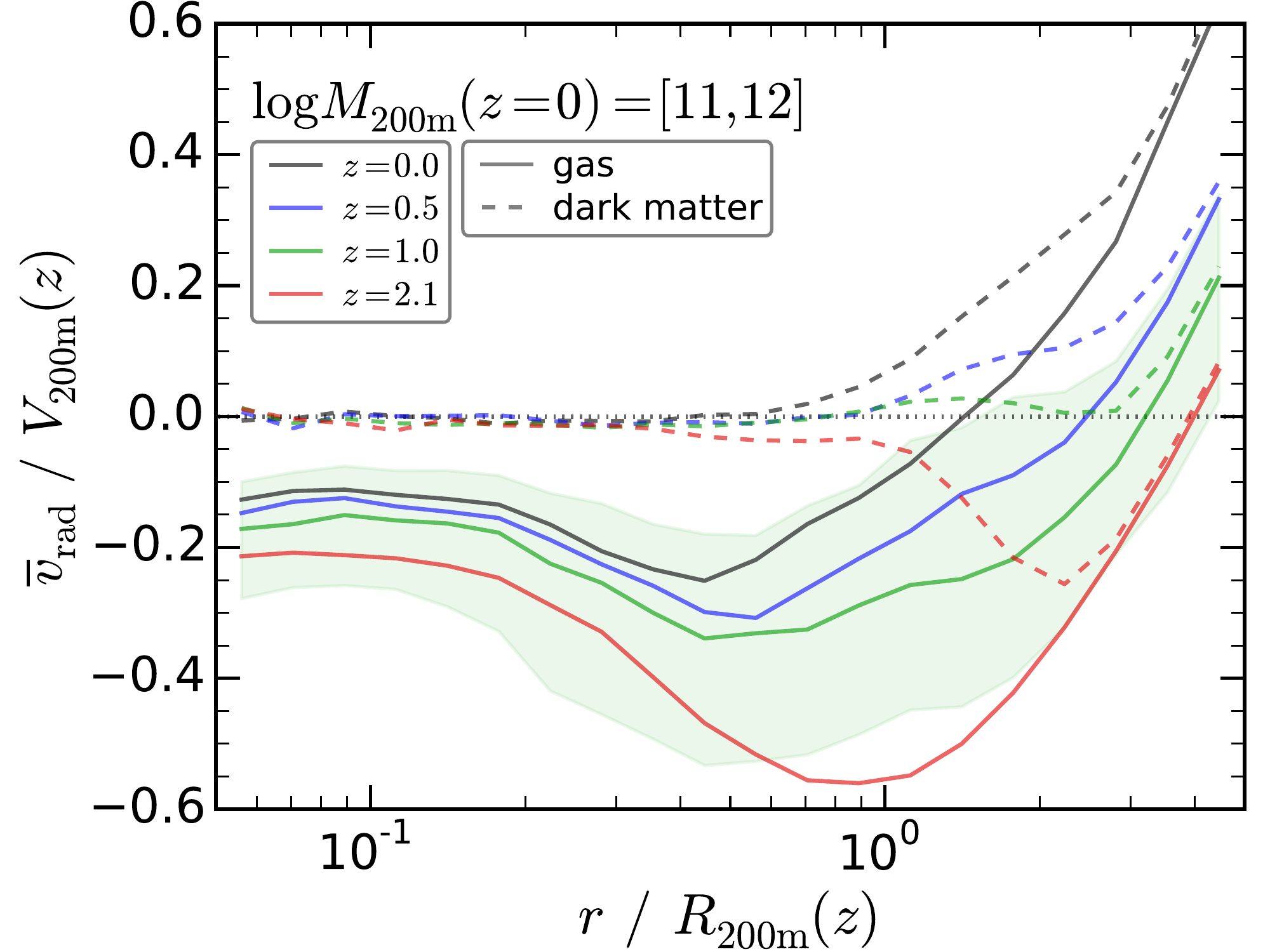}
\includegraphics[width = \figurewidth \columnwidth]{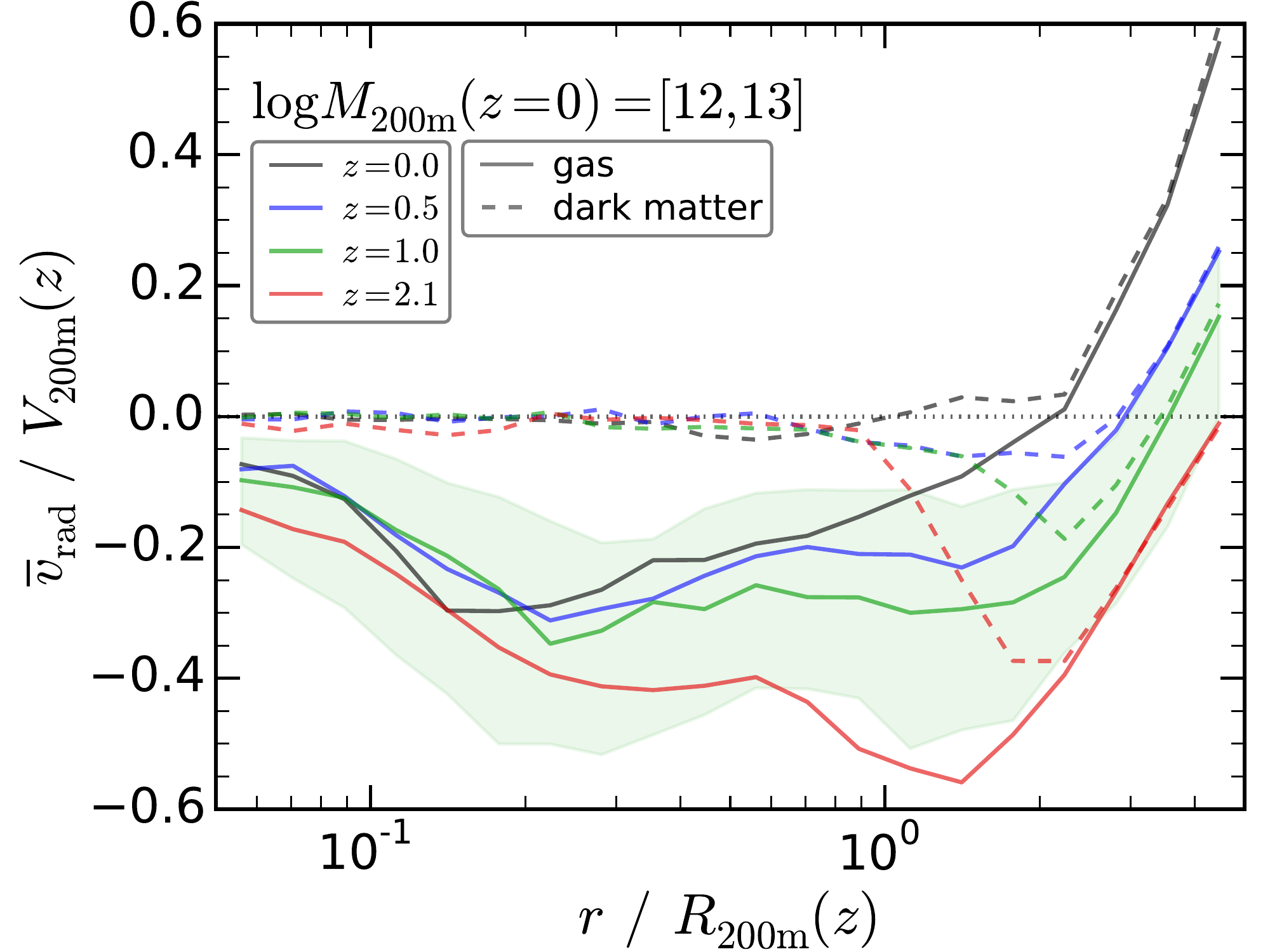}
\caption{
\textit{From the simulation with star formation and feedback}: same as \autoref{fig:profile_sf}, except that we scale $r$ by each halo's $\rthm(z)$.
Additionally, we scale $\rho(r / \rthm)$ to the cosmic $\rho_{\rm matter}(z)$ and $\vradave(r / \rthm(z))$ to each halo's $\vthm(z) = \sqrt{G \mthm(z) / \rthm(z)}$ to examine self-similarity.
For all quantities scaling profiles to $r / \rthm$ preserves a strong amount of self-similarity, meaning that the evolution, if not the absolute value, of $\rthm(z)$ captures the physical scales of cosmic accretion.
}
\label{fig:profile_virial_sf}
\end{figure*}

\renewcommand{\figurewidth}{1.0}
\begin{figure}
\centering
{\large Simulation with Star Formation \& Feedback} \\
\includegraphics[width = \figurewidth \columnwidth]{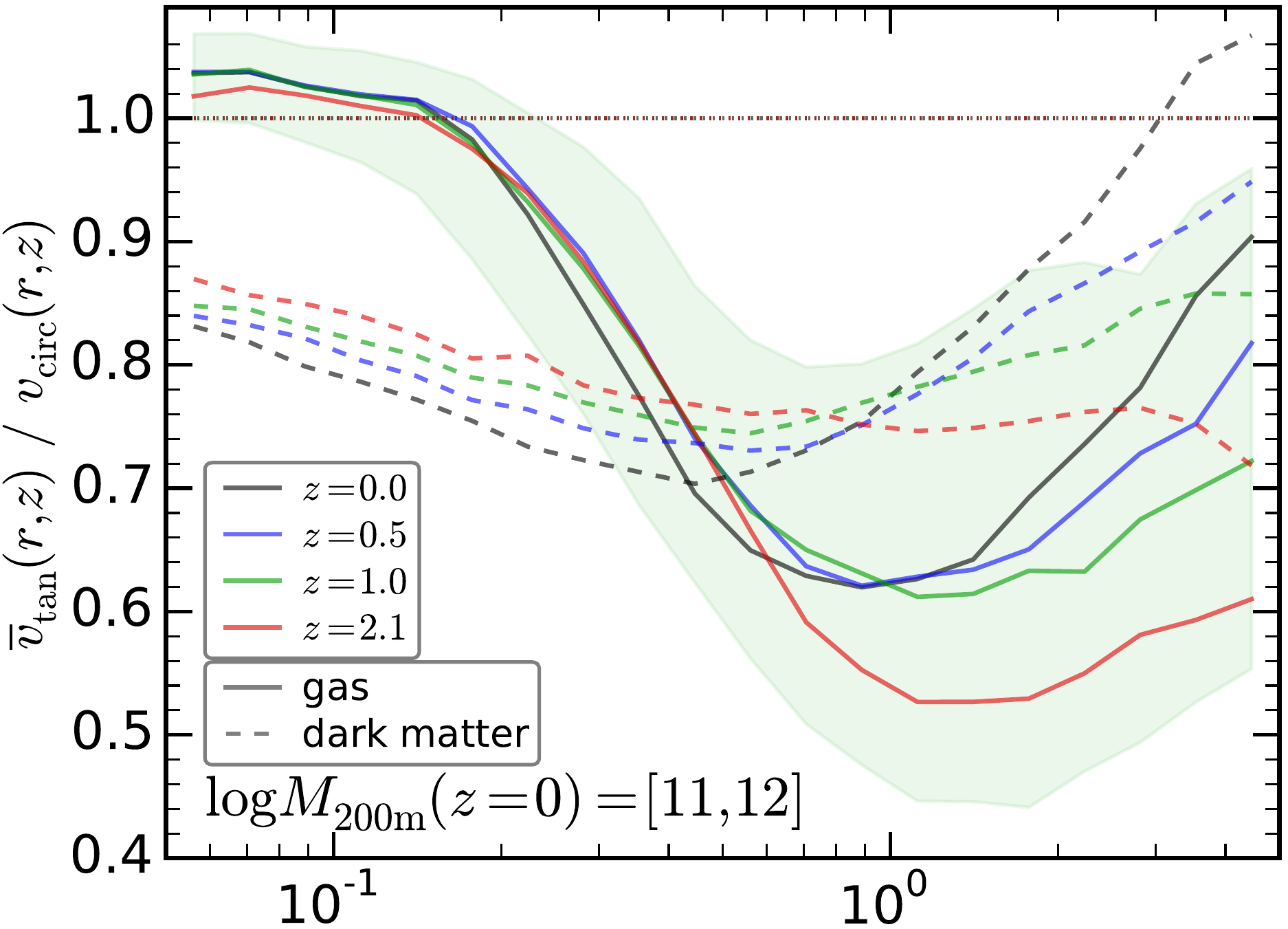}
\includegraphics[width = \figurewidth \columnwidth]{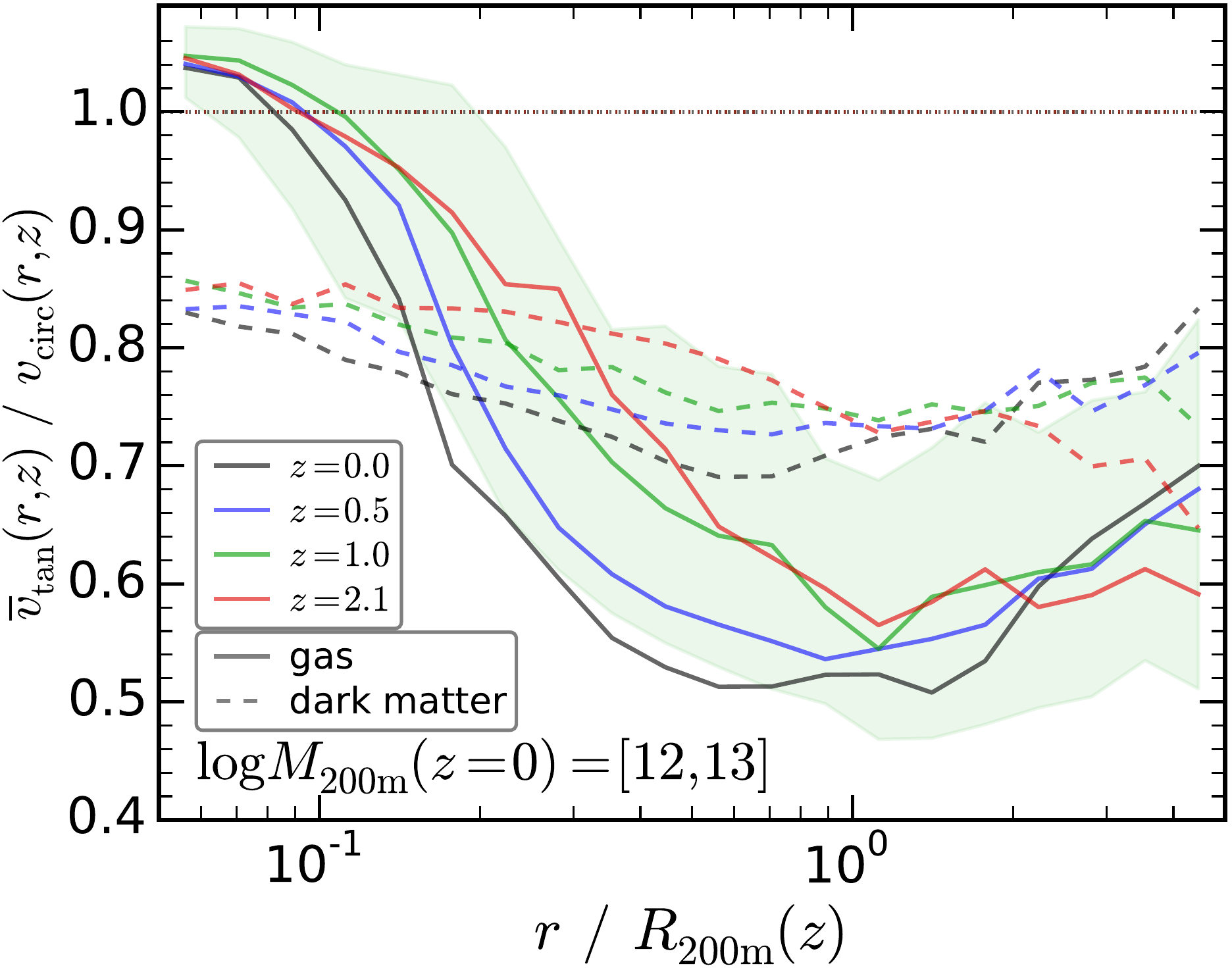}
\caption{
\textit{From the simulation with star formation and feedback}: same as \autoref{fig:angular_momentum_profile_sf}, except that we scale $r$ by each halo's $\rthm(z)$.
While this scaling breaks the self-similarity of the angular-momentum support of dark matter, it leads to a nearly self-similar onset of angular-momentum support for baryons at $r / \rthm = 0.1 - 0.17$, depending on mass.
Thus, the physical extent of rotationally supported gas responds to $\rthm(z)$.
}
\label{fig:angular_momentum_profile_virial_sf}
\end{figure}

Until now, we have shown that many features in the profiles of both baryons and dark matter are coincident with some fraction of a halo's $\rthm(z)$.
We now examine the relation to $\rthm(z)$ in more detail in order to understand to what extent it is a physically meaningful radius.
\autoref{fig:profile_virial_sf} shows the same profiles as in \autoref{fig:profile_sf}, but scaled by each halo's $\rthm(z)$.
While we are most interested in the degree to which this scaling to $r / \rthm$ captures physical features in the profile, we also scale $\rho(r)$ to the cosmic $\rho_{\rm matter}(z)$ and $\vradave(r)$ to each halo's $\vthm(z)$, to examine more clearly the degree of self-similarity in the profile.
Thus, our primary interest is whether any features in the profiles occur at a fixed $r / \rthm$, and our secondary interest is whether the profiles normalized by virial values are self-similar.

For both baryons and dark matter, $\rho(r / \rthm) / \rho_{\rm matter}(z)$ is self-similar to within a factor of $\sim 2$, with a somewhat stronger degree of self-similarity at $r \gtrsim \rthm$ than at smaller $r$ \citep[see also][for similar results for massive galaxy clusters]{Lau2015}.
Thus, $\rhodark$ at $r \lesssim 2 \, \rthm$ is more invariant over time at fixed $r$ than at fixed $r / \rthm(z)$, a result of the dissipationless nature of dark matter.
However, $\rhodark$ at $r \gtrsim 2 \, \rthm$ and $\rhogas$ at all $r$ are more invariant at fixed $r / \rthm(z)$, because they respond to continued physical cosmic accretion.

$\fbaryon$ shows similar invariance at fixed $r / \rthm(z)$, with deviations of $\sim 10\%$ at $r \gtrsim 0.5 \, \rthm(z)$.
$\fbaryon$ shows larger deviations at smaller $r$, where regulation by gas cooling and angular-momentum support are more important.
The $r$ of the minimum of $\fbaryon$ occurs at $r \approx (0.2 - 0.6) \rthm$, being smaller for higher-mass halos.
At $r$ beyond this minimum, $\fbaryon(r)$ clearly responds to $\rthm(z)$.

Considering $\vradave(r / \rthm) / \vthm(z)$, $\rinfall \approx 2 \, \rthm$ for dark matter.
Scaled to $r / \rthm$, this shows more clearly that $\rthm(z)$ roughly tracks the inner edge of the infall region for dark matter, where $\vradave = 0$.
For gas, the velocity profile shows less self-similarity, though in all cases $\rinfall = (0.2 - 1) \rthm$.
Neither $\vradavegas(r / \rthm) / \vthm(z)$ nor $\vradavedark(r  / \rthm) / \vthm(z)$ are particularly self-similar in terms of normalization, because the infall velocities of both components decrease over time, while we find that $\vthm$ for these halos is nearly static (it increases slightly with time as their potential deepens somewhat).
However, these scaled velocity profiles are somewhat more self-similar than the non-scaled (physical) profiles.

Finally, \autoref{fig:angular_momentum_profile_virial_sf} shows the scaled profile of angular-momentum support, that is, $\vtanave(r / \rthm) / \vcirc(r / \rthm)$.
Recall from \autoref{fig:angular_momentum_profile_sf} that the level of angular-momentum support for dark matter is nearly constant over time at fixed $r$ ($\lesssim 2 \, \rthm$), similar to $\rhodark$, while the level of angular-momentum support for gas grows significantly at fixed $r$, and thus the $r$ out to which gas experiences strong rotational support increases with time.
However, \autoref{fig:angular_momentum_profile_virial_sf} shows that radial extent of angular-momentum supported gas is fixed at $r / \rthm \approx 0.1$, that is, it responds to $\rthm(z)$.
(Similar results persist for our simulation with only radiative cooling.)
This is likely because, for a given halo, cosmic accretion at later times comes in with a higher impact parameter and thus a higher specific angular momentum \citep{Wetzel2011, Kimm2011, Pichon2011, Stewart2013}.
For dark matter, accreting mass is deposited (in a time-average sense) at large $r$, so it does not significantly change the angular momentum at small $r$.
However, as gas cools and advects to smaller $r$, it largely conserves angular momentum, so it continues to advect efficiently until it becomes strongly rotationally supported, which occurs at larger $r$ over time as the accretion at $\rthm(z)$ has higher specific angular momentum.

Thus, we conclude that, $\rthm(z)$ does have physical meaning, given its correlation with these profiles, for all properties except $\rhodark(r)$ at $r < 2 \, \rthm(z)$.
This does not mean necessarily that $\rthm(z)$ captures the exact location of any physically meaningful feature.
Indeed, the above figures show that the exact $r$ depends on the property in question.
Rather, the scaling of $\rthm$ with $z$ captures the \textit{relative} radius at which physical features occur.
We will pursue a more detailed investigation of various virial scalings in future work.

\subsection{Physical mass growth and accretion}

\renewcommand{\figurewidth}{1.0}
\begin{figure*}
\centering
{\large Simulation with Star Formation \& Feedback} \\
\includegraphics[width = \figurewidth \columnwidth]{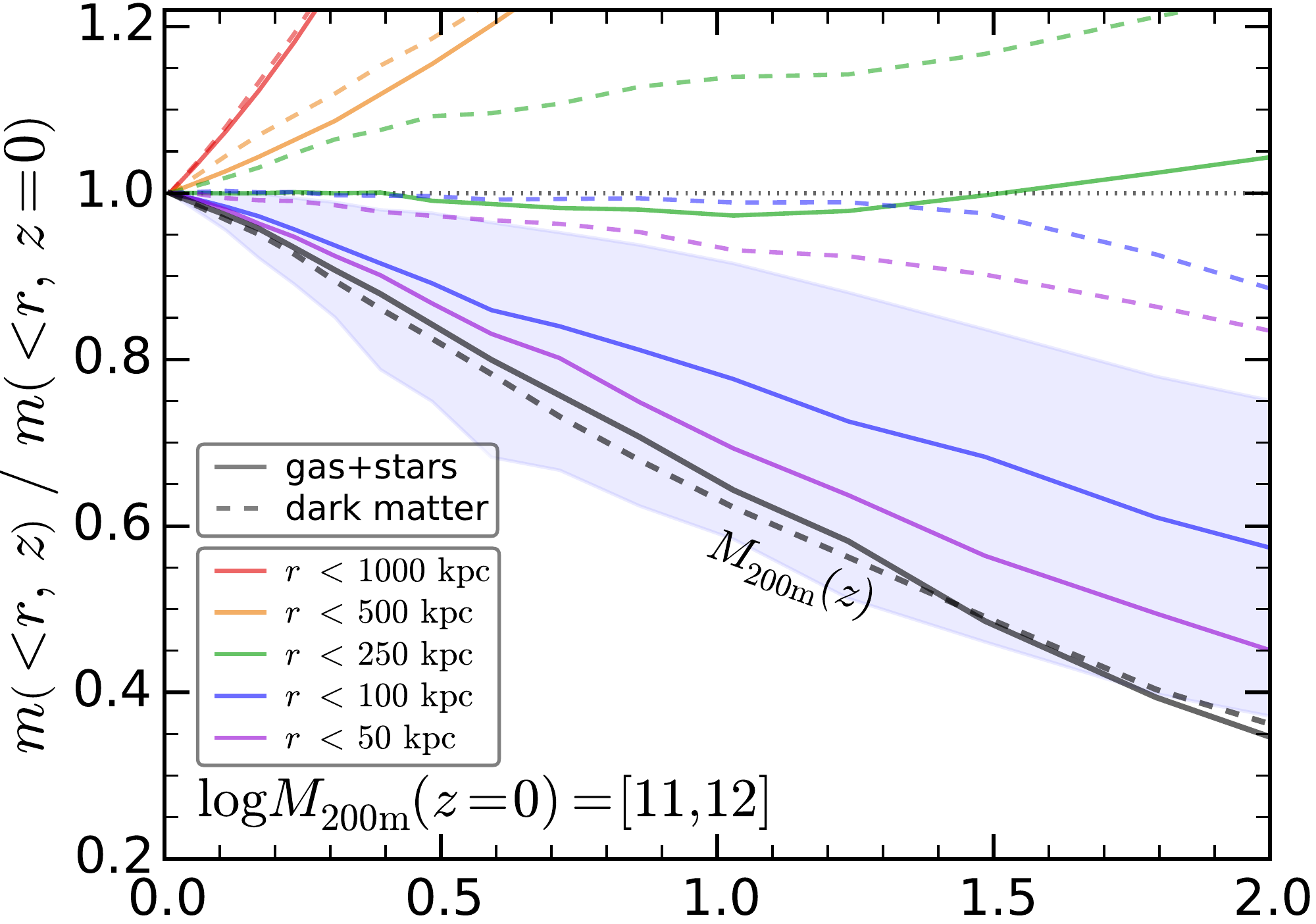}
\includegraphics[width = \figurewidth \columnwidth]{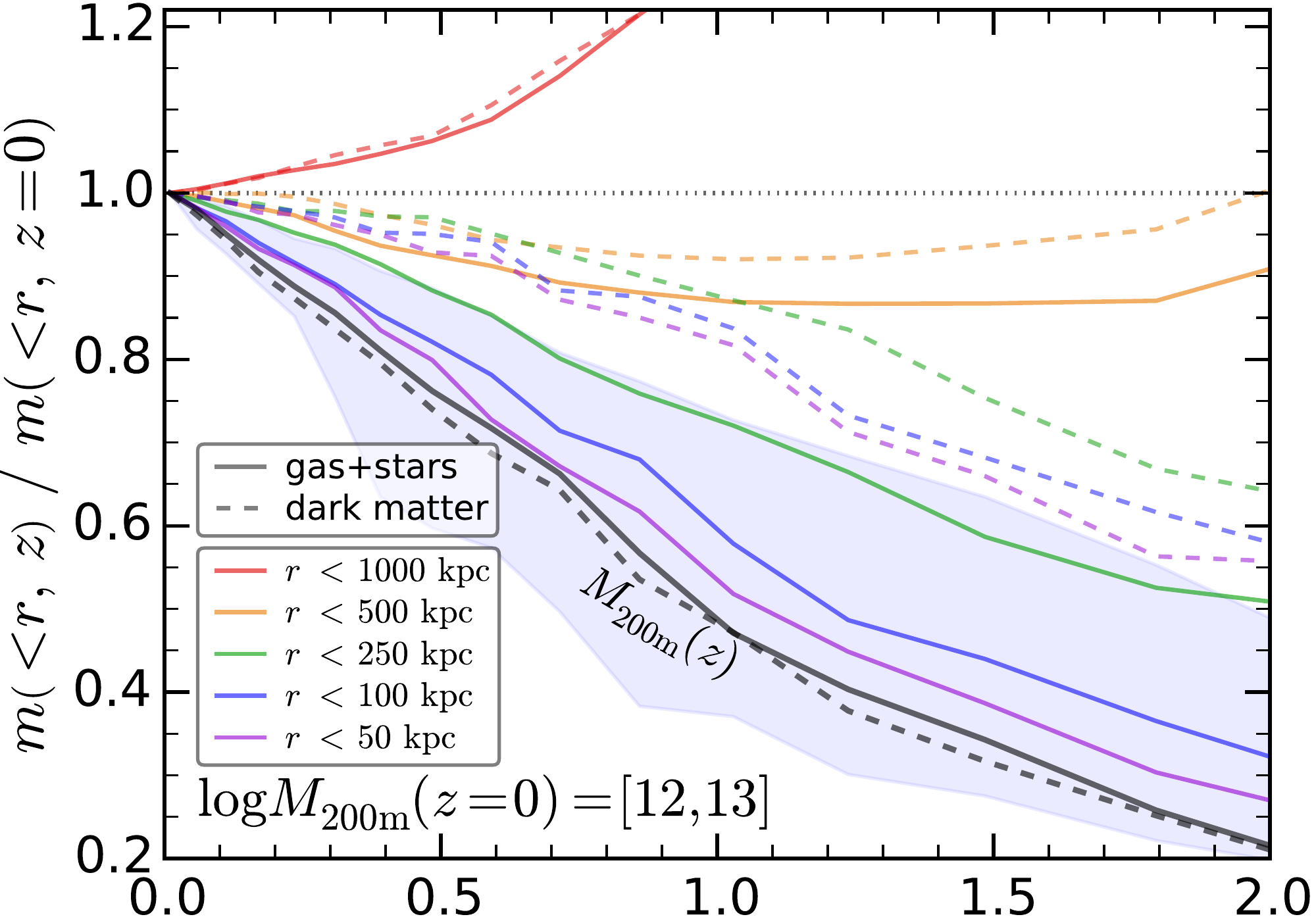}
\includegraphics[width = \figurewidth \columnwidth]{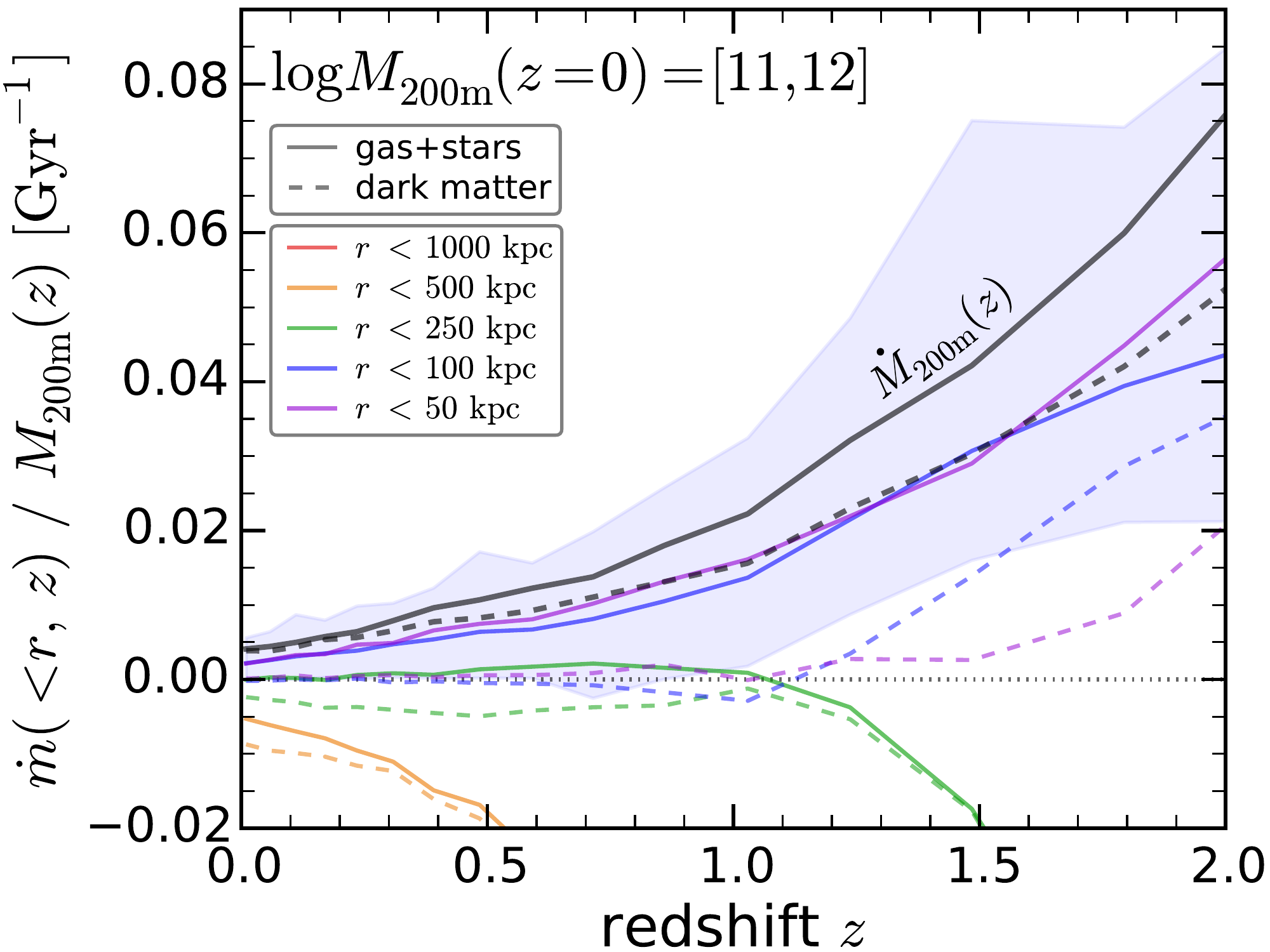}
\includegraphics[width = \figurewidth \columnwidth]{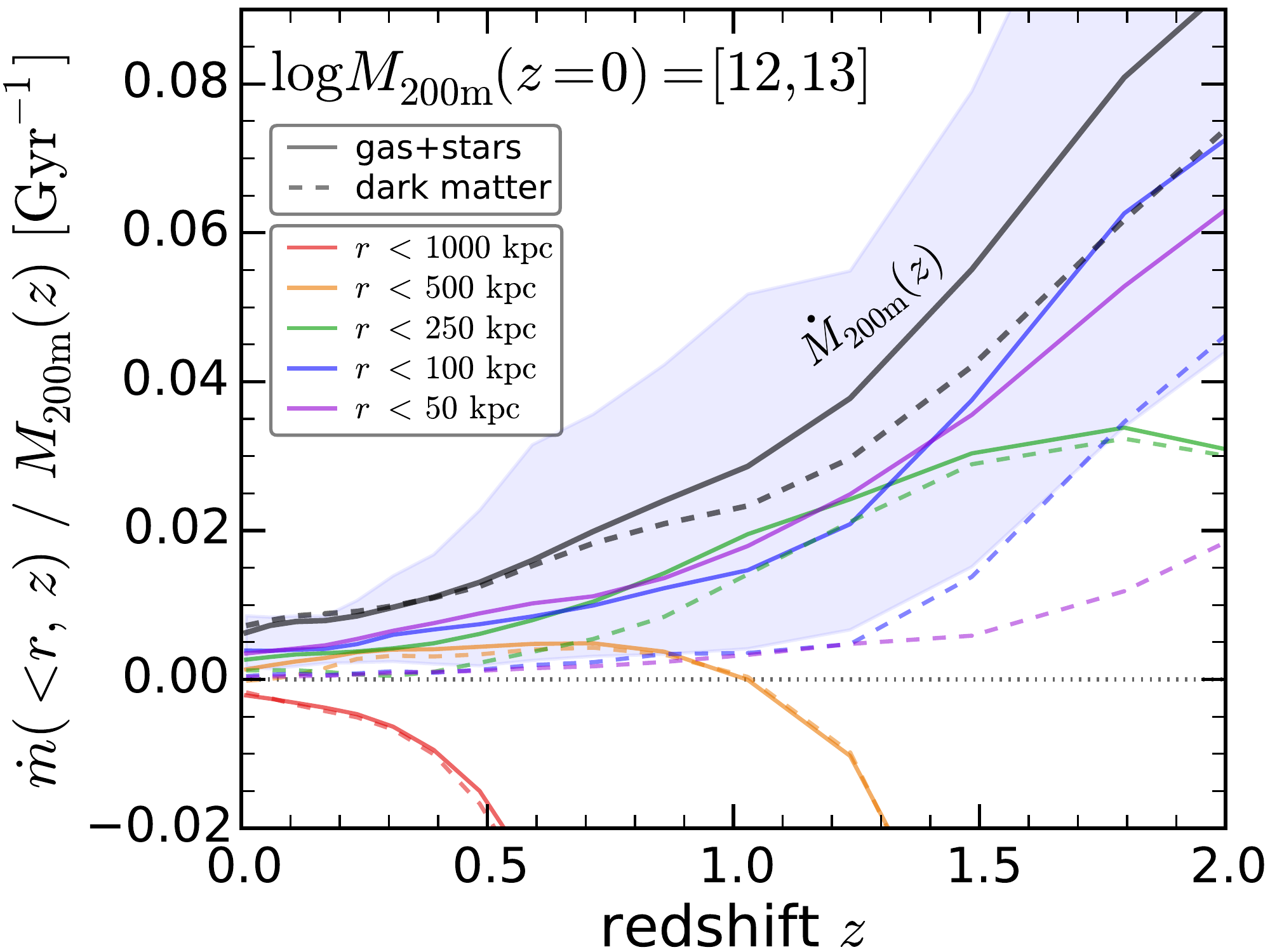}
\caption{
\textit{From the simulation with star formation and feedback}: mass growth and accretion rate histories at fixed physical radius, $r$, for halos selected in two bins of $\mthm$ at $z = 0$, for baryons (gas + stars; solid) and dark matter (dashed).
Shaded region shows the 68\% scatter for baryon mass at $r < 93 \kpc$.
Thick black curve shows the \textit{inferred} growth/accretion of each component as measured within an evolving $\mthm(z) = m(< \rthm(z))$.
For reference, the range of median $\rthm(z)$ for halo progenitors from $z = 0$ to 2 is $185 - 55$, $390 - 90 \kpc$ physical for $\mthm(z = 0) = 10 ^ {11 - 12}$, $10 ^ {12 - 13} \msun$, respectively.
\textbf{Top row}: mass of each component, $m$, within fixed $r$, normalized to $m(< r, z = 0)$, similar to \autoref{fig:mass_v_redshift_dm}.
At large $r$, the fractional growth (decline) of baryon mass closely tracks that of dark matter.
However, at $r \sim \rthm(z)$, baryonic mass growth starts to decouple from that of dark matter, and at the smallest $r$, the baryonic mass growth is significant and closely reflects $\mthm(z)$.
\textbf{Bottom row}: specific accretion rate, $\dot{m}(r, z) / \mthm(z)$, at fixed $r$.
For dark matter, we multiply by $\omegabaryon / \omegadarkmatter$ to compare with baryons more easily.
At a given $r$, the rate for dark matter increases, crosses zero (corresponding to turn-around), reaches a maximum (at $r \approx 2 \, \rthm$), and quickly declines to zero (static).
Baryons behave similarly at large $r$; after turn-around, their accretion rate remains significant at all smaller $r$ and $z$, and their evolution closely tracks that at $\rthm(z)$, though at an overall reduced rate.
}
\label{fig:mass_v_redshift_sf}
\end{figure*}

Finally, \autoref{fig:mass_v_redshift_sf} summarizes this section and the most important results of this paper by showing the mass growth and specific accretion rate over time at fixed $r$, similar to \autoref{fig:mass_v_redshift_dm}, but for both baryons and dark matter.

\autoref{fig:mass_v_redshift_sf} (top row) shows the median ratio $m(< r, z) / m(< r, z = 0)$ at various $r$ separately for both baryons (solid) and dark matter (dashed), for the same halos as in \autoref{fig:profile_sf}.
Scaling by $m(< r, z = 0)$ of each component allows easy comparison of baryons and dark matter at different $r$ in the same panel.
The panel also shows both $\mdark(< \rthm(z))$ and $\mbaryon(< \rthm(z))$ as dashed and solid thick black curves.

In the presence of star formation and feedback, the (lack of) growth of dark matter is qualitatively similar to \autoref{fig:mass_v_redshift_dm}.
At large $r$, $\mdark(< r)$ declines over time, following the expanding universe.
After a shell reaches turn-around, the enclosed mass grows over time, but this growth quickly stalls because of the dissipationless nature of dark matter.
Cosmic accretion deposits dark-matter mass at larger $r$ in a shell-like manner, with little growth of $\mdark(< r)$ at smaller $r$.

Baryon mass closely tracks that of dark matter at large $r$.
However, after turn-around, dissipative cooling causes baryon mass to decouple from dark matter, as $\mbaryon(< r)$ grows significantly over time at fixed $r$.
Even at the smallest $r$, the relative growth rate of baryon mass mimics the relative growth rate of $\mthm(z)$ (thick curves), so the accretion at $\rthm(z)$ regulates baryonic mass growth at smaller $r$.

For our lowest-mass bin, note the relative flatness of mass growth at $r = 200 \kpc$.
At $z \lesssim 1$, this $r$ corresponds to the turn-around radius, $r_{\rm ta}$, which as \autoref{fig:profile_sf} showed, does not grow for these low-mass halos.
\autoref{fig:mass_v_redshift_sf} shows more explicitly that $\mbaryon(< r = 200 \kpc)$ does not change at $z \lesssim 1$, nor will it change in the future, because it corresponds to the asymptotic baryon mass of these low-mass halos.

Comparing the absolute growth at different $r$ in \autoref{fig:mass_v_redshift_sf} (top row) is difficult, because we normalize growth at each $r$ by a different $m(< r, z = 0)$.
Thus, \autoref{fig:mass_v_redshift_sf} (bottom row) shows the specific accretion rate at fixed $r$.
Here, we measure the accretion rate by differencing $m(< r)$ at fixed $r$ across adjacent snapshots, and we show the specific accretion rate, $\dot{m}(r, z) / \mthm(z)$, to allow easier comparison across mass bins (the results do not change qualitatively if instead we examine the absolute accretion rate).
Finally, we multiply the specific accretion rate for dark matter by the cosmic $\omegabaryon / \omegadarkmatter$ to allow easy comparison.
Again, thick black curves show the specific accretion rates inferred via $\dot{M}_{\rm 200m}(z) / \mthm(z)$ for baryons (solid) and dark matter (dashed).

At large $r$, the specific accretion rate is negative, corresponding to uncollapsed mass that is expanding with the universe.
After turn-around, the specific accretion rate becomes positive.
At all $r < r_{\rm ta}$, including $\rthm(z)$, the rate declines over time, as caused by the declining infall velocity and density of accreting matter, as demonstrated in previous sections.
At fixed $r$, the specific accretion rate of dark matter declines over time and then stalls near zero.
Before stalling, the rate of decline approximately tracks that at $\rthm(z)$, so the mass flux at $\rthm(z)$ sets the accretion rate at a given $r$ for some time.
For baryons, after turn-around, at essentially all fixed $r$, the accretion rate tracks that at $\rthm(z)$, though lower by a factor of $\sim 2$, depending on mass.
Thus, the gas accretion rate at $\rthm(z)$ governs that at \textit{all} smaller $r$, though at a reduced rate, because gas cooling is not instantaneous, and gas experiences significant rotational support, especially at smaller $r$.

The accretion rate inferred from $\dot{M}_{\rm 200m}(z)$ is not quite synonymous for dark matter and baryons (dashed and solid thick black curves), because $\rthm$ represents a radius at which the dynamics of the two components start to decouple, as we demonstrated above.
(We checked that using a larger $r$ yields accretion rates that are more synonymous.)
For this reason, the better agreement of baryonic accretion rates at fixed $r$ with $\dot{M}_{\rm 200m}(z)$ of dark matter at $\mthm(z = 0) = 10 ^ {11 - 12} \msun$ is mostly a coincidence, though perhaps a fortuitous one, given that baryonic accretion rates often are inferred from dark-matter simulations via $\left( \omegabaryon / \omegamatter \right) \dot{M}_{\rm 200m}(z)$.

We conclude that the accretion rate and mass growth of both dark matter and baryons are governed by the cosmic accretion rate at $r \gtrsim \rthm(z)$.
However, at $r \lesssim \rthm(z)$, these components differ significantly.
The accretion rate and mass growth of dark matter stalls, because it is dissipationless.
Because gas cools, the baryonic accretion rate and mass growth continue down to all $r$, at a rate that tracks that at $\rthm(z)$, but with a lower absolute value because of the additional physics of finite cooling efficiency and angular-momentum support.

\section{Summary and Discussion}
\label{sec:summary_discussion}

\subsection{Summary of Results}
\label{sec:summary}

In this work, we examined the \textit{physical} nature of cosmic accretion and mass growth in halos with $\mthm(z = 0) = 10 ^ {11 - 14} \msun$.
While cosmic accretion and mass growth are typically measured using some evolving virial relation, such as $\mthm(z)$, we examined the \textit{physical} growth of both dark matter and baryons at fixed physical radii over time at $z < 2$, including the relationship of this growth to $\mthm(z)$.
We presented and analyzed a suite of cosmological simulations, {\it Omega25}, that incorporate both dark matter and gas dynamics with differing treatments of gas cooling, star formation, and thermal feedback (though absent strong stellar winds) to explore systematically the physics that governs the accretion of dark matter and baryons into halos and their galaxies.

We summarize our main results, first for the cosmic accretion of \textit{dark matter}:

\begin{enumerate}

\item \textit{Physically meaningful cosmic accretion occurs for dark matter at $z \gtrsim 1$ (depending on mass).}
Over time, the declining average density of matter in the universe, combined with the stronger dynamical effects of dark energy and large-scale tidal motions for accretion at larger physical radii, reduce the infall velocity of accreting matter and thus the accretion/growth rate.

\item \textit{At $z \lesssim 1$ (depending on mass), halos with $\mthm < 10 ^ {13} \msun$ do not experience significant physical accretion/growth of dark matter at \textit{any} radius.}
This is because dark matter is dissipationless, so it is deposited (in a time-average sense) at $r \gtrsim \rthm(z)$ in a shell-like manner, not within the halo.

\item \textit{The most physically meaningful radius to measure cosmic accretion and mass growth of dark matter is at the radius of maximum average infall velocity, $\rinfall$.}
Physically, $\rinfall \approx \rsplashback$, the splashback (or secondary turn-around) radius of  (collisionless) dark matter.
For halos in our mass range, $\rinfall \approx 2 \, \rthm$, though there is no meaningful $\rinfall$ for low-mass halos at $z \lesssim 1$.
While commonly used, $\rthm$ represents an incomplete census of the mass that has passed through the halo, though $\rthm$ approximately corresponds to the outer edge of where the profile is most static.

\end{enumerate}

Additionally, we summarize our main results for the cosmic accretion of \textit{baryons}:

\begin{enumerate}

\item \textit{Physically meaningful cosmic accretion and mass growth of baryons persists at all radii and across all redshifts.}
The difference between dark matter and gas arises not because gas is collisional, but rather because gas can cool radiatively and advect to smaller radii.
While dark-matter growth is subject to pseudo-evolution, baryon growth is not.

\item \textit{The physical accretion rate of baryons at all radii inside of the halo roughly tracks the accretion rate into the halo measured at $r \gtrsim \rthm(z)$.}
Though the rate at smaller radii has a somewhat lower normalization as governed by the efficiency of gas cooling and angular-momentum support.

\item \textit{Accreting gas becomes strongly rotationally supported at $r \approx 0.1 \, \rthm$, independent of redshift.}
Thus, the rate of gas inflow into the galaxy is regulated by angular-momentum support.

\item \textit{Inflowing gas starts to decouple from dark matter at $\rsplashback > \rthm$}.
For halos in our mass range, $\rsplashback \approx 2 \, \rthm$.
This is the smallest radius where the specific accretion rates are the same for both baryons and dark matter.

\item \textit{The physical size of the region that sources cosmic accretion into low-mass halos ($\mthm < 10 ^ {12} \msun$) does not grow at $z < 1$.}
Matter decouples from the cosmic expansion and starts to fall into the halo at the turn-around radius, $r_{\rm ta} > \rthm$, but this radius stops increasing for low-mass halos at $z < 1$.
In this sense, low-mass halos have decoupled from the cosmic background for most of their history, but physically meaningful infall of gas to smaller $r$ does persists because of gas cooling.

\end{enumerate}

\subsection{Discussion}
\label{sec:discussion}

\subsubsection{Physical significance of the virial radius}

We have shown that scaling profiles of dark matter and baryons to $r / \rthm(z)$ preserves a strong degree of self-similarity in $\rho$, $\fbaryon$, $\vradave$, and $\vtanave(r) / \vcirc(r)$.
While this argues for the meaningfulness of the scaling of $\rthm$ with $z$, this does not argue necessarily for the absolute value of $\rthm(z)$.
Our results, and those of other works \citep{Busha2005, AnderhaldenDiemand2011, DiemerKravtsov2014, Adhikari2014, Lau2015}, indicate that commonly used virial radii, such as $\rthm$, represent an incomplete census of the physical mass associated with halos.
Furthermore, other commonly used definitions, such as $\rthc$ and $\rvir$ \citep{BryanNorman1998} are even smaller than $\rthm$.

Given that dark matter and baryons are continuous density fields, that cosmic accretion is triaxial and clumpy, and that halos are not fully relaxed ``virialized'' systems, any definition of a virial radius/boundary for a halo is at some level an oversimplification.
That said, our results suggest that $\rsplashback \approx 2 \, \rthm$ is the most physically meaningful radius to measure the mass growth and accretion of dark matter for halos in our mass range.
Additionally, beyond just the splashback of dark matter, this radius also corresponds to where a significant fraction of (observable) galaxies around massive groups/clusters are splashback satellite galaxies that passed through the host \citep[for example,][]{Wetzel2014}.

Moreover, $\rsplashback \approx 2 \, \rthm$ is the most natural radius for comparing dark matter with baryons because this is where their average infall dynamics starts to decouple.
For baryons, their collisional and dissipational dynamics mean that there is no shell crossing (with the exception of any gas that remains bound in massive subhalos), therefore there is no analogous $\rsplashback$.
In sufficiently high-mass halos, gas can experience an analogous virial-shock radius, though this can occur at a range of radii with respect to $\rthm$, depending on mass and $z$ \citep{DekelBirnboim2006}.

Thus, we suggest that $r \approx 2 \, \rthm$ has many advantages over $\rthm$ (or even smaller radii) for measuring the physically meaningful extent, and thus the cosmic accretion rate and mass growth, of halos in our mass range.
Nevertheless, the exact value of $\rsplashback$ and thus $\rinfall$, with respect to $\rthm$, depends on the specific accretion rate and therefore on halo mass and $z$, such that $\rsplashback$ and $\rinfall$ occur more typically at $r \sim 1.5 \, \rthm$ in massive galaxy clusters \citep{DiemerKravtsov2014, Adhikari2014, Lau2015}.
In a follow-up analysis, we will examine a broader range of virial definitions to examine which best capture physical scales of density, velocity, and thermal profiles for dark matter and gas, including dependence across a more comprehensive range of halo mass and $z$.

\subsubsection{Implications for the growth and size of galaxies}

As discussed in the introduction, galaxies obey tight scaling relations with their host halos, including both mass and size.
This tightness over cosmic time may seem surprising in the context of the pseudo-evolution of halo mass, given that dark-matter mass and density do not change at small $r$, on scales of the galaxy.
However, our results provide some insights into this relation.
Baryons continue to cool and advect to scales of the galaxy at a rate that roughly tracks the accretion rate at $\rthm(z)$.
Thus, we expect that $\mthm(z) = m(< \rthm(z))$ \textit{approximately} captures the collapsed mass of the halo, which also represents the baryons available to feed the galaxy.
In detail, our results imply that $m(< 2 \, \rthm(z))$ may provide a better correlation with the properties of the galaxy.
If true, this means that galaxy growth at $z < 2$ is at some level limited and/or regulated by cosmic accretion at $r \approx \rthm$, despite the added complexity of feedback and wind recycling.
This picture is supported at least indirectly by the similarity of the decline in halo accretion rates and galaxy star-formation rates at $z < 2$ \citep[for example,][]{Bouche2010, Lilly2013}, though it requires more investigation in the context of simulations with more detailed and physically motivated feedback models.

While we have shown that the accretion/growth rates of dark matter and baryons are decoupled at small $r$ within a halo, our results do support the methodology of using accretion rates from dark-matter simulations to estimate baryonic accretion rates into halos, provided that this is measured at $r \gtrsim \rthm$, or more optimally, at $r \approx 2 \, \rthm$.
Furthermore, while the accretion rates are decoupled at small $r$, they may still correlate with each other, in the sense that the halos with the highest accretion rates of dark matter may also have the highest rates for baryons.
If true, this may shed light on recent ``age matching'' models that assume a tight correlation between star formation in galaxies at $z \sim 0$ and the formation timescales (effectively measured at high $z$) of their host halos \citep{HearinWatson2013, Watson2015}.
In future work, we will examine such correlations between rates of baryons and dark matter across cosmic time in more detail.

Finally, our result that gas typically becomes strongly rotationally supported at $r \approx 0.1 \, \rthm$, independent of redshift, has interesting implications for galaxy size growth.
This suggests that galactic disks are fed by torquing or viscous advection from much more extended pseudo-disks of rotationally supported gas.
Indeed, several analyses of cosmological zoom-in simulations of individual $\sim L_*$ galaxies found that they had extended thick and/or warped disks of cool accreting gas out to $\sim 50 \kpc$ and beyond \citep{Agertz2009, Roskar2010, Stewart2011, Danovich2015}.
Furthermore, observations of nearby galaxies show that disks of cold atomic gas often extend many times beyond the size of the stellar disk \citep[for example,][]{Walter2008}.
If the resultant size of the galactic disk is governed by the size of the extended gaseous pseudo-disk, then this could explain the tight correlation between the observed size of a galaxy and the inferred size of its host halo as noted in \citet{Kravtsov2013}.
Nevertheless, we emphasize that the size of the stellar disk is much smaller at $r \approx 0.01 \, \rthm$, so such a connection requires more detailed analysis using higher-resolution simulations.

\subsubsection{Impact of stronger feedback}

Our results do not change substantially in comparing simulations with only cooling to those with star formation and thermal feedback.
However, we do not include more detailed radiative and/or momentum feedback, so our simulations do not drive strong galactic outflows that are encapsulated in recent high-resolution simulations with more detailed treatments \citep[for example,][]{Hopkins2014a}.
In principle, strong stellar winds could suppress gas accretion rates below what we find in this work.
\citet{vandeVoort2011} examined cosmic accretion rates into halos in cosmological simulations using models for stellar and black-hole feedback in which winds are driven at a variety of fixed values for gas particle velocities, finding that the specific accretion rates into halos were broadly similar for baryons and dark matter, and that their feedback models change baryonic accretion rates only slightly for halos $\gtrsim 10 ^ {11} \msun$, though feedback yielded stronger reduction in gas accretion rates at smaller $r$.
Using a different implementation of fixed-velocity wind models in a cosmological simulation, \citet{FaucherGiguere2011} found that winds can reduce the rates of infalling gas, particularly in low-mass halos, by up to a factor of 2 for their strongest wind model, again with stronger reduction at smaller $r$.
More recently, \citet{Woods2014} used a combination of delayed cooling supernova and early stellar feedback in a handful of cosmological zoom-in halos of mass $\sim 10 ^ {12} \msun$ at $z = 0$, finding that gas accretion rates into halos did not change with their feedback implementation.
Similarly, \citet{Nelson2015} examined gas accretion rates in cosmological simulations that generate galactic winds at a velocity that scales with the local velocity dispersion of dark matter, with additional thermal and ionizing feedback from black holes, and found that feedback reduced gas accretion rates only slightly at $\rvir$ as compared with no feedback, but that feedback increased the gas inflow rates at $0.25 \, \rvir$ as a result recycling from the winds.

Overall, these works suggest that galactic winds do not substantially affect baryonic accretion rates at and beyond $\rthm$ but that they do drive gas recycling that modulates inflow/outflow rates near the core of the halo.
Furthermore, it is not clear how much galactic winds, launched from the galaxy with a low impact parameter and thus a low angular momentum, would change the angular-momentum distribution of extended halo gas.
However, we emphasize that almost all previous works examined winds with phenomenologically tuned prescriptions for wind velocities and its coupling (or lack thereof) with the surrounding halo gas.
Such analyses should be revisited with more realistic and comprehensive treatments of stellar feedback \citep{Muratov2015}.

\section*{Acknowledgments}

We thank Andrey Kravtsov, Surhud More, Phil Hopkins, Frank van den Bosch, Andrew Hearin, Erik Tollerud, Erwin Lau, and Benedikt Diemer for useful discussions and/or comments on an early draft.
We also thank the reviewer for useful comments.
A.R.W. acknowledges the hospitality and stimulating environment of the Aspen Center for Physics, supported in part by the NSF.
A.R.W. also gratefully acknowledges support from the Moore Center for Theoretical Cosmology and Physics at Caltech.
This work was supported in part by NSF grants AST-1412768 \& 1009811, NASA ATP grant NNX11AE07G, NASA Chandra grants GO213004B and TM4-15007X, the Research Corporation, and by the facilities and staff of the Yale University Faculty of Arts and Sciences High Performance Computing Center.


\end{document}